\documentclass[twocolumn,showpacs,amsfonts,aps,prc,floatfix,superscriptaddress]{revtex4-1}

\usepackage[colorlinks=true, pdfstartview=FitV, linkcolor=red, citecolor=blue, urlcolor=blue]{hyperref}
\usepackage{amsmath}
\usepackage{bm}
\usepackage{epsfig}
\usepackage{color}
\usepackage{epstopdf}
\usepackage{graphicx}
\usepackage[normalem]{ulem}

\begin{document}

\title{
Equation of state at finite densities for QCD matter in nuclear collisions
}

\author{Akihiko Monnai}
\affiliation{KEK Theory Center, Institute of Particle and Nuclear Studies, \\
High Energy Accelerator Research Organization, Tsukuba, Ibaraki 305-0801, Japan}

\author{Bj\"orn Schenke}
\affiliation{Physics Department, Brookhaven National Laboratory, Upton, NY 11973, USA}

\author{Chun Shen}
\affiliation{Department of Physics and Astronomy, Wayne State University, Detroit, Michigan, USA}
\affiliation{RIKEN BNL Research Center, Brookhaven National Laboratory, Upton, NY 11973, USA}

\date{\today}

\begin{abstract}
We construct the QCD equation of state at finite chemical potentials including net baryon, electric charge, and strangeness, based on the conserved charge susceptibilities determined from lattice QCD simulations and the equation of state of the hadron resonance gas model. For the application to relativistic heavy ion collisions we consider the situation of strangeness neutrality and matter with a fixed electric charge-to-baryon ratio, resembling that of heavy nuclei. The importance of finite electric charge and strangeness chemical potentials for particle production in heavy ion collisions is demonstrated using hydrodynamic simulations.
\end{abstract}


\maketitle

\section{Introduction}
\label{sec1}

The nearly perfect fluidity of the quantum chromo dynamic (QCD) matter discovered in heavy ion collisions at the BNL Relativistic Heavy Ion Collider (RHIC) and CERN Large Hadron Collider (LHC) has provided us with rare opportunities to experimentally explore the nuclear equation of state, which encodes the fundamental thermodynamic properties of nuclear matter. The system created in these collisions is expected to be deconfined from hadronic matter to the quark-gluon plasma (QGP) above approximately two trillion degrees Kelvin. 

The theoretical study of the non-perturbative QCD equation of state dates back to the MIT bag model \cite{Chodos:1974je,Chodos:1974pn} where hadrons are assumed to be located in a bag embedded in the QCD vacuum characterized by the bag constant. This prescription introduces confinement phenomenologically. Also, several model approaches, such as the potential model \cite{DeRujula:1975qlm} and the Nambu-Jona-Lasinio model \cite{Nambu:1961tp,Nambu:1961fr}, have been proposed to understand the thermodynamic properties of QCD. A more comprehensive picture became available with the advent of first-principle calculations based on lattice QCD techniques. The quark-hadron phase transition is found to be a crossover by (2+1)-flavor lattice QCD calculations, in contrast to the SU(3) pure gauge case where a first-order phase transition is predicted \cite{Brown:1990ev,AliKhan:2000wou,Aoki:2006we}. 
Recent lattice QCD simulations with a physical pion mass have been able to provide us with the realistic equation of state at finite temperatures and vanishing chemical potentials \cite{Borsanyi:2013bia, Bazavov:2014pvz}. 

At finite densities the equation of state is not well known, owing to the fermion sign problem of the first principle method (for a review see \cite{deForcrand:2010ys}). Several techniques have been developed in lattice QCD, including the Taylor expansion method \cite{Gavai:2001fr,Allton:2002zi}, the imaginary chemical potential method \cite{deForcrand:2002hgr,DElia:2002tig,Gunther:2016vcp}, Lefschetz thimble decomposition \cite{Pham:1983,Witten:2010cx}, and the complex Langevin method \cite{Parisi:1984cs,Klauder:1985,Ambjorn:1985iw}, but so far no complete calculations are available at larger chemical potentials. There can be non-trivial structures in the $\mu_B$-$T$ QCD phase diagram other than the QGP phase \cite{Fukushima:2010bq}; it is conjectured that there is a critical point based on the chiral model that predicts that the quark-hadron crossover becomes a first-order transition at some finite temperature and chemical potential \cite{Asakawa:1989bq}. Beam energy scan (BES) programs are being performed to explore finite-density QCD matter and determine its detailed phase structure at RHIC and the CERN Super Proton Synchrotron (SPS), and are planned at several facilities including the GSI Facility for Antiproton and Ion Research (FAIR), JINR Nuclotron-based Ion Collider fAсility (NICA) and JAEA/KEK Japan Proton Accelerator Research Complex (J-PARC). 

From the viewpoint of hydrodynamic modeling of relativistic nuclear collisions, the equation of state is needed to close the set of equations of motion, by characterizing the thermodynamic properties of the system. The equation of state at vanishing density obtained from lattice QCD calculations has been employed in comparisons of hydrodynamic simulations with experimental data from heavy-ion collisions \cite{Pratt:2015zsa,Sangaline:2015isa,Bernhard:2016tnd,Pang:2016vdc,Monnai:2017cbv,Paquet:2017mny}, where bulk observables are generally well reproduced. For quantitative predictions and analysis of the BES experimental data, an equation of state at finite chemical potentials is needed as input to hydrodynamic models. Several works have been devoted to the construction of such quantitative models of the finite-density equation of state, including Refs.~\cite{Nonaka:2004pg,Bluhm:2004xn,Bluhm:2007nu,Steinheimer:2010ib,Huovinen:2011xc,Hempel:2013tfa,Albright:2014gva,Albright:2015uua,Rougemont:2017tlu,Critelli:2017oub,Vovchenko:2017gkg,Parotto:2018pwx,Vovchenko:2018zgt,Fu:2018qsk,Fu:2018swz,Motornenko:2018hjw,Plumberg:2018fxo}.

In this work, we present a framework to construct a QCD equation of state model (\textsc{neos}) with multiple charges: net baryon ($B$), strangeness ($S$) and electric charge ($Q$) based on state-of-the-art lattice QCD \cite{Borsanyi:2011sw,Bellwied:2015lba,Borsanyi:2018grb,Bazavov:2012jq, Ding:2015fca, Bazavov:2017dus} and hadron resonance gas results. A version of the equation of state, which only has baryon chemical potential, has previously been introduced and used in Refs.~\cite{Denicol:2015nhu,Monnai:2015sca,Shen:2017ruz,Shen:2017bsr,Denicol:2018wdp,Shen:2018pty,Gale:2018vuh}. We numerically calculate the equation of state with conditions on the conserved charges close to those in relativistic heavy-ion collisions. This analysis is expected to be relevant in mid to low energy heavy-ion collisions and at forward rapidity where the fragments of shattered nuclei are relatively abundant \cite{Li:2018fow,Monnai:2012jc}. The presence and interplay of the chemical potentials are expected to play an important role in for example the hadron chemistry or particle abundances (see e.g. \cite{Andronic:2005yp}), and the separation of hadron and anti-hadron flow observables \cite{Karpenko:2015xea,Hatta:2015era,Hatta:2015hca}.  

The multi-dimensional phase diagram has been studied \cite{Toublan:2004ks,Xu:2011pz,Kamikado:2012bt,Ueda:2013sia,Barducci:2004tt,Nishida:2003fb,Son:2000xc} and some conjecture non-trivial phase structures. In this study, we consider a crossover equation of state as a baseline for the application to relativistic nuclear collisions.

The paper is organized as follows. In Sec.~\ref{sec2}, the construction of finite-density equations of state based on the Taylor expansion method for lattice QCD and the hadron resonance gas is presented. The numerical evaluation of the hybrid equation of state is performed in Sec.~\ref{sec3}, where the strangeness neutrality condition and the fixed charge-to-baryon ratio of nuclei are taken into account. In Sec.~\ref{sec4}, particle ratios are estimated in hydrodynamic simulations assuming different conditions on the charge content of the system. Sec.~\ref{sec5} presents conclusions and discussions. Natural units $c = \hbar = k_B = 1$ and the Minkowski metric $g^{\mu \nu} = \mathrm{diag}(+,-,-,-)$ are used.

\section{The equation of state}
\label{sec2}

Based on the Taylor expansion method \cite{Gavai:2001fr,Allton:2002zi} we employ lattice QCD results of the conserved charge susceptibilities to construct the equation of state in the QGP phase. In the hadronic phase we use the equation of state of a hadron resonance gas, because the Taylor expansion method is not reliable at low temperatures. The use of a non-interacting resonance gas model is partly motivated by the good agreement between thermodynamic quantities at vanishing chemical potential, including susceptibilities, from lattice QCD and the hadron resonance gas. Also, the Cooper-Frye prescription \cite{Cooper:1974mv} of kinetic freeze-out requires that the hydrodynamic equation of state precisely matches that of the kinetic theory description of the hadron resonance gas on the freeze-out hypersurface for the successful conservation of energy-momentum and all charges. If at low temperatures the lattice result was used instead of the hadron resonance gas model, the truncation of the Taylor expansion at finite order would lead to an underestimation of the pressure in the hadronic phase, because higher order susceptibilities can be large for the hadron resonance gas in the Boltzmann limit, \textit{e.g.}, $\chi_{2n}^B/\chi_{2}^B = 1$.

\subsection{Lattice QCD equation of state}

We consider the Taylor expansion method of lattice QCD as mentioned earlier. For the three-flavor QCD system, the expansion of the pressure around the vanishing density limit reads
\begin{eqnarray}
\frac{P}{T^4} &=& \frac{P_0}{T^4} + \sum_{l,m,n} \frac{\chi^{B,Q,S}_{l,m,n}}{l!m!n!} \bigg( \frac{\mu_B}{T} \bigg)^{l}  \bigg( \frac{\mu_Q}{T} \bigg)^{m}  \bigg( \frac{\mu_S}{T} \bigg)^{n}, 
\label{Psus}
\end{eqnarray}
where $P$ is the pressure, $P_0$ is the pressure at vanishing chemical potentials, $T$ is the temperature, and $\mu_{B,Q,S}$ are the chemical potentials of baryon number, electric charge, and strangeness, respectively. $\chi_{l,m,n}^{B,Q,S}$ is the ($l$+$m$+$n$)-th order susceptibility defined at vanishing chemical potentials:
\begin{equation}
  \chi_{l,m,n}^{B,Q,S} = \left.\frac{\partial^l \partial^m \partial^n P(T,\mu_B,\mu_Q,\mu_S)/T^4}{\partial(\mu_B/T)^l\partial(\mu_Q/T)^m\partial(\mu_S/T)^n}\right|_{\mu_{B,Q,S}=0}.
\end{equation}
The number $l$+$m$+$n$ should be even, owing to the matter-antimatter symmetry. The expansion is valid only when the fugacity $\mu_{B,Q,S}/T$ is sufficiently small. The lattice QCD results are parametrically extrapolated to high temperatures under the condition that they do not violate the Stefan-Boltzmann limits. See Appendix \ref{sec:A}.

\subsection{Hadron resonance gas equation of state}

The hydrostatic pressure of the hadron resonance gas can be written as
\begin{eqnarray}
P &=& \pm T \sum_i \int \frac{g_i d^3p}{(2\pi)^3} \ln [1 \pm e^{-(E_i-\mu_i)/T} ]\nonumber\\
&=& \sum_i \sum_k (\mp1)^{k+1} \frac{1}{k^2} \frac{g_i}{2\pi^2} m_i^2 T^2 e^{k\mu_i/T} K_2\bigg(\frac{k m_i}{T}\bigg), \label{eq:P_had}
\end{eqnarray}
where $i$ is the index for particle species, $g_i$ is the degeneracy, $m_i$ the particle's mass, and $K_2(x)$ is the modified Bessel function of the second kind. The index $k$ describes the expansion of quantum distributions around the classical ones. It is generally sufficient to take into account the contributions of $k\leq3$ for pions, $k \leq 2$ for kaons and $k = 1$ for the heavier particles. The upper signs are for fermions and the lower signs for bosons. The hadronic chemical potential is $\mu_i$ = $B_i \mu_B + Q_i \mu_Q + S_i \mu_S$ where $B_i$, $Q_i$, and $S_i$ are the quantum numbers of net baryon, electric charge, and strangeness, respectively.

\subsection{Hybrid equation of state}

The complete nuclear equation of state is constructed by connecting the pressure of the lattice QCD equation of state to that of the hadron resonance gas model \cite{Huovinen:2009yb}
\begin{eqnarray}
\frac{P}{T^4} &=& \frac{1}{2}[1- f(T,\mu_J)] \frac{P_{\mathrm{had}}(T,\mu_J)}{T^4} \nonumber \\
&+& \frac{1}{2}[1+ f(T,\mu_J)] \frac{P_{\mathrm{lat}}(T_s,\mu_J)}{T_s^4} , \label{eq:econ}
 \end{eqnarray}
where $J = \{B,Q,S\}$. Here $f(T,\mu_J)$ is an arbitrary function for the connection of the two functions which satisfies $f \to 1$ when $T \gg T_c$ and $f \to 0$ when $T \ll T_c$, where $T_c$ is the connecting temperature.
In this work we choose $f$ to be a hyperbolic tangent, defined in \eqref{eq:cfunction}.
A temperature shift $T_s (T, \mu_J)$ is introduced phenomenologically to preserve the monotonicity conditions of thermodynamic variables at larger chemical potentials. Since $T_s$ is generally a function of $T$ and $\mu_J$, one can define 
\begin{eqnarray}
\tilde{P}_{\mathrm{lat}}(T,\mu_J) = P_{\mathrm{lat}}(T_s,\mu_J) \times \frac{T^4}{T_s^4},
 \end{eqnarray}
which is the shifted QGP equation of state. Here $\tilde{P}_{\mathrm{lat}}$ should reduce to $P_{\mathrm{lat}}$ at small chemical potentials. While in this work we will not use a temperature shift, the shifting temperature is used in similar constructions of the equation of state in previous works ~\cite{Denicol:2015nhu,Monnai:2015sca,Shen:2017ruz,Shen:2017bsr,Denicol:2018wdp,Shen:2018pty,Gale:2018vuh}.

We require that the thermodynamic variables monotonously increase as functions of $T$ and $\mu_J$, respectively, as
\begin{eqnarray}
\frac{\partial^2 P}{\partial T^2} &=& \frac{\partial s}{\partial T} > 0, \label{eq:tcc1} \\ 
\frac{\partial^2 P}{\partial \mu_J^2} &=& \frac{\partial n_J}{\partial \mu_J} > 0 . \label{eq:tcc2}
\end{eqnarray}
Those conditions may be trivially satisfied for the hadron resonance gas or lattice QCD equation of state when $\mu_J/T$ is not large, but the connection procedure can make it non-trivial. The conditions can be expressed as
\begin{eqnarray}
\frac{\partial^2 P}{\partial T^2} &=& \frac{1}{2}[1- f(T,\mu_J)] \frac{\partial s_{\mathrm{had}}(T,\mu_J)}{\partial T} \nonumber \\
&+& \frac{1}{2} [1+ f(T,\mu_J)] \frac{\partial \tilde{s}_{\mathrm{lat}}(T,\mu_J)}{\partial T} \nonumber \\
&+& \sum_{J} \frac{\partial f(T,\mu_J)}{\partial T} [ \tilde{s}_{\mathrm{lat}}(T,\mu_J) - s_{\mathrm{had}}(T,\mu_J)] \nonumber \\
&+& \frac{1}{2} \sum_{J} \frac{\partial^2 f(T,\mu_J)}{\partial T^2} [ \tilde{P}_{\mathrm{lat}}(T,\mu_J) - P_{\mathrm{had}}(T,\mu_J)] \nonumber \\
&>& 0.
\end{eqnarray}
\begin{eqnarray}
\frac{\partial^2 P}{\partial \mu_J^2} &=& \frac{1}{2}[1- f(T,\mu_J)] \frac{\partial n^J_{\mathrm{had}}(T,\mu_J)}{\partial \mu_J} \nonumber \\
&+& \frac{1}{2} [1+ f(T,\mu_J)] \frac{\partial \tilde{n}^J_{\mathrm{lat}}(T,\mu_J)}{\partial \mu_J} \nonumber \\
&+& \frac{1}{2} \sum_{J} \frac{\partial f(T,\mu_J)}{\partial \mu_J} [ \tilde{n}^J_{\mathrm{lat}}(T,\mu_J) - n^J_{\mathrm{had}}(T,\mu_J)] \nonumber \\
&+& \frac{1}{2} \sum_{J} \frac{\partial^2 f(T,\mu_J)}{\partial \mu_J^2} [ \tilde{P}_{\mathrm{lat}}(T,\mu_J) - P_{\mathrm{had}}(T,\mu_J)] \nonumber \\
&>& 0 .
\end{eqnarray}
Assuming that the thermodynamic quantities on the lattice QCD side are larger than those on the hadron resonance gas side, the conditions are still not trivially satisfied as the second order derivatives of $f$ can be negative. These conditions will be handled numerically in the next section.

The thermodynamic variables, the entropy density $s$, the conserved charge densities $n_J$, the energy density $e$, and the sound velocity $c_s$ are given as
\begin{eqnarray}
s &=& \left. \frac{\partial P}{\partial T} \right|_{\mu_{J}}, \\
n_{J} &=& \left. \frac{\partial P}{\partial \mu_J} \right|_{T, \mu_K}, \\
e &=& Ts - P + \sum_J \mu_J n_J, \\
c_s^2 &=& \left. \frac{\partial P}{\partial e} \right|_{n_{J}} + \sum_J \frac{n_J}{e+P} \left. \frac{\partial P}{\partial n_J} \right|_{e, n_K}, \label{eq:cs2}
\end{eqnarray}
respectively, using the standard thermodynamic relations. Here $\{J,K\} = B,Q,S$ and $J\neq K$.

\subsection{Multiple charges in nuclear collisions}

A standard nucleus is made of protons and neutrons so the averaged density of strangeness is zero, which may be expressed as $n_S(T, \mu_B,\mu_Q,\mu_S) = 0$. However, neglecting electric charge for the moment, the conventional choice of $\mu_S = 0$ leads to $n_S \neq 0$, because the strangeness density is dependent on $\mu_B$ as the strange quark carries both net baryon number and strangeness. Thus, in the presence of a finite net-baryon number, $\mu_S$ should generally be non-vanishing, so $n_S = 0$ can be fulfilled. The condition, of course, can in principle be locally broken in the presence of geometrical fluctuations or diffusion processes. The equation of state with the strangeness neutrality condition can be expressed in terms of $T$ and $\mu_B$ because $\mu_S = \mu_S (T, \mu_B)$. 

The electric charge density $n_Q(T, \mu_B,\mu_Q,\mu_S)$ is non-vanishing in nuclei as $n_Q = (Z/A) n_B$, where $Z$ is the proton number and $A$ is the nucleon number. The list of $Z/A$ ratios of the nuclei used in collider experiments is shown in Table~\ref{table:nucl}. The typical ratio for heavy nuclei such as Au or Pb is $Z/A \approx 0.4$. The precise $n_Q$ dependence is expected to become more important when comparing collisions of isobar systems.

\begin{table}
\begin{tabular}{c|c}
\hline \hline
Nucleus & $Z/A$  \\ \hline
$^{1}_{1}$H & 1.000 \\ \hline
$^{2}_{1}$H & 0.500 \\ \hline
$^{3}_{2}$He & 0.667 \\ \hline
$^{27}_{13}$Al & 0.481 \\ \hline
$^{63}_{29}$Cu & 0.460 \\ \hline
$^{96}_{40}$Zr & 0.417 \\ \hline
$^{96}_{44}$Ru & 0.458 \\ \hline
$^{127}_{54}$Xe & 0.425 \\ \hline
$^{197}_{\ 79}$Au & 0.401 \\ \hline
$^{208}_{\ 82}$Pb & 0.394 \\ \hline
$^{238}_{\ 92}$U & 0.387 \\ \hline
\hline
\end{tabular}
\caption{Ratios of protons to nucleons $Z/A$ for the nuclei used in the collider experiments at RHIC and LHC.}
\label{table:nucl}
\end{table}

\section{Numerical construction}
\label{sec3}

In this section, we numerically construct the hybrid QCD equation of state at finite densities (\ref{eq:econ}). On the lattice QCD side, we make use of one of the latest (2+1)-flavor calculations of the equation of state at vanishing chemical potentials \cite{Bazavov:2014pvz} and the diagonal and off-diagonal susceptibilities up to the fourth order \cite{Bazavov:2012jq, Ding:2015fca, Bazavov:2017dus, Sharma}. In addition, we introduce some of the most relevant sixth order susceptibilities, to allow for a proper matching of all quantities, as discussed later. The specific functional forms of parametrization are summarized in Appendix~\ref{sec:B}.

All hadron resonances from the particle data group \cite{Tanabashi:2018oca} with $u$, $d$, and $s$ constituent quark components and masses smaller than 2 GeV are taken into account in the resonance gas model. We use 
\begin{equation}\label{eq:cfunction}
    f(T,\mu_B) = \tanh [(T-T_c(\mu_B))/\Delta T_c]\,,
\end{equation} 
where $T_c(\mu_B) = 0.16\ \mathrm{GeV} - 0.4\times(0.139\ \mathrm{GeV}^{-1} \mu_B^2 + 0.053\ \mathrm{GeV}^{-3} \mu_B^4)$ motivated by and modified from the chemical freeze-out curve \cite{Cleymans:2005xv}. The dependence of the connecting temperature on strangeness and electric charge chemical potentials is assumed to be weak and neglected for the moment. The connecting width is $\Delta T_c = 0.1 T_c(0)$. Here, we choose not to perform a temperature shift and use $T_s=T$. 
It should be noted that this is not a unique choice of the parameters, but the thermodynamic monotonicity conditions (\ref{eq:tcc1}) and (\ref{eq:tcc2}) leave a rather narrow window for the possible parameter values. The smooth matching leads to an equation of state with a crossover transition.
Implementation of a QCD critical point and the first-order phase transition is also possible for different choices of $f$. It will be discussed elsewhere as the location of the critical point and the critical behavior near it are currently not well known. For a possible approach to include a critical point see \cite{Critelli:2017oub,Parotto:2018pwx,Plumberg:2018fxo}.

The sixth-order susceptibilities should be relevant near and below the crossover transition. The term involving $\chi_6^B$ is naively expected to give the largest contribution to the pressure and the net baryon number because of the hierarchy in the chemical potentials $\mu_B > \mu_S > \mu_Q$ in nuclear collisions. The strangeness density and the electric charge density are not directly sensitive to $\chi_6^B$, because they are derivatives with respect to $\mu_S$ or $\mu_Q$, respectively, implying that the terms involving $\chi_{5,1}^{B,S}$ and $\chi_{5,1}^{B,Q}$ will be the important ones for them. We introduce those three susceptibilities in a phenomenological approach so that $n_B$, $\mu_S$, and $\mu_Q$ are smooth functions of $T$ and $\mu_B$, and that the results of the hadron resonance gas model are preserved below $T_c$, because of the relatively large uncertainties in the current lattice calculations of higher order susceptibilities. We find that the effects of the sixth-order susceptibilities are limited to the large chemical potential regions near the crossover transition.

For the strangeness and electric charges, we consider three cases: (i) $\mu_S = \mu_Q = 0$, (ii) $n_S = 0$ and $\mu_Q = 0$, and (iii) $n_S = 0$ and $n_Q = 0.4 n_B$. They are reffered to as \textsc{neos} B, \textsc{neos} BS and \textsc{neos} BQS, respectively. The first is the commonly used scenario in which one assumes that the net baryon chemical potential is the only non-vanishing one in the system. The second imposes the strangeness neutrality condition but neglects the electric charge chemical potential. The third is the most realistic scenario for the collision of heavy nuclei where $Z/A \sim 0.4$ \cite{Bazavov:2017dus}. It is also straightforward to calculate the equation of state as functions of $\mu_B$, $\mu_Q$, and $\mu_S$ for more general systems. 

\subsection{Vanishing strangeness and electric charge chemical potentials $\mu_S = \mu_Q = 0$}

\begin{figure}[tb]
\includegraphics[width=3.4in]{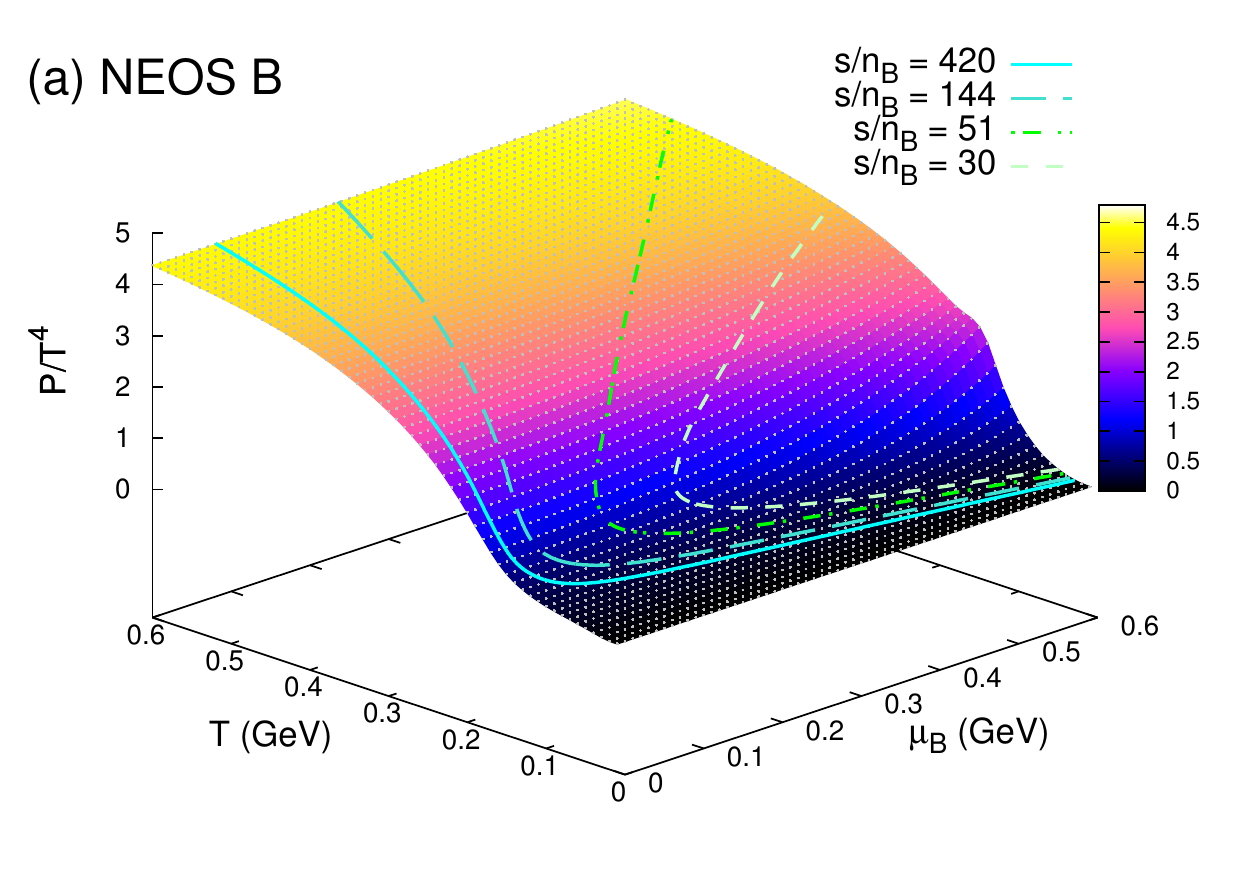}
\includegraphics[width=3.4in]{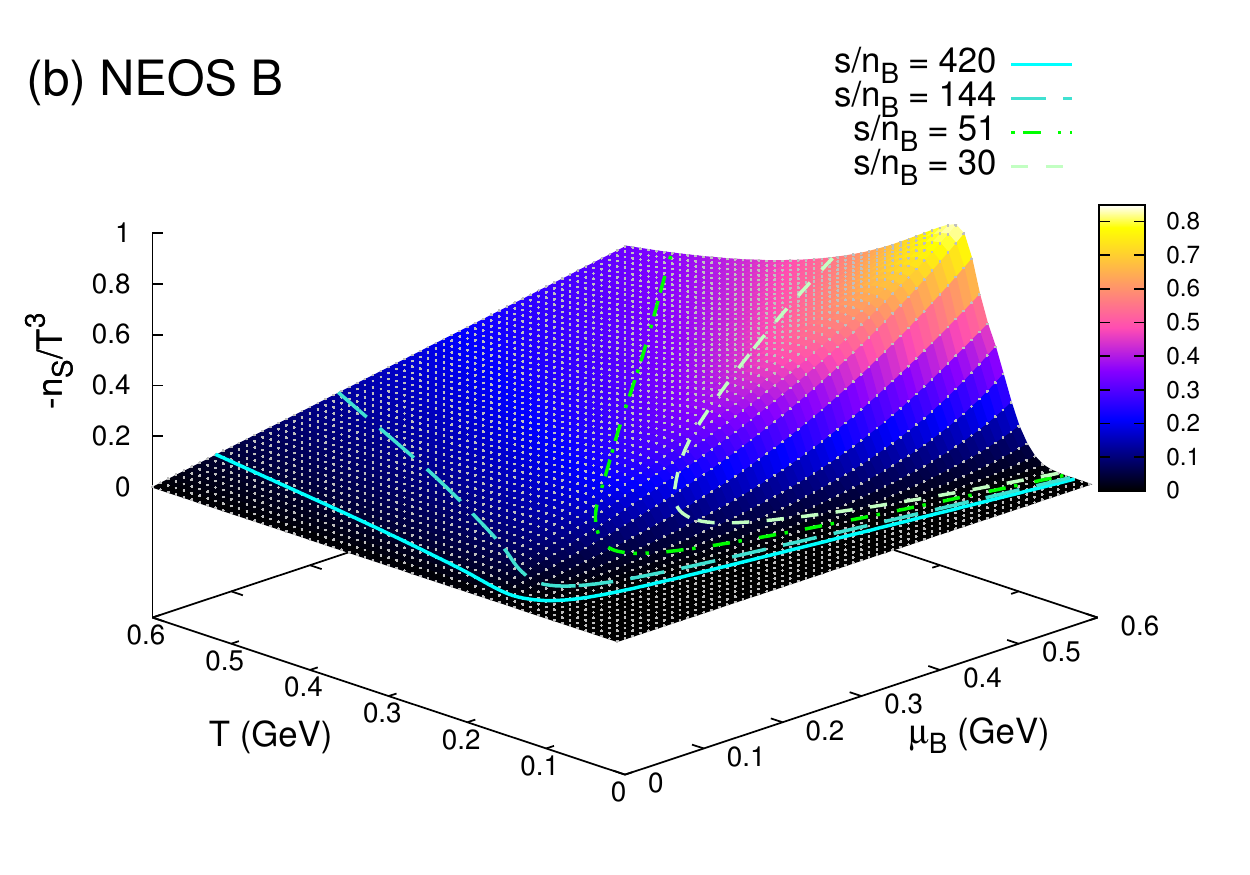}
\caption{(Color online) (a) The dimensionless pressure $P/T^4$ and (b) the dimensionless strangeness density $n_S/T^3$ as functions of $T$ and $\mu_B$ where $\mu_S = \mu_Q = 0$. The solid, long-dashed, dash-dotted, and short-dashed lines are the trajectories for constant $s/n_B$ = 420, 144, 51, and 30, respectively.}
\label{fig:b}
\end{figure}

First, the case where $\mu_S = \mu_Q = 0$ is investigated. The pressure of the resulting equation of state is plotted in Fig.~\ref{fig:b} (a). One can see the monotonous increase of $P$ as a function of $T$ or $\mu_B$. The equation of state reduces to that of lattice QCD at $\mu_B = 0$ at the vanishing density limit. The constant entropy density over net baryon density lines, $s/n_B$ = 420, 144, 51, and 30, are plotted to illustrate the most relevant regions for the BES programs. They correspond to Au+Au collisions at $\sqrt{s_{NN}} = $ 200, 62.4, 19.6, and 14.5 AGeV, respectively \cite{Gunther:2016vcp}. Note that the ratio is roughly constant during the time evolution in nuclear collisions when the entropy and the net baryon number are conserved, which is the case for the nearly-perfect fluid. The trajectory of $s/n_B$ is a straight line at higher temperatures where the system is relatively close to conformal, because $s/n_B \sim T/\mu_B$. It turns around near the crossover towards lower temperatures as pions begin to dominate over protons because of the mass difference, and large baryon chemical potential is required to have protons at lower temperatures for keeping the $s/n_B$ ratio fixed. In the limit $T\to 0$, the chemical potential approaches the proton mass. 

As discussed earlier, the condition $\mu_S = 0$, which is often assumed in nuclear collision analyses, leads to a non-vanishing strangeness density $n_S$. The value of $-n_S/T^3$ is shown in Fig.~\ref{fig:b} (b). Positive baryon chemical potential leads to negative strangeness density because the $s$ quark has a negative strangeness chemical potential. The high temperature behavior can be understood as $n_S/T^3 \sim \chi_{1,1}^{B,S} \mu_B/T \sim - \mu_B/3T$ (\ref{eq:chi11}). The strangeness density is relatively small at lower temperatures because kaons, the lightest strange hadrons, have net baryon number zero. 

\subsection{Strangeness neutrality $n_S = 0$ and vanishing electric charge chemical potential $\mu_Q = 0$}

\begin{figure}[tb]
\includegraphics[width=3.4in]{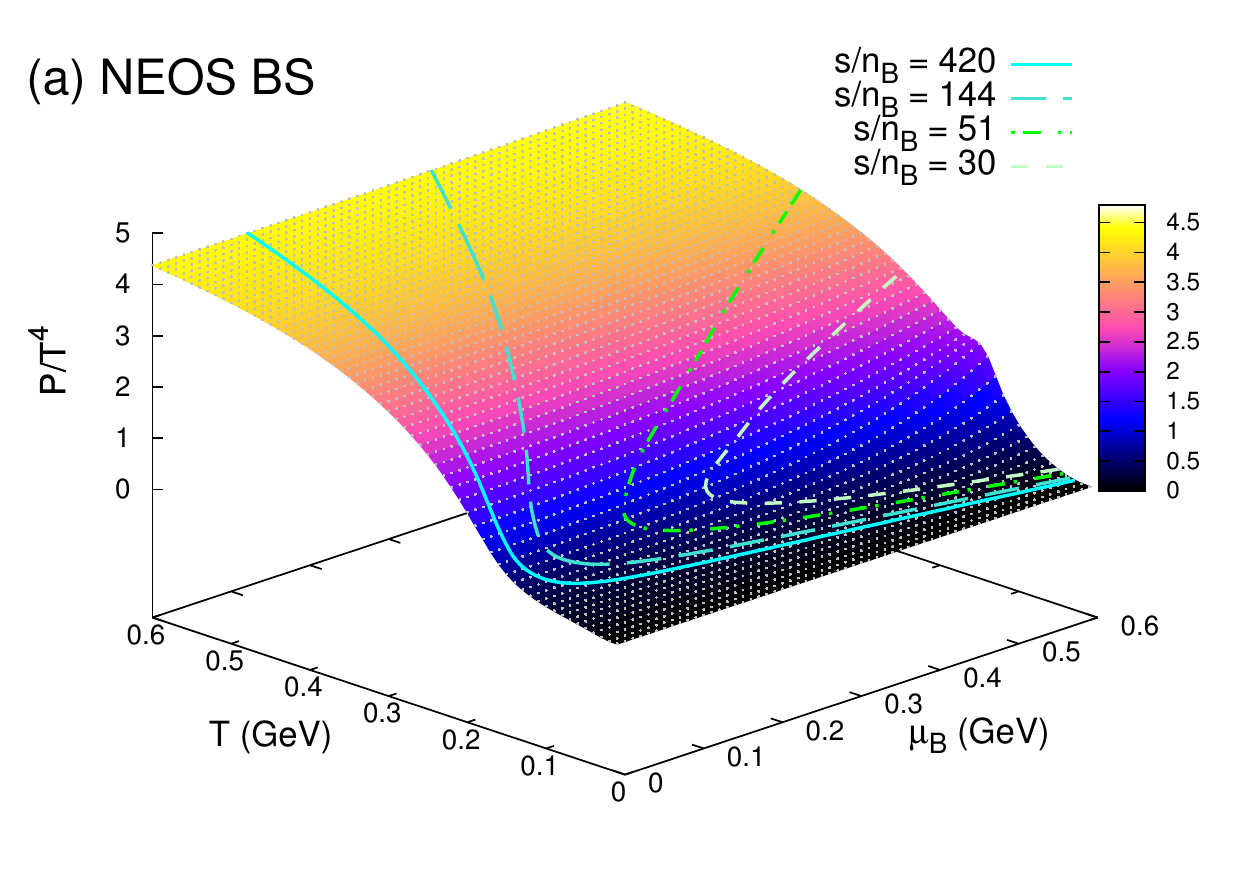}
\includegraphics[width=3.4in]{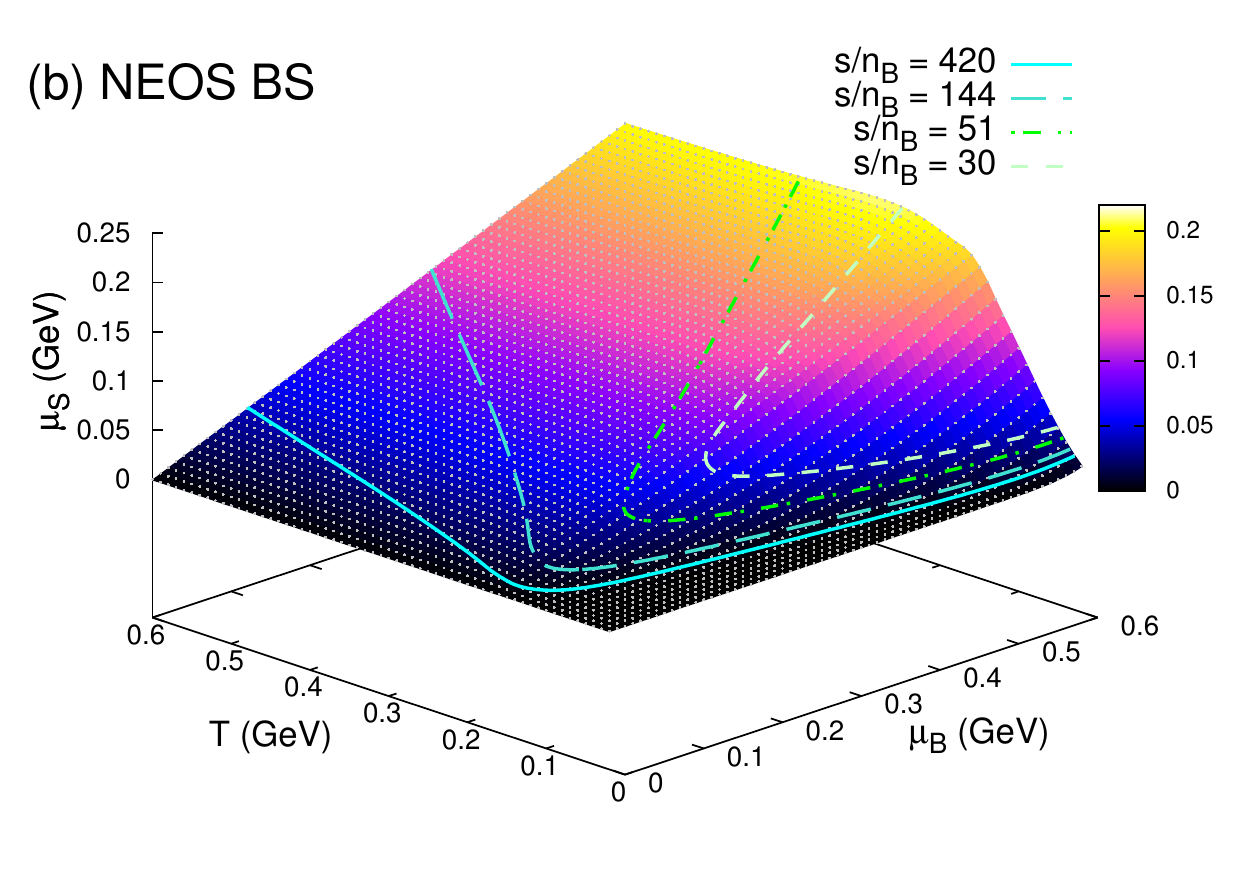}
\caption{(Color online) (a) The dimensionless pressure $P/T^4$ and (b) the strangeness chemical potential $\mu_S$ as functions of $T$ and $\mu_B$ where $n_S = 0$ and $\mu_Q = 0$. The solid, long-dashed, dash-dotted, and short-dashed lines are the trajectories for constant $s/n_B$ = 420, 144, 51, and 30, respectively. }
\label{fig:bs}
\end{figure}

\begin{figure}[tb]
\includegraphics[width=3.2in]{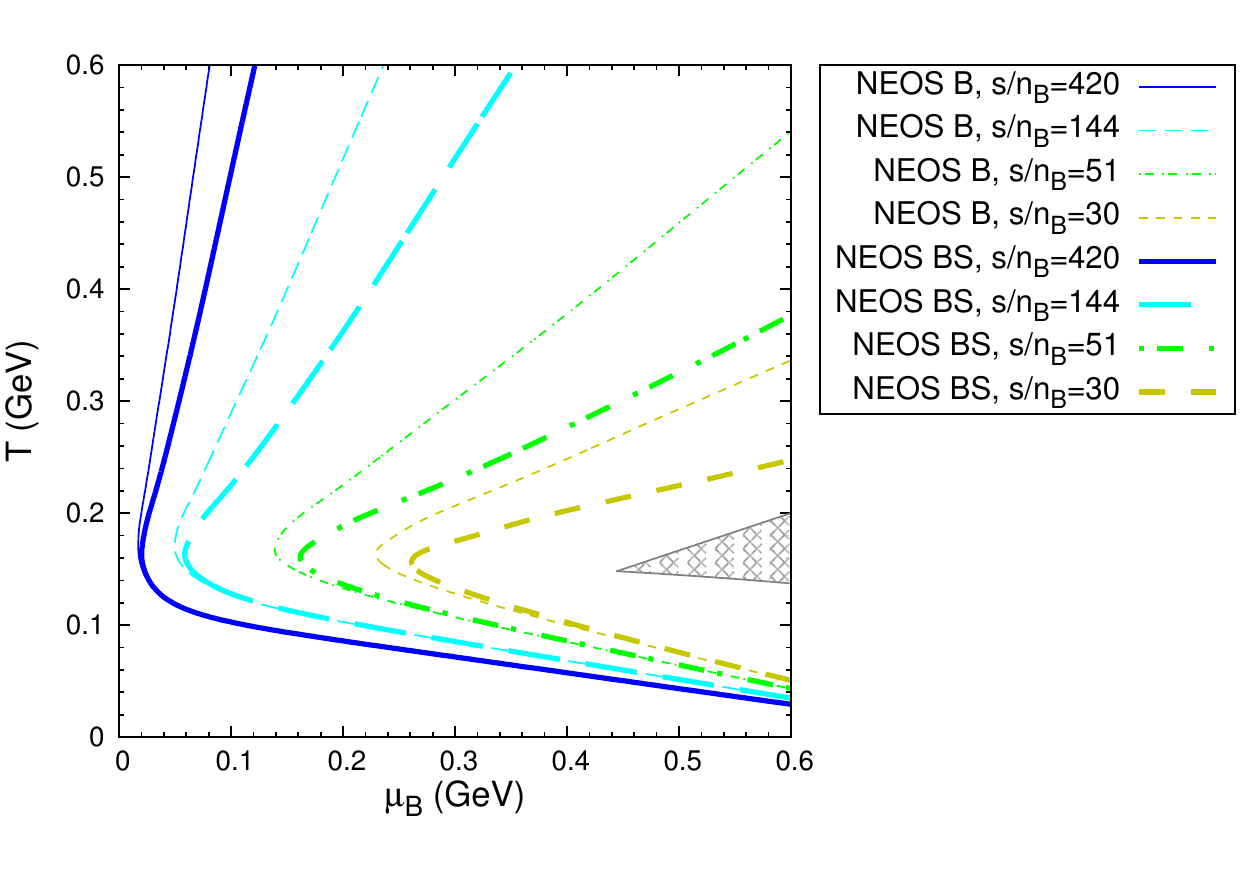}
\caption{(Color online) The comparison of the trajectories for constant $s/n_B$ = 420 (solid line), 144 (long-dashed line), 51 (dash-dotted line), and 30 (short-dashed line) lines between \textsc{neos} B and \textsc{neos} BS denoted by narrow and thick lines, respectively. The gray area shows the region where $\mu_B/T > 3$ above $T_c$.}
\label{fig:b_vs_bs}
\end{figure}

We next study the case where $n_S = 0$ and $\mu_Q = 0$. The pressure is shown as a function of $T$ and $\mu_B$ in Fig.~\ref{fig:bs} (a).
The $n_Q/n_B$ ratio is arbitrary, and approaches $n_Q/n_B \sim 0.5$ in the parton gas limit (\ref{eq:pg-q}). One can see that the equation of state is modified at larger baryon chemical potentials compared with that of the $\mu_S = 0$ case. The constant $s/n_B$ lines are also shifted to larger $\mu_B$ (Fig.~\ref{fig:bs} (b)) because the strangeness neutrality implies $\mu_S \sim \mu_B/3$ at high temperatures. For clarity we show the projections of the constant $s/n_B$ lines onto the $\mu_B$-$T$ plane in  Fig.~\ref{fig:b_vs_bs}. Here one can see that $\mu_B$ has to be about 3/2 times larger for a given $n_B$ when $\mu_S\neq 0$. The gray area in the figure shows a conjectured region $\mu_B/T > 3$ where the Taylor expansion method of lattice QCD is not well-defined. A larger value of $\mu_B$ will result in a larger thermodynamic force $\nabla^\mu (\mu_B/T)$ for the net baryon diffusion current \cite{Denicol:2018wdp}. It will have an important effect on constraining the net baryon diffusion constant in future phenomenological studies. While initial strangeness fluctuations and a strangeness diffusion current can  break strangeness neutrality locally, this should not diminish the effect of enhanced $\mu_B$ on the net baryon diffusion, since it is a sub-leading effect and the strangeness is still globally conserved at zero. 

Importantly, $\mu_s$ is non-zero at freeze-out, which will affect results on particle-antiparticle ratios of strange hadrons in hydrodynamic models, as is the case in thermal models \cite{Andronic:2005yp}. The potentially large effect of the strangeness neutrality condition is also discussed in Ref.~\cite{Bazavov:2012vg}.

It should be noted that while we have now imposed more realistic conditions compared to the previous case, the thermodynamic properties of the QCD system itself remain the same, we merely look at different slices of the multi-dimensional equation of state.

\subsection{Strangeness neutrality $n_S = 0$ and fixed electric charge-to-baryon ratio $n_Q = 0.4 n_B$}

\begin{figure}[tb]
\includegraphics[width=3.4in]{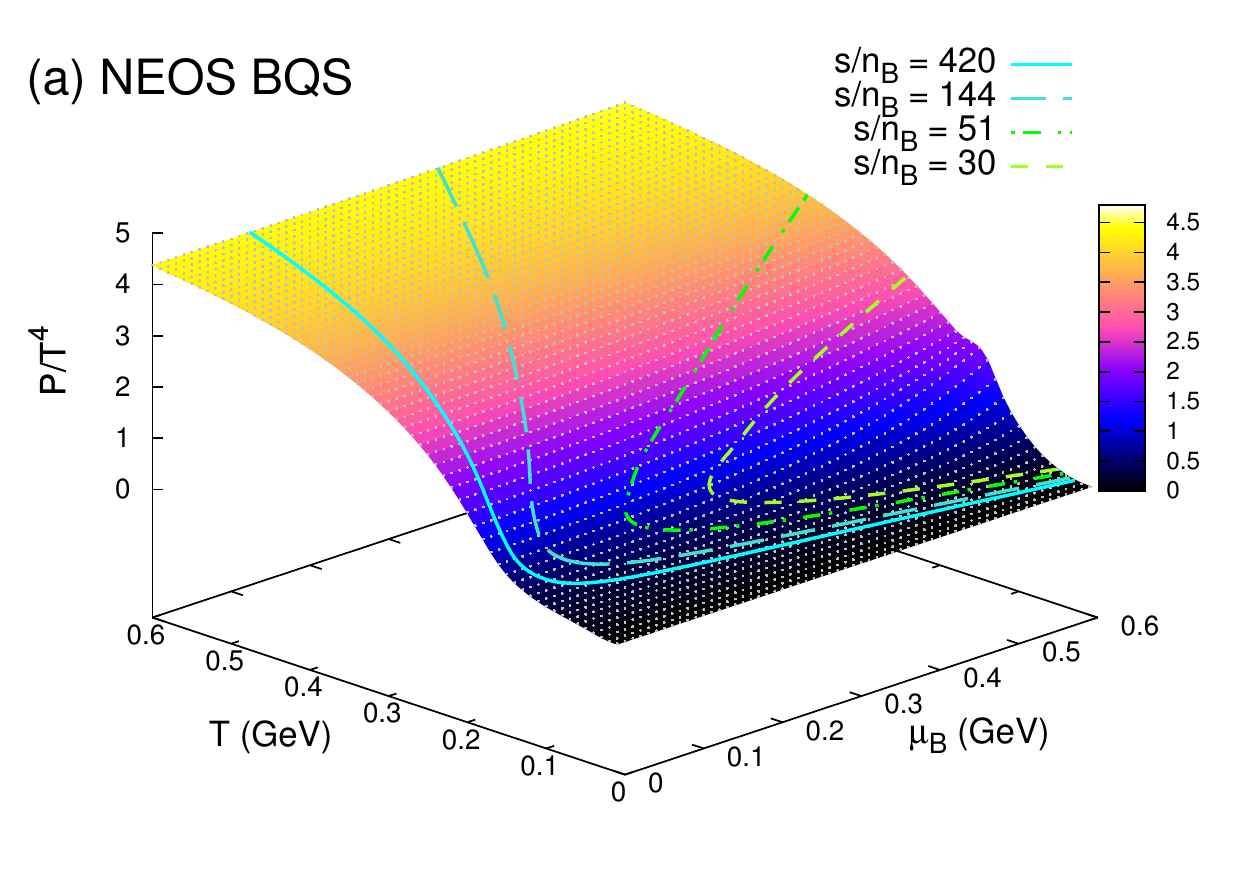}
\includegraphics[width=3.4in]{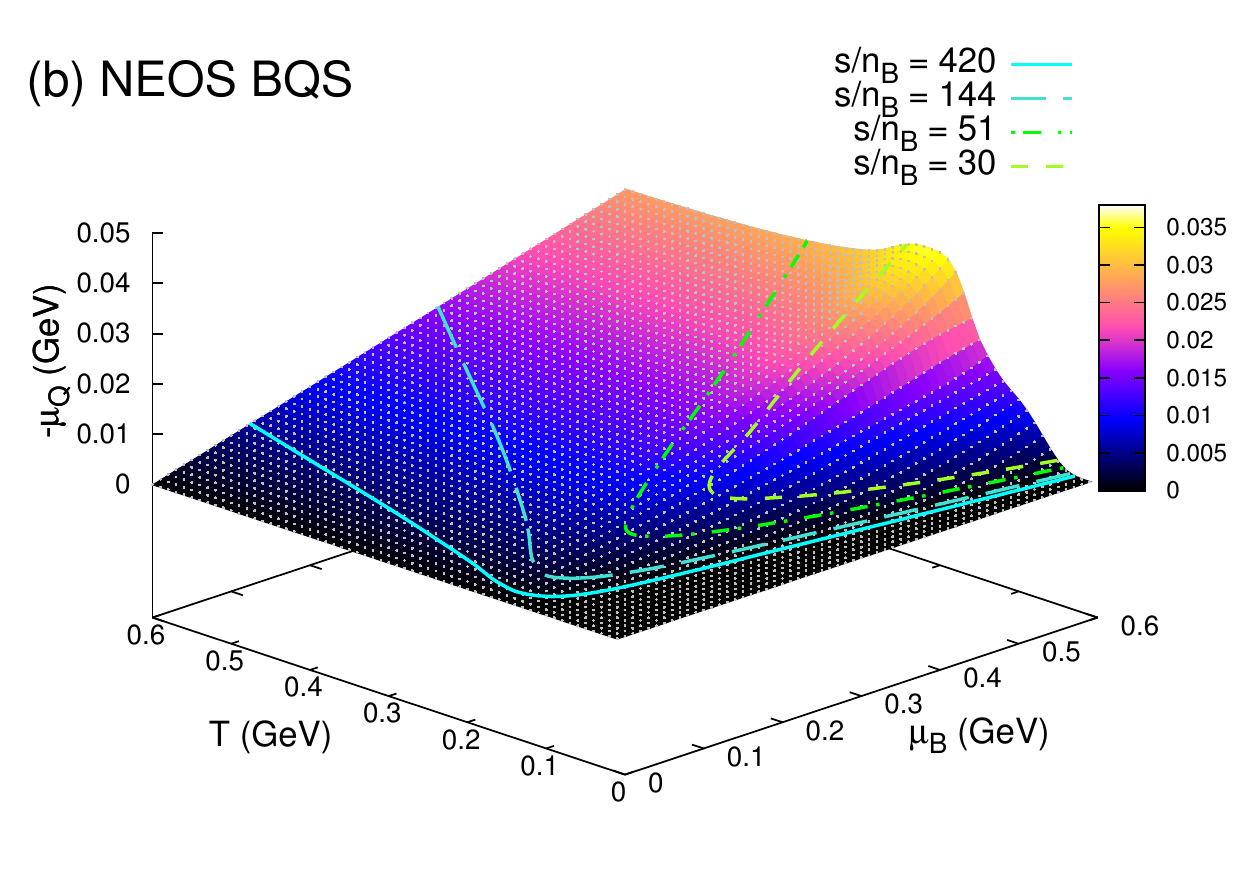}
\caption{(Color online) (a) The dimensionless pressure $P/T^4$ and (b) the electric charge chemical potential $\mu_Q$ as functions of $T$ and $\mu_B$ where $n_S = 0$ and $n_Q = 0.4 n_B$. The solid, long-dashed, dash-dotted, and short-dashed lines are the trajectories for constant $s/n_B$ = 420, 144, 51, and 30, respectively.}
\label{fig:bqs}
\end{figure}

Finally, we investigate the case where $n_S = 0$ and $n_Q = 0.4 n_B$. This is the setup most relevant to Au+Au and Pb+Pb collisions. The dimensionless pressure $P/T^4$ is plotted in Fig.~\ref{fig:bqs} (a). The difference from the previous case is small in this setup but should be meaningful for correctly understanding particle-antiparticle ratios of charged particles. 

The electric charge chemical potential shown in Fig.~\ref{fig:bqs} (b) is negative, owing to the interplay of multiple conserved charges. Since the number of neutrons is larger than that of protons in heavy nuclei, $d$ quarks are slightly more abundant than $u$ quarks in the QGP phase and $\pi^-$ more abundant than $\pi^+$ in the hadronic phase. While the overall system is positively charged, a negative electric chemical potential is needed for describing this situation. $\mu_Q$ becomes positive for the system of $^{3}$He since $Z/A  > 1/2$. This would have to be taken into account for the collisions involving such nuclei. 

It should be noted that $\mu_Q$ is small and is rather sensitive to the fine structure of the equation of state, including higher-order susceptibilities, at large chemical potentials. This implies that improvement in the lattice QCD calculations, including higher order susceptibilities, will be important in quantitative analyses.

\subsection{Discussion}

We have constructed the nuclear equation of state under several different conditions. We now study the differences between the different scenarios in more detail.

\begin{figure}[tb]
\includegraphics[width=3.3in]{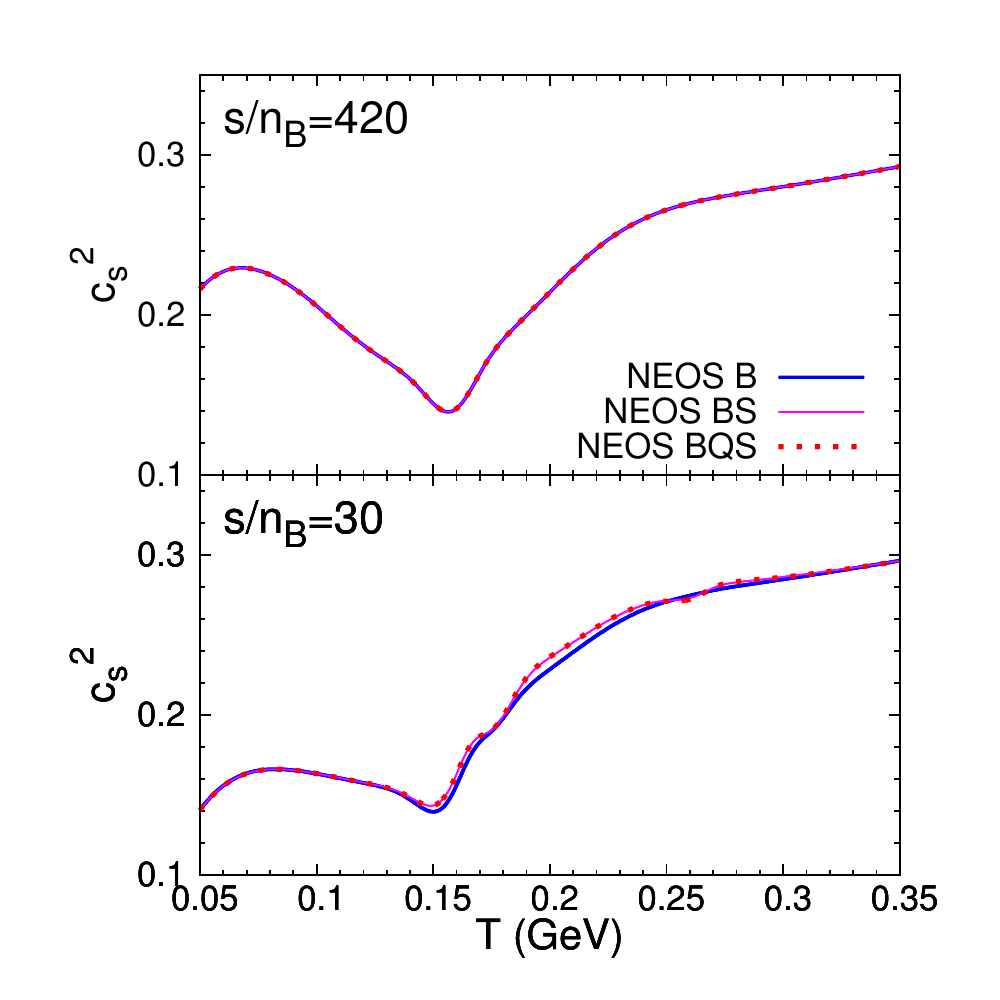}
\caption{(Color online) The squared sound velocity as a function of the temperature along the constant trajectories of $\mu_S = \mu_Q = 0$ (thick solid line), $n_S = 0$ and $\mu_Q = 0$ (thin solid line), and $n_S = 0$ and $n_Q = 0.4 n_B$ (thick dotted line) under the conditions $s/n_B$ = 420 (top) and 30 (bottom).}
\label{fig:b-cs2}
\end{figure}

The sound velocities of the equation of state under the three different conditions along two constant $s/n_B$ lines are plotted in Fig.~\ref{fig:b-cs2}. One can see that finite-density effects are visible comparing the sound velocities of $s/n_B = 420$ and $30$.
Around the crossover temperature, the EoS becomes soft and $c_s^2$ has a minimum. The location of the minimum shifts towards lower temperatures as the net baryon density increases. 
Also, the sound velocity becomes larger in the QGP phase and smaller in the hadronic phase at larger chemical potentials. This is because the net baryon contribution in $c_s^2$ (\ref{eq:cs2}) is positive for the former phase and negative for the latter phase. At higher temperatures, it starts to approach the Stefan-Boltzmann limit $c_s^2 = 1/3$. For the three presented equations of state, $c_s^2$ reaches 94.8 \% of the value the Stefan-Boltzmann limit at $T=0.6$ GeV and 97.2 \% at $T=0.8$ GeV for $s/n_B = 420$.

Comparing \textsc{neos} B to \textsc{neos} BS, the strangeness neutrality condition is found to slightly enhance the sound velocity in the QGP phase. It should be noted that if one neglected the derivatives involving $n_S$ in the calculation of $c_s^2$ (\ref{eq:cs2}) for \textsc{neos} B, the sound velocity would be further underestimated than our current result, because
\begin{eqnarray}
c_s^2 &\neq& \left. \frac{\partial P}{\partial e} \right|_{n_{B}} + \frac{n_B}{e+P} \left. \frac{\partial P}{\partial n_B} \right|_{e} ,
\end{eqnarray}
when $\mu_S = 0$, \textit{i.e.}, $n_S \neq 0$, which again highlights the importance of adequate treatment of the multiple conserved charges. 
The effects of the fixed charge to baryon ratio on the sound velocity is almost negligible. Since the effect of the electric chemical potential is not large, the difference in the sound velocity is also not large when one neglects the derivatives involving $\mu_Q$ and $n_Q$ in the definition (\ref{eq:cs2}).

\begin{figure}[tb]
\includegraphics[width=3.4in]{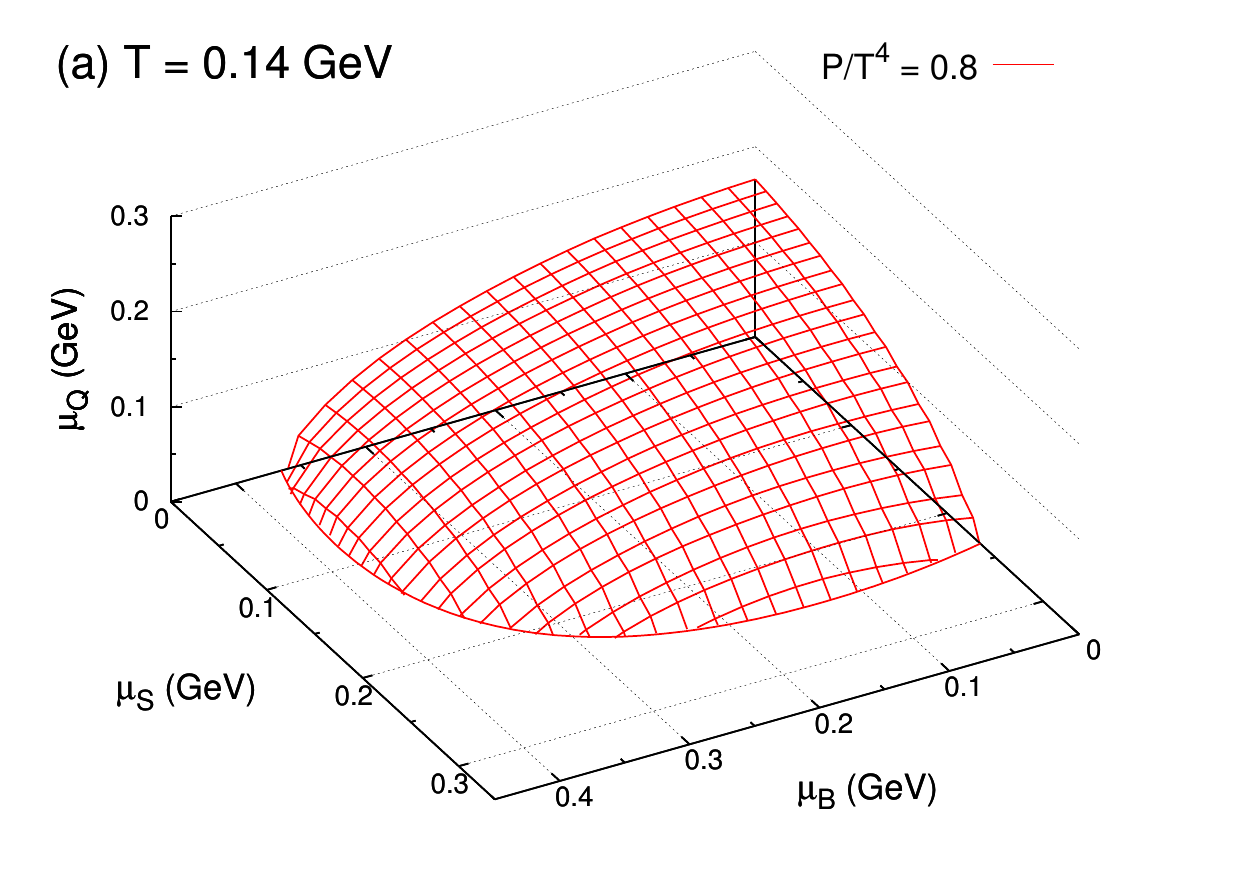}
\includegraphics[width=3.4in]{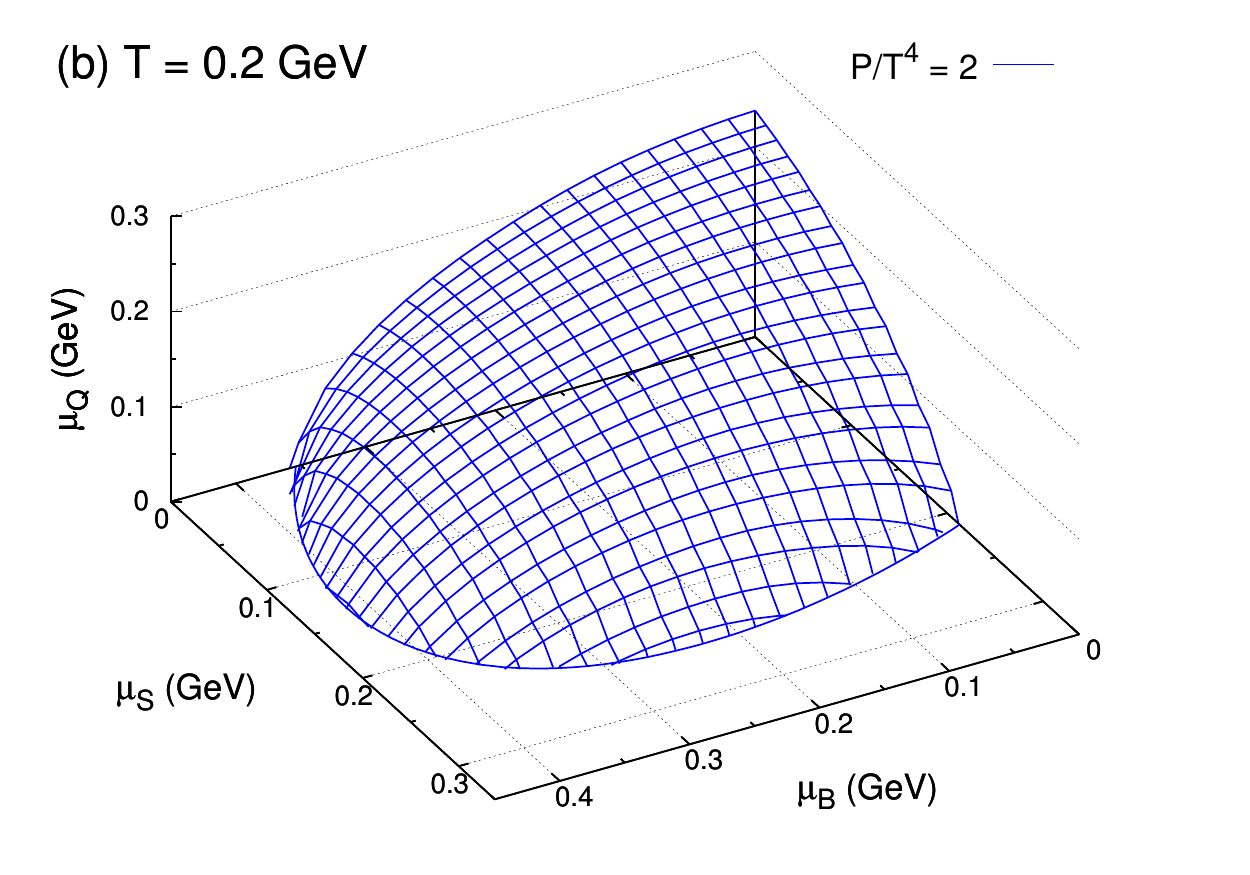}
\caption{(Color online) The constant pressure plane as a function of $\mu_B$, $\mu_Q$, and $\mu_S$ in (a) the hadronic phase at $P/T^4 = 0.8$ and $T = 0.14$ GeV and (b) the QGP phase at $P/T^4 = 2$ and $T = 0.2$ GeV.}
\label{fig:bqs_P}
\end{figure}

We next focus on the interplay of the multiple conserved charges and plot an arbitrary constant pressure plane in the chemical potential $\mu_B$-$\mu_S$-$\mu_Q$ space at a constant temperature in the hadronic phase in Figure~\ref{fig:bqs_P} (a). This quantifies the chemical potential dependences of this thermodynamic quantity. For demonstration, we choose $P/T^4 = 0.8$ and $T = 0.14$ GeV. The intercepts of each axis, defined implicitly as $P(\mu_B^\mathrm{int},0,0) = P(0,\mu_Q^\mathrm{int},0) = P(0,0,\mu_S^\mathrm{int})$, are ordered as $\mu_B^\mathrm{int} > \mu_S^\mathrm{int} > \mu_Q^\mathrm{int}$, partly reflecting the fact that the lightest particles that carry those charges are protons, kaons, and pions in the hadronic phase, respectively (\ref{eq:P_had}). 

Figure~\ref{fig:bqs_P} (b) presents the same in the QGP phase. Here $P/T^4 = 2$ and $T = 0.2$ GeV are considered. The ordering of the intercepts can be seen to be $\mu_B^\mathrm{int} > \mu_Q^\mathrm{int} > \mu_S^\mathrm{int}$ in the QGP phase. This is consistent with the na\"ive expectation that $\mu_B/3 \sim 2\mu_Q/3 \sim \mu_S$ in the massless parton gas limit (\ref{Psus2})-(\ref{mus}). The intercept $\mu_S^\mathrm{int}$ is slightly larger owing to the fact that it is still close to the crossover transition and that the strange quarks have a non-negligible mass. $\mu_B$ takes a maximum value at some positive finite $\mu_S$ because $s$ quarks have positive net baryon number and negative strangeness (\ref{mus}), \textit{i.e.}, the leading-order off-diagonal susceptibility between the net baryon and strangeness is negative. This is not the case for the cross-coupling between the electric charge and the net baryon or strangeness.

\begin{figure}[tb]
\includegraphics[width=3.4in]{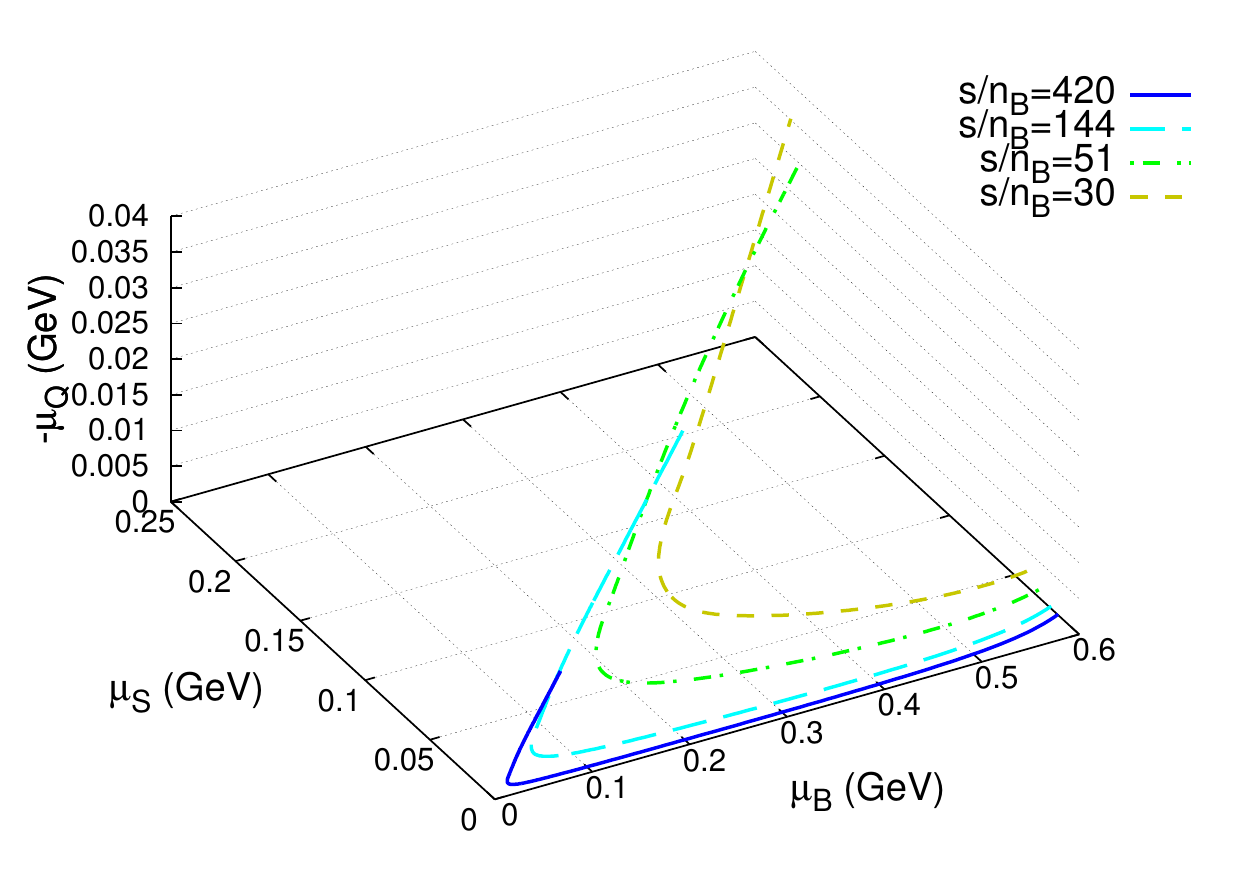}
\caption{(Color online) The solid, long-dashed, dash-dotted, and short-dashed lines are the trajectories for constant $s/n_B$ = 420, 144, 51, and 30, respectively, as functions of $\mu_B$, $\mu_Q$, and $\mu_S$.}
\label{fig:bqs_snB}
\end{figure}

Constant $s/n_B$ lines for the case that $n_S = 0$ and $n_Q = 0.4 n_B$ are plotted in Fig.~\ref{fig:bqs_snB} to illustrate the typical range of the chemical potentials covered by heavy-ion collider experiments. The trajectories coincide at high temperatures because $s/n_B \sim T/\mu_B$ and the conditions on $n_S$ and $n_Q$ make $\mu_S$ and $\mu_Q$ roughy proportional to $\mu_B$. The trajectories slightly bend towards the larger strangeness chemical potential at large $\mu_B$ and small $\mu_S$ regions, which correspond to low temperatures below $T \sim 0.1$ GeV. This could be caused by the suppression of kaons compared with pions owing to the mass difference. The behavior can also be seen in Fig.~\ref{fig:bs}~(b). The bending does not occur for the charge chemical potential because pions, the lightest hadrons, have electric charge. It is note-worthy that broader ranges may be explored in actual collider events since the system is geometrically fluctuating and large local variation of the entropy-to-conserved-charge ratios can occur. 

We note that in the region where the lattice QCD contribution dominates, the validity of our parametrization is limited to the range where $\mu_B/T$ is sufficiently small. From a practical point of view, for the application to nuclear collisions, however, these regions are not expected to much affect the bulk physics, because most of the fluid elements do not go through the large $\mu_B/T$ regions near $T_c$. This can be seen in the constant $s/n_B$ lines shown in Figs.~\ref{fig:b} through \ref{fig:bqs}. 

It would also be interesting to compare our results with the ones obtained by other lattice QCD approaches to finite density regions, such as the one from the imaginary chemical potential method \cite{Vovchenko:2017xad}.

\section{Application to nuclear collisions}
\label{sec4}

To study the effects of imposing strangeness neutrality and a realistic charge-to-baryon ratio on observables in heavy ion collisions, we perform hybrid model calculations of Pb+Pb collisions at center of mass energy $\sqrt{s}=17.3\,{\rm A GeV}$ involving viscous hydrodynamic simulations with the three different equations of state described above and a microscopic hadronic afterburner. A more detailed description of the hybrid model is given in \cite{Schenke:2019ruo}. We compute particle yields and compare particle ratios to experimental data from the Super Proton Synchrotron (SPS) \cite{Afanasiev:2002mx,Alt:2004kq,Alt:2006dk,Alt:2007aa,Alt:2008qm,Alt:2008iv} (compiled in \cite{NA49data}).

For the initial state, we use the dynamical model presented in \cite{Shen:2017bsr}, which provides fluctuating distributions of net baryon and energy-momentum densities in three spatial dimensions. The 3+1D hydrodynamic simulation \textsc{Music}  \cite{Schenke:2010nt,Schenke:2010rr,Schenke:2011bn} is run here with zero bulk viscosity and a constant shear viscosity to entropy density ratio of $\eta/s=0.08$.
We switch from hydrodynamics to the hadron cascade UrQMD \cite{Bass:1998ca,Bleicher:1999xi} at a switching energy density $e_{\rm sw}$, whose value we vary below. 

\begin{figure}[tb]
\includegraphics[width=3.4in]{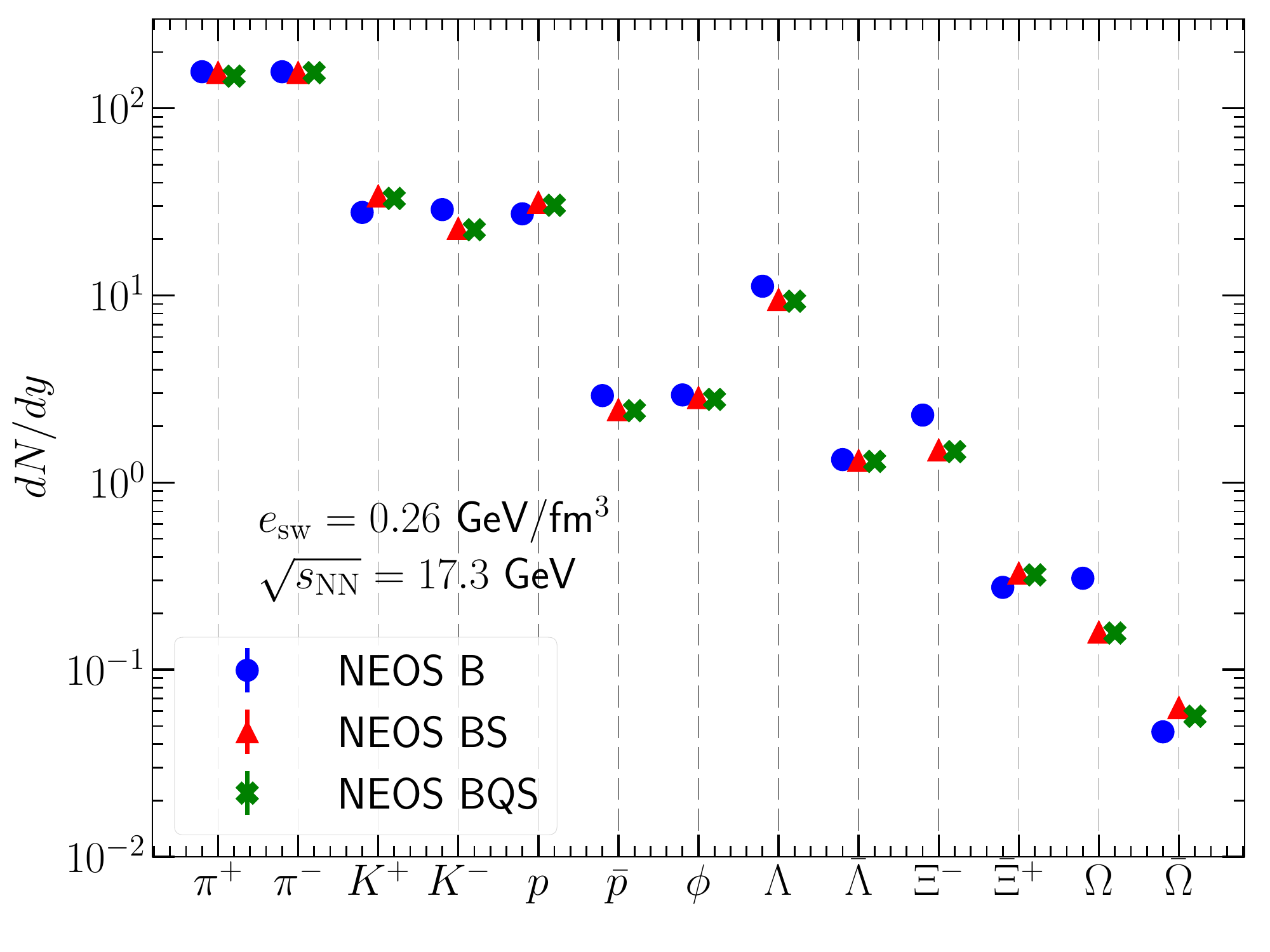}
\includegraphics[width=3.4in]{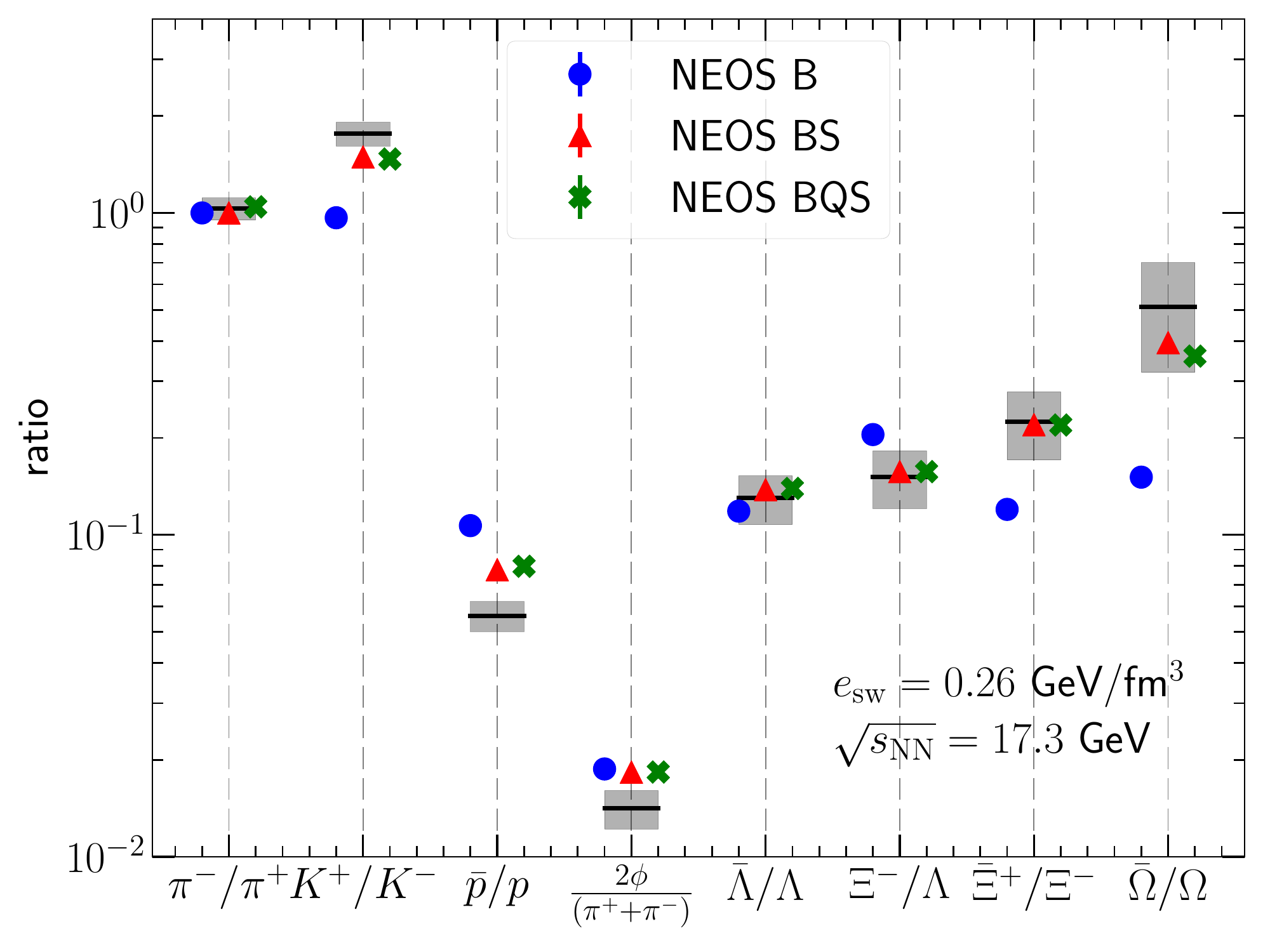}
\caption{(Color online) Top panel: Particle yields from central Pb+Pb collisions at $\sqrt{s}=17.3\,{\rm AGeV}$ determined using a hybrid calculation consisting of hydrodynamics with the indicated equations of state and a hadronic afterburner. Bottom panel: Particle ratios for the three different equations of state from the same calculation compared to experimental data \cite{Afanasiev:2002mx,Alt:2004kq,Alt:2006dk,Alt:2007aa,Alt:2008qm,Alt:2008iv} (compiled in \cite{NA49data}).}
\label{fig:yieldsEOS}
\end{figure}

In Fig.\,\ref{fig:yieldsEOS} we show the particle yields (top) and particle ratios (bottom) from these simulations using $e_{\rm sw}=0.26\,{\rm GeV/fm}^3$, and the \textsc{neos} equations of state with different constraints on strangeness and electric charge. One can see that imposing strangeness neutrality has a visible effect - mainly on the strange and anti-strange particle yields. This effect is amplified in the particle ratios. Yields of particles with positive strangeness are increased while those of particles with negative strangeness are decreased, which is due to the finite positive strangeness chemical potential present in \textsc{neos} BS (and \textsc{neos} BQS).

The agreement between the theoretical calculations and experimental data is improved for most particles with strangeness in {\sc neos} BS.
In the meson sector where $\mu_B$ is absent, the ratio of $K^+$ over $K^-$ gets enhanced by the strangeness neutrality condition and agreement with experimental data at SPS energy improves.

Protons, and to a lesser degree anti-protons, are modified, because in the presence of $\mu_S$, the baryon chemical potential $\mu_B$ also changes. The small change for pions is likely due to the modification of resonance abundances when going from \textsc{neos} B to \textsc{neos} BS.

Introducing the constraint on the electric charge by using \textsc{neos} BQS, we find only very mild modifications of the particle yields. The negative $\mu_Q$ leads to a slight increase of negative relative to positive charged particles, as can be best seen in the plot of the particle ratios as the difference between the points for \textsc{neos} BS and \textsc{neos} BQS. The introduction of $\mu_Q$ can explain at least qualitatively that $\pi^-$ are slightly more abundant than $\pi^+$. The ratio $\bar{\Omega}/\Omega$ behaves in the opposite way. Possibly changes of $\mu_B$ and $\mu_S$ when introducing $\mu_Q$ could contribute to this behavior. We note that the main effect of baryon--anti-baryon annihilation within UrQMD is the reduction of the anti-proton yield by approximately 35\% at $\sqrt{s_\mathrm{NN}} = 17.3$\,GeV. Yields of $\bar{\Lambda}$ and  $\bar{\Omega}$ are reduced by 25\% and 20\%, respectively.

\begin{figure}[tb]
\includegraphics[width=3.4in]{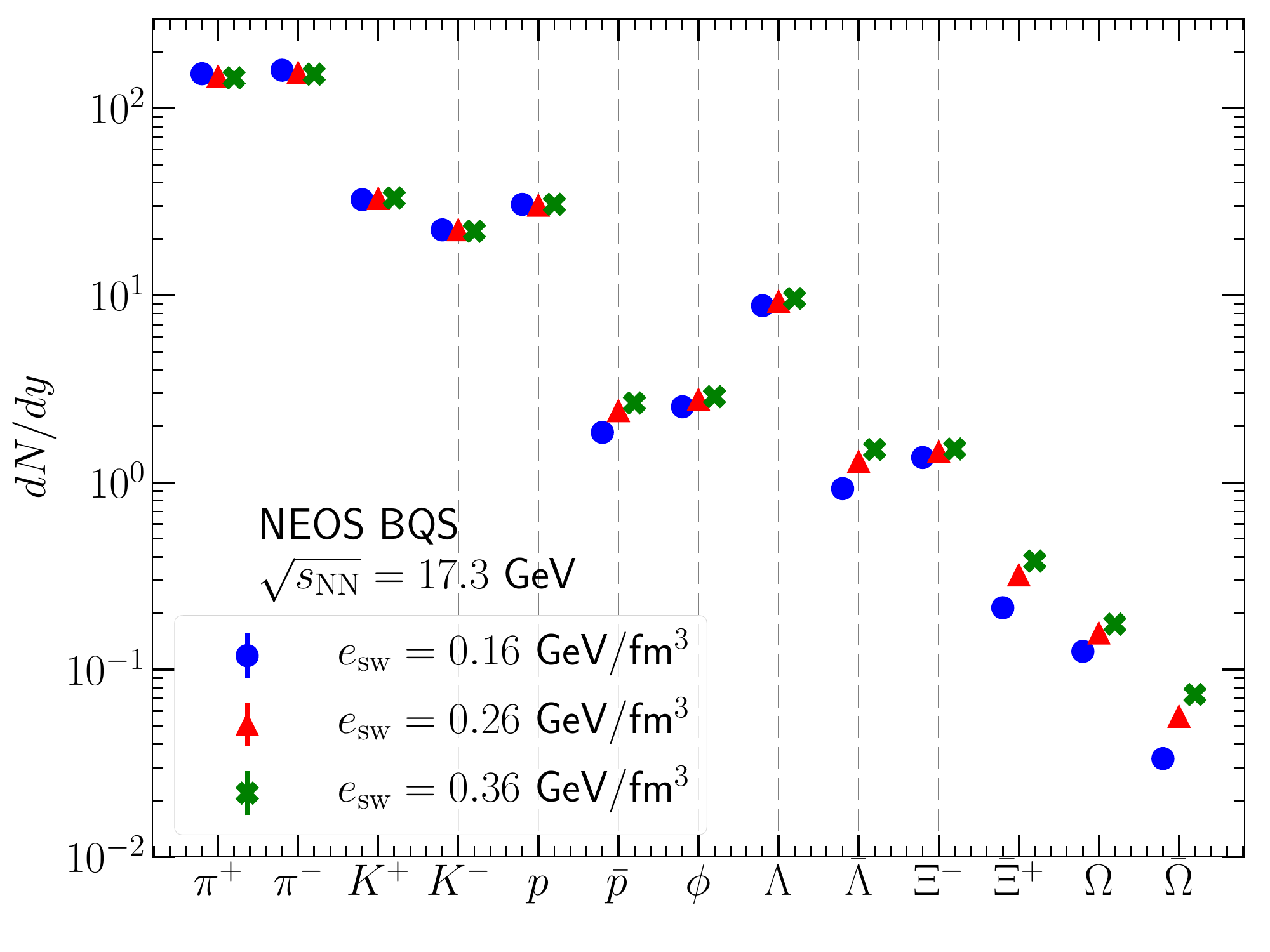}
\includegraphics[width=3.4in]{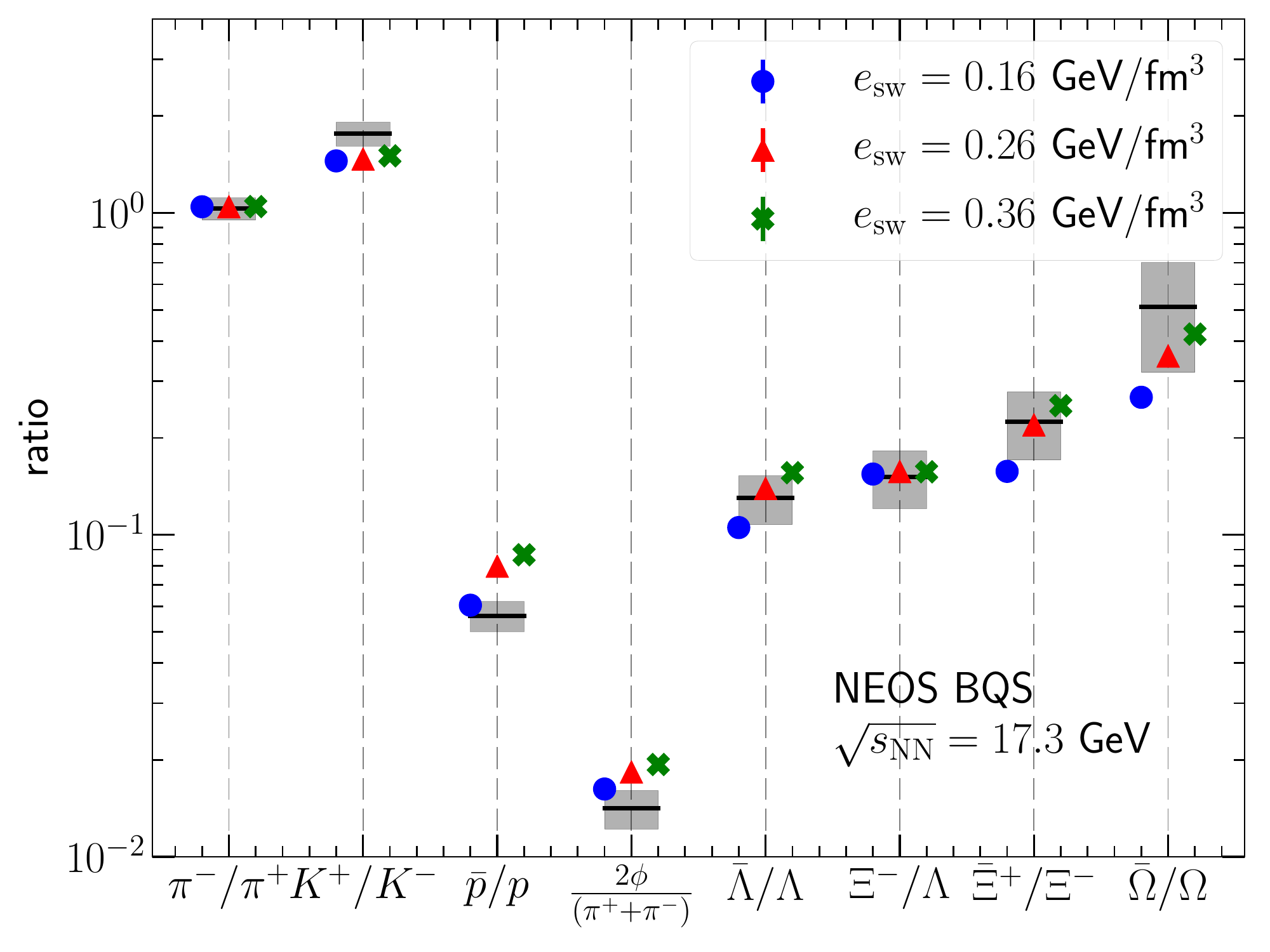}
\caption{(Color online) Top panel: Particle yields from central Pb+Pb collisions at $\sqrt{s}=17.3\,{\rm AGeV}$ determined using a hybrid calculation consisting of hydrodynamics with the \textsc{neos} BQS equation of state and a hadronic afterburner, using three different switching energy densities. Bottom panel: Particle ratios for the three different switching energy densities from the same calculation compared to experimental data \cite{Afanasiev:2002mx,Alt:2004kq,Alt:2006dk,Alt:2007aa,Alt:2008qm,Alt:2008iv} (compiled in \cite{NA49data}).}
\label{fig:yieldseden}
\end{figure}

In Fig.\,\ref{fig:yieldseden} we study the effect of different switching energy densities on particle yields (top) and ratios (bottom) for \textsc{neos} BQS. Using a lower switching energy density means assuming that the system is fully chemically equilibrated down to lower temperatures. 
Anti-baryons are most sensitive to the switching energy density. The reason could be that lower $e_{\rm sw}$ means lower temperature at switching, which goes along with a larger baryon chemical potential (see Fig.\,\ref{fig:b_vs_bs}). Lower temperature tends to decrease heavier particles' yields, while the larger baryon chemical potential will lead to more baryons, weakening the effect of lower temperature, and to less anti-baryons, adding to the effect. 
We find that the experimental data on particle ratios prefers a switching temperature between $e_{\rm sw}=0.16$ and $0.36\,{\rm GeV/fm}^3$.

\section{Conclusions}
\label{sec5}

By matching the Taylor expanded lattice QCD equation of state in the high-temperature region to that of a hadron resonance gas model at low temperature, we have constructed the QCD equation of state at finite net baryon, electric charge, and strangeness chemical potentials. We employ phenomenologically motivated sixth order susceptibilities to allow for a smooth matching that respects the thermodynamic monotonicity conditions. 
The equation of state is designed to be used in simulations of collisions of heavy nuclei such as Au or Pb in a wide range of collision energies explored by the beam energy scan programs.

The equation of state without strangeness chemical potential leads to the violation of the strangeness neutrality condition, which should hold in heavy ion collisions. A positive finite strangeness chemical potential is observed when the condition $n_S = 0$ is kept. Also enforcing the approximate relation between electric charge and baryon number for heavy nuclei, $n_Q/n_B = 0.4$, introduces a negative finite electric chemical potential. These constraints should be important for understanding the difference between particle yields and flow observables of particles and anti-particles within hydrodynamic models of heavy-ion collisions.

We have presented the effect of enforcing different constraints on strangeness and electric charge on the particle yields and ratios in $\sqrt{s}=17.3\,{\rm A GeV}$ Pb+Pb collisions. While strange and anti-strange particles are most affected by these constraints, modifications of non-strange particles were also observed, mostly driven by the finite $\mu_S$, less so the finite, negative, $\mu_Q$. This is understood by the fact that the introduction of $\mu_S$ (and $\mu_Q$) will also alter $\mu_B$.

These results are also important because they imply that we do not explore the $\mu_B$-$T$ plane in the BES experiments, but a certain slice in the $\mu_B$-$\mu_Q$-$\mu_S$-$T$ hyper-plane. This can affect the search of the QCD critical point because the traditional critical point at $\mu_Q = \mu_S = 0$ may not be reached. 

Further importance may arise with regard to isobar collisions. The equation of state can be different for different isobar collision systems, which should be taken into account for correctly understanding the background signals to the chiral magnetic effect. A similar discussion is applicable to small systems, where light nuclei such as proton, deuteron, or $^3$He tend to have larger $Z/A$ ratios, and the sign of the electric charge chemical potential can be flipped. It is also possible to perform event-by-event hydrodynamic analyses distinguishing protons and neutrons.

Future prospects for model improvements include introduction of the full sixth order susceptibilities from lattice QCD calculations. Although they are vanishing in the high temperature limit, they could play a non-trivial role near the crossover at larger chemical potentials. 

Our equation of state model \textsc{neos} is publicly available \cite{neos}.

\begin{acknowledgments}
The authors thank Frithjof Karsch, Swagato Mukherjee, and Sayantan Sharma for useful discussion. AM is supported by JSPS KAKENHI Grant Number JP19K14722. BPS is supported under DOE Contract No. DE-SC0012704. CS is supported under DOE Contract No. DE-SC0013460. This research used resources of the National Energy Research Scientific Computing Center, which is supported by the Office of Science of the U.S. Department of Energy under Contract No. DE-AC02-05CH11231. This work is supported in part by the U.S. Department of Energy, Office of Science, Office of Nuclear Physics, within the framework of the Beam Energy Scan Theory (BEST) Topical Collaboration.
\end{acknowledgments}

\appendix

\section{The Stefan-Boltzmann limit}
\label{sec:A}

For the massless non-interacting system of $u$, $d$ and $s$, the analytical expression of the pressure reads
\begin{eqnarray}
\frac{P}{T^4} &=& \frac{8\pi^2}{45} + \frac{7\pi^2}{60} N_f + \frac{1}{2} \sum_{f=u,d,s} \bigg( \frac{\mu_f}{T} \bigg)^2 \nonumber \\
&+& \frac{1}{4\pi^2} \sum_{f=u,d,s}  \bigg( \frac{\mu_f}{T} \bigg)^4 , \label{Psus2}
\end{eqnarray}
where $N_f = 3$ is the number of flavors. The speed of sound (\ref{eq:cs2}) is independent of the temperature and chemical potentials, $c_s^2 = 1/3$. It is note-worthy that the sixth and higher order susceptibilities are vanishing in this case. The susceptibilities of the conserved charges are given using Eqs.~(\ref{Psus}), (\ref{Psus2}) and 
\begin{eqnarray}
\mu_u &=& \frac{1}{3} \mu_B + \frac{2}{3} \mu_Q, \label{muu}\\
\mu_d &=& \frac{1}{3} \mu_B - \frac{1}{3} \mu_Q, \label{mud}\\
\mu_s &=& \frac{1}{3} \mu_B - \frac{1}{3} \mu_Q - \mu_S. \label{mus}
\end{eqnarray} 
The second-order susceptibilities in the Stefan-Boltzmann limit are 
\begin{eqnarray}
\chi_2^B &=& \frac{1}{3}, \ \ 
\chi_2^Q = \frac{2}{3}, \ \
\chi_2^S = 1, \label{eq:chi2} \\
\chi_{1,1}^{B,Q} &=& 0, \ \
\chi_{1,1}^{B,S} = - \frac{1}{3}, \ \
\chi_{1,1}^{Q,S} = \frac{1}{3},  \label{eq:chi11}
\end{eqnarray}
and the fourth-order ones are
\begin{eqnarray}
\chi_4^B &=& \frac{2}{9 \pi^2}, \ \chi_4^Q = \frac{4}{3 \pi^2}, \ \chi_4^S = \frac{6}{\pi^2}, \\
\chi_{3,1}^{B,S} &=& -\frac{2}{9 \pi^2}, \ \chi_{2,2}^{B,S} = \frac{2}{3 \pi^2}, \ \chi_{1,3}^{B,S} = -\frac{2}{\pi^2},  \\
\chi_{3,1}^{B,Q} &=& 0, \ \chi_{2,2}^{B,Q} = \frac{4}{9 \pi^2}, \ \chi_{1,3}^{B,Q} = \frac{4}{9\pi^2}, \\
\chi_{3,1}^{Q,S} &=& \frac{2}{9\pi^2}, \ \chi_{2,2}^{Q,S} = \frac{2}{3 \pi^2}, \ \chi_{1,3}^{Q,S} = \frac{2}{\pi^2}, \\
\chi_{2,1,1}^{B,Q,S} &=& \frac{2}{9 \pi^2}, \ \chi_{1,2,1}^{B,Q,S} = -\frac{2}{9 \pi^2}, \ \chi_{1,1,2}^{B,Q,S} = - \frac{2}{3 \pi^2}. \label{eq:chi4} \nonumber \\
\end{eqnarray}
They are used as anchors for the high temperature behaviors of the equation of state, where lattice QCD data are scarce, so that the basic thermodynamic features are not violated in the large $T$ limit.

One can analytically solve the linearized equations for charge densities $n_B$, $n_Q$, and $n_S$ obtained by keeping the second order diagonal and off-diagonal susceptibilities:
\begin{eqnarray}
\begin{pmatrix} 
n_B \\ n_Q \\ n_S
\end{pmatrix}
= T^2
\begin{pmatrix} 
\chi_2^B & \chi_{1,1}^{B,Q} & \chi_{1,1}^{B,S} \\ \chi_{1,1}^{B,Q} & \chi_2^Q & \chi_{1,1}^{Q,S} \\ \chi_{1,1}^{B,S} & \chi_{1,1}^{Q,S} & \chi_2^S
\end{pmatrix}
\begin{pmatrix} 
\mu_B \\ \mu_Q \\ \mu_S
\end{pmatrix}.
\end{eqnarray}
The solutions are 
\begin{eqnarray}
\mu_B &=& (5 n_B-n_Q+2n_S)/T^2, \\
\mu_Q &=& (- n_B+2n_Q+n_S)/T^2, \label{eq:pg-q}\\
\mu_S &=& (2 n_B - n_Q + 2n_S)/T^2,
\end{eqnarray}
in the Stefan-Boltzmann limit. In the case where $n_S = 0$ and $n_Q = 0.4 n_B$, those can be expressed as $\mu_B = 4.6 n_B/T^2$, $\mu_Q = - 0.2 n_B/T^2$, and $\mu_S = 1.6 n_B/T^2$. It is worth mentioning that the sign of $\mu_Q$ is rather sensitive to the proportionality constant between the net baryon and charge densities and turns positive at $n_Q = 0.5 n_B$.

\section{Parametrizations of susceptibilities}
\label{sec:B}

The parametrizations of the diagonal and off-diagonal susceptibilities at zero chemical potentials in the regime above $T_c$ are presented here. The second-order susceptibilities are parametrized as
\begin{eqnarray} 
\chi_2 &=& h_0 \bigg(1-\frac{h_1}{T^2} \bigg) g^{+}_1 g^{+}_2 + h_2 T^{n} g^{-}_2 ,
\end{eqnarray}
where
\begin{eqnarray} 
g_i^{\pm}(T_i,\Delta T_i) = \frac{1}{2} \bigg[1\pm \tanh \bigg(\frac{T-T_i}{\Delta T_i}\bigg) \bigg] .
\end{eqnarray}
The parameters are listed in Table~\ref{table:chi2}. 

The functional forms for the fourth-order susceptibilities are
\begin{eqnarray} 
\chi_4 &=& \bigg(h_3+\frac{h_4}{T}+\frac{h_5}{T^2} \bigg) g^{+}_3 + h_6 \chi_2 g^{-}_3 ,
\end{eqnarray}
where the base $\chi_2$ is chosen for purely parametric purposes. The individual parameters can be found in Table~\ref{table:chi4}. The Stefan-Boltzmann limits are used to regulate the high temperature behavior of the parametrizations. It is note-worthy that the $h_3$ values are typically not the exact Stefan-Boltzmann values because they are parameters for the fitting which is valid conservatively up to around 600 MeV, and the convergence of the fitting functions can be rather slow. Also, the lattice data itself does not approach the limit at around $3T_c$ in some cases. The fitting can be further improved when lattice QCD data become available for a wider temperature range.

The sixth-order susceptibilities used are 
\begin{eqnarray} 
\chi_6^B &=& h_7 g^{+}_4 g^{-}_5 g^{-}_6 g^{-}_7 + h_8 g^{+}_4 g^{+}_5 g^{-}_6 g^{-}_7 \nonumber \\
&+& h_9 g^{+}_4 g^{+}_5 g^{+}_6 g^{-}_7 , \\
\chi_{5,1}^{B,Q} &=& h_7 g^{+}_4 g^{-}_5 g^{-}_6 g^{-}_7 g^{-}_8 + h_8 g^{+}_4 g^{+}_5 g^{-}_6 g^{-}_7 g^{-}_8 \nonumber \\
&+& h_9 g^{+}_4 g^{+}_5 g^{+}_6 g^{-}_7 g^{-}_8 + h_{10} g^{+}_4 g^{+}_5 g^{+}_6 g^{+}_7 g^{-}_8 , \\
\chi_{5,1}^{B,S} &=& h_7 g^{+}_4 g^{+}_5 g^{-}_6 g^{-}_7 g^{-}_8 + h_8 g^{+}_4 g^{+}_5 g^{+}_6 g^{-}_7 g^{-}_8 \nonumber \\
&+& h_9 g^{+}_4 g^{+}_5 g^{+}_6 g^{+}_7 g^{-}_8 ,
\end{eqnarray}
where the parameter coefficients are listed in Table~\ref{table:chi6}. Unlike the second- and fourth order ones, they are not based on lattice QCD but determined phenomenologically from the thermodynamic conditions as mentioned in the main text. It should thus be noted that they are effectively contaminated by the contributions of higher order susceptibilities and can be different from those obtained in lattice calculations.

\onecolumngrid

\begin{table}
\begin{tabular}{c|c|c|c|c|c|c|c|c}
\hline \hline
Susceptibility & $h_0$ & $h_1$ (GeV$^2$) & $T_1$ (GeV)& $\Delta T_1$ (GeV) & $T_2$ (GeV) & $\Delta T_2$ (GeV) & $h_2$ (GeV$^{-n}$) & $n$  \\ \hline
$\chi_2^B$ & $3.37\times 10^{-1}$ & $9.65\times 10^{-3}$ & $1.73\times 10^{-1}$ & $2.13\times 10^{-2}$ & $1.69\times 10^{-1}$ & $1.57\times 10^{-2}$ & $3.42\times 10^{5}$ & $8$  \\ \hline
$\chi_2^Q$ & $6.71\times 10^{-1}$ & $6.45\times 10^{-3}$ & $1.40\times 10^{-1}$ & $2.87\times 10^{-2}$ & $1.00\times 10^{-1}$ & $1.00\times 10^{-3}$ & $1.00\times 10^{4}$ & $6$  \\ \hline
$\chi_2^S$ & $1.02\times 10^{0}$ & $1.55\times 10^{-2}$ & $1.69\times 10^{-1}$ & $3.35\times 10^{-2}$ & $1.52\times 10^{-1}$ & $3.17\times 10^{-2}$ & $1.24\times 10^{6}$ & $8$  \\ \hline
$\chi_{1,1}^{B,Q}$ & $1.97 \times 10^{-4}$ & $-2.09\times 10^{0}$ & $1.34\times 10^{-1}$ & $7.28\times 10^{-2}$ & $1.59\times 10^{-1}$ & $2.80\times 10^{-2}$ & $9.00 \times 10^{4}$ & $8$  \\ \hline
$\chi_{1,1}^{B,S}$ & $-3.38\times 10^{-1}$ & $1.28\times 10^{-2}$ & $1.64\times 10^{-1}$ & $3.16\times 10^{-2}$ & $9.99\times 10^{-2}$ & $2.74\times 10^{-2}$ & $-3.42\times 10^{5}$ & $8$  \\ \hline
$\chi_{1,1}^{Q,S}$ & $3.39 \times 10^{-1}$ & $1.46 \times 10^{-2}$ & $1.79 \times 10^{-1}$ & $3.33\times 10^{-2}$ & $1.60 \times 10^{-1}$ & $2.90 \times 10^{-2}$ & $4.34 \times 10^{5}$ & $8$  \\ \hline
\hline
\end{tabular}
\caption{The list of parameters used for the parametrization of the second-order susceptibilities.}
\label{table:chi2}
\end{table}

\begin{table}
\begin{tabular}{c|c|c|c|c|c|c|c}
\hline \hline
Susceptibility & $h_3$ & $h_4$ (GeV)& $h_5$ (GeV$^2$)& $T_3$ (GeV) & $\Delta T_3$ (GeV) & $h_6$ & $\chi_2$  \\ \hline
$\chi_4^B$ & $1.45 \times 10^{-2}$ & $2.49\times 10^{-3}$ & $0$ & $1.62\times 10^{-1}$ & $2.27\times 10^{-2}$ & $1.00\times 10^{0}$ & $\chi_2^B$  \\ \hline
$\chi_4^Q$ & $1.35 \times 10^{-1}$ & $0$ & $0$ & $1.61 \times 10^{-1}$ & $1.74 \times 10^{-2}$ & $1.25\times 10^{0}$ & $\chi_2^Q$  \\ \hline
$\chi_4^S$ & $6.36 \times 10^{-1}$ & $-1.12\times 10^{-1}$ & $2.09 \times 10^{-2}$ & $1.65 \times 10^{-1}$ & $1.93 \times 10^{-2}$ & $8.85\times 10^{-1}$ & $\chi_2^S$  \\ \hline
$\chi_{3,1}^{B,Q}$ & $0$ & $0$ & $0$ & $1.63 \times 10^{-1}$ & $1.16 \times 10^{-2}$ & $9.96 \times 10^{-1}$ & $\chi_{1,1}^{B,Q}$  \\ \hline
$\chi_{2,2}^{B,Q}$ & $4.42 \times 10^{-2}$ & $1.31 \times 10^{-3}$ & $-4.79\times 10^{-4}$ & $1.59 \times 10^{-1}$ & $1.42 \times 10^{-2}$ & $7.95 \times 10^{-1}$ & $\chi_{2}^{B}$  \\ \hline
$\chi_{1,3}^{B,Q}$ & $4.25 \times 10^{-2}$ & $4.54 \times 10^{-3}$ & $-1.91\times 10^{-3}$ & $1.58 \times 10^{-1}$ & $1.70 \times 10^{-2}$ & $8.79 \times 10^{-1}$ & $\chi_{2}^{B}$  \\ \hline
$\chi_{3,1}^{B,S}$ & $-2.87 \times 10^{-2}$ & $7.93 \times 10^{-3}$ & $-1.90\times 10^{-3}$ & $1.62 \times 10^{-1}$ & $2.18 \times 10^{-2}$ & $6.60 \times 10^{-1}$ & $\chi_{1,1}^{B,S}$  \\ \hline
$\chi_{2,2}^{B,S}$ & $7.87 \times 10^{-2}$ & $-1.35 \times 10^{-2}$ & $2.60 \times 10^{-3}$ & $1.68 \times 10^{-1}$ & $2.46 \times 10^{-2}$ & $-8.80 \times 10^{-1}$ & $\chi_{1,1}^{B,S}$  \\ \hline
$\chi_{1,3}^{B,S}$ & $-2.04 \times 10^{-1}$ & $1.85 \times 10^{-3}$ & $-7.88 \times 10^{-4}$ & $1.62 \times 10^{-1}$ & $1.97 \times 10^{-2}$ & $9.85 \times 10^{-1}$ & $\chi_{1,1}^{B,S}$  \\ \hline
$\chi_{3,1}^{Q,S}$ & $2.31 \times 10^{-2}$ & $-9.73 \times 10^{-4}$ & $3.42 \times 10^{-4}$ & $1.60 \times 10^{-1}$ & $3.06 \times 10^{-2}$ & $1.08 \times 10^{0}$ & $\chi_{2}^{B}$  \\ \hline
$\chi_{2,2}^{Q,S}$ & $6.88 \times 10^{-2}$ & $-2.24 \times 10^{-3}$ & $9.64 \times 10^{-4}$ & $1.63 \times 10^{-1}$ & $2.60 \times 10^{-2}$ & $1.12 \times 10^{0}$ & $\chi_{2}^{B}$  \\ \hline
$\chi_{1,3}^{Q,S}$ & $2.02 \times 10^{-1}$ & $1.04 \times 10^{-3}$ & $-6.41 \times 10^{-4}$ & $1.80 \times 10^{-1}$ & $3.08 \times 10^{-2}$ & $1.16 \times 10^{0}$ & $\chi_{2}^{B}$  \\ \hline
$\chi_{2,1,1}^{B,Q,S}$ & $2.24 \times 10^{-2}$ & $9.45 \times 10^{-5}$ & $-2.33 \times 10^{-5}$ & $1.62 \times 10^{-1}$ & $1.30 \times 10^{-2}$ & $5.81 \times 10^{-2}$ & $\chi_{2}^{B}$  \\ \hline
$\chi_{1,2,1}^{B,Q,S}$ & $-2.30 \times 10^{-2}$ & $1.00 \times 10^{-3}$ & $-4.84 \times 10^{-4}$ & $1.54 \times 10^{-1}$ & $1.51 \times 10^{-2}$ & $-1.39 \times 10^{-1}$ & $\chi_{2}^{B}$  \\ \hline
$\chi_{1,1,2}^{B,Q,S}$ & $-6.72 \times 10^{-2}$ & $-6.89 \times 10^{-4}$ & $3.00 \times 10^{-4}$ & $1.63 \times 10^{-1}$ & $1.66 \times 10^{-2}$ & $-1.07 \times 10^{-1}$ & $\chi_{2}^{B}$  \\ \hline
\hline
\end{tabular}
\caption{The list of parameters used for the parametrization of the fourth-order susceptibilities.}
\label{table:chi4}
\end{table}

\begin{table}
\begin{tabular}{c|c|c|c|c|c|c|c}
\hline \hline
Susceptibility & $h_7$ & $h_8$ & $h_9$ & $h_{10}$ & $T_4$ (GeV)& $\Delta T_4$ (GeV) & $T_5$ (GeV) \\ \hline
$\chi_6^B$ & $7.54 \times 10^{-2}$ & $2.70 \times 10^{-2}$ & $-1.64\times 10^{-2}$ & - & $1.27 \times 10^{-1}$ & $1.73\times 10^{-2}$ & $1.57 \times 10^{-1}$  \\ \hline
$\chi_{5,1}^{B,Q}$ & $2.59 \times 10^{-2}$ & $1.39 \times 10^{-2}$ & $1.81\times 10^{-2}$ & $8.73 \times 10^{-4}$ & $1.21 \times 10^{-1}$ & $1.12\times 10^{-2}$ & $1.52 \times 10^{-1}$  \\ \hline
$\chi_{5,1}^{B,S}$ & $-5.52 \times 10^{-2}$ & $4.38 \times 10^{-3}$ & $-6.94\times 10^{-3}$ & - & $1.00 \times 10^{-1}$ & $7.50\times 10^{-3}$ & $1.52 \times 10^{-1}$  \\ \hline \hline
& $\Delta T_5$ (GeV) & $T_6$ (GeV) & $\Delta T_6$ (GeV) & $T_7$ (GeV) & $\Delta T_7$ (GeV) & $T_8$ (GeV) & $\Delta T_8$ (GeV) \\ \hline
$\chi_6^B$  & $1.09 \times 10^{-2}$ & $2.17 \times 10^{-1}$ & $5.12 \times 10^{-2}$ & $2.63 \times 10^{-1}$ & $1.43\times 10^{-2}$ & - & -  \\ \hline 
$\chi_{5,1}^{B,Q}$ & $1.11 \times 10^{-2}$ & $1.64 \times 10^{-1}$ & $7.24 \times 10^{-3}$ & $1.96 \times 10^{-1}$ & $2.58 \times 10^{-2}$ & $2.49 \times 10^{-1}$ & $1.55 \times 10^{-2}$  \\ \hline
$\chi_{5,1}^{B,S}$ & $1.20 \times 10^{-2}$ & $1.34 \times 10^{-1}$ & $1.07 \times 10^{-2}$ & $1.72 \times 10^{-1}$ & $1.13 \times 10^{-2}$ & $2.02 \times 10^{-1}$ & $1.81 \times 10^{-2}$  \\ \hline \hline
\end{tabular}
\caption{The list of parameters used for the parametrization of the sixth-order susceptibilities.}
\label{table:chi6}
\end{table}

\twocolumngrid

\bibliography{neos}

\begin{thebibliography}{93}%
\makeatletter
\providecommand \@ifxundefined [1]{%
 \@ifx{#1\undefined}
}%
\providecommand \@ifnum [1]{%
 \ifnum #1\expandafter \@firstoftwo
 \else \expandafter \@secondoftwo
 \fi
}%
\providecommand \@ifx [1]{%
 \ifx #1\expandafter \@firstoftwo
 \else \expandafter \@secondoftwo
 \fi
}%
\providecommand \natexlab [1]{#1}%
\providecommand \enquote  [1]{``#1''}%
\providecommand \bibnamefont  [1]{#1}%
\providecommand \bibfnamefont [1]{#1}%
\providecommand \citenamefont [1]{#1}%
\providecommand \href@noop [0]{\@secondoftwo}%
\providecommand \href [0]{\begingroup \@sanitize@url \@href}%
\providecommand \@href[1]{\@@startlink{#1}\@@href}%
\providecommand \@@href[1]{\endgroup#1\@@endlink}%
\providecommand \@sanitize@url [0]{\catcode `\\12\catcode `\$12\catcode
  `\&12\catcode `\#12\catcode `\^12\catcode `\_12\catcode `\%12\relax}%
\providecommand \@@startlink[1]{}%
\providecommand \@@endlink[0]{}%
\providecommand \url  [0]{\begingroup\@sanitize@url \@url }%
\providecommand \@url [1]{\endgroup\@href {#1}{\urlprefix }}%
\providecommand \urlprefix  [0]{URL }%
\providecommand \Eprint [0]{\href }%
\providecommand \doibase [0]{http://dx.doi.org/}%
\providecommand \selectlanguage [0]{\@gobble}%
\providecommand \bibinfo  [0]{\@secondoftwo}%
\providecommand \bibfield  [0]{\@secondoftwo}%
\providecommand \translation [1]{[#1]}%
\providecommand \BibitemOpen [0]{}%
\providecommand \bibitemStop [0]{}%
\providecommand \bibitemNoStop [0]{.\EOS\space}%
\providecommand \EOS [0]{\spacefactor3000\relax}%
\providecommand \BibitemShut  [1]{\csname bibitem#1\endcsname}%
\let\auto@bib@innerbib\@empty
\bibitem [{\citenamefont {Chodos}\ \emph
  {et~al.}(1974{\natexlab{a}})\citenamefont {Chodos}, \citenamefont {Jaffe},
  \citenamefont {Johnson}, \citenamefont {Thorn},\ and\ \citenamefont
  {Weisskopf}}]{Chodos:1974je}%
  \BibitemOpen
  \bibfield  {author} {\bibinfo {author} {\bibfnamefont {A.}~\bibnamefont
  {Chodos}}, \bibinfo {author} {\bibfnamefont {R.~L.}\ \bibnamefont {Jaffe}},
  \bibinfo {author} {\bibfnamefont {K.}~\bibnamefont {Johnson}}, \bibinfo
  {author} {\bibfnamefont {C.~B.}\ \bibnamefont {Thorn}}, \ and\ \bibinfo
  {author} {\bibfnamefont {V.~F.}\ \bibnamefont {Weisskopf}},\ }\href {\doibase
  10.1103/PhysRevD.9.3471} {\bibfield  {journal} {\bibinfo  {journal} {Phys.
  Rev.}\ }\textbf {\bibinfo {volume} {D9}},\ \bibinfo {pages} {3471} (\bibinfo
  {year} {1974}{\natexlab{a}})}\BibitemShut {NoStop}%
\bibitem [{\citenamefont {Chodos}\ \emph
  {et~al.}(1974{\natexlab{b}})\citenamefont {Chodos}, \citenamefont {Jaffe},
  \citenamefont {Johnson},\ and\ \citenamefont {Thorn}}]{Chodos:1974pn}%
  \BibitemOpen
  \bibfield  {author} {\bibinfo {author} {\bibfnamefont {A.}~\bibnamefont
  {Chodos}}, \bibinfo {author} {\bibfnamefont {R.~L.}\ \bibnamefont {Jaffe}},
  \bibinfo {author} {\bibfnamefont {K.}~\bibnamefont {Johnson}}, \ and\
  \bibinfo {author} {\bibfnamefont {C.~B.}\ \bibnamefont {Thorn}},\ }\href
  {\doibase 10.1103/PhysRevD.10.2599} {\bibfield  {journal} {\bibinfo
  {journal} {Phys. Rev.}\ }\textbf {\bibinfo {volume} {D10}},\ \bibinfo {pages}
  {2599} (\bibinfo {year} {1974}{\natexlab{b}})}\BibitemShut {NoStop}%
\bibitem [{\citenamefont {De~Rujula}\ \emph {et~al.}(1975)\citenamefont
  {De~Rujula}, \citenamefont {Georgi},\ and\ \citenamefont
  {Glashow}}]{DeRujula:1975qlm}%
  \BibitemOpen
  \bibfield  {author} {\bibinfo {author} {\bibfnamefont {A.}~\bibnamefont
  {De~Rujula}}, \bibinfo {author} {\bibfnamefont {H.}~\bibnamefont {Georgi}}, \
  and\ \bibinfo {author} {\bibfnamefont {S.~L.}\ \bibnamefont {Glashow}},\
  }\href {\doibase 10.1103/PhysRevD.12.147} {\bibfield  {journal} {\bibinfo
  {journal} {Phys. Rev.}\ }\textbf {\bibinfo {volume} {D12}},\ \bibinfo {pages}
  {147} (\bibinfo {year} {1975})}\BibitemShut {NoStop}%
\bibitem [{\citenamefont {Nambu}\ and\ \citenamefont
  {Jona-Lasinio}(1961{\natexlab{a}})}]{Nambu:1961tp}%
  \BibitemOpen
  \bibfield  {author} {\bibinfo {author} {\bibfnamefont {Y.}~\bibnamefont
  {Nambu}}\ and\ \bibinfo {author} {\bibfnamefont {G.}~\bibnamefont
  {Jona-Lasinio}},\ }\href {\doibase 10.1103/PhysRev.122.345} {\bibfield
  {journal} {\bibinfo  {journal} {Phys. Rev.}\ }\textbf {\bibinfo {volume}
  {122}},\ \bibinfo {pages} {345} (\bibinfo {year} {1961}{\natexlab{a}})},\
  \bibinfo {note} {[,127(1961)]}\BibitemShut {NoStop}%
\bibitem [{\citenamefont {Nambu}\ and\ \citenamefont
  {Jona-Lasinio}(1961{\natexlab{b}})}]{Nambu:1961fr}%
  \BibitemOpen
  \bibfield  {author} {\bibinfo {author} {\bibfnamefont {Y.}~\bibnamefont
  {Nambu}}\ and\ \bibinfo {author} {\bibfnamefont {G.}~\bibnamefont
  {Jona-Lasinio}},\ }\href {\doibase 10.1103/PhysRev.124.246} {\bibfield
  {journal} {\bibinfo  {journal} {Phys. Rev.}\ }\textbf {\bibinfo {volume}
  {124}},\ \bibinfo {pages} {246} (\bibinfo {year} {1961}{\natexlab{b}})},\
  \bibinfo {note} {[,141(1961)]}\BibitemShut {NoStop}%
\bibitem [{\citenamefont {Brown}\ \emph {et~al.}(1990)\citenamefont {Brown},
  \citenamefont {Butler}, \citenamefont {Chen}, \citenamefont {Christ},
  \citenamefont {Dong}, \citenamefont {Schaffer}, \citenamefont {Unger},\ and\
  \citenamefont {Vaccarino}}]{Brown:1990ev}%
  \BibitemOpen
  \bibfield  {author} {\bibinfo {author} {\bibfnamefont {F.~R.}\ \bibnamefont
  {Brown}}, \bibinfo {author} {\bibfnamefont {F.~P.}\ \bibnamefont {Butler}},
  \bibinfo {author} {\bibfnamefont {H.}~\bibnamefont {Chen}}, \bibinfo {author}
  {\bibfnamefont {N.~H.}\ \bibnamefont {Christ}}, \bibinfo {author}
  {\bibfnamefont {Z.-h.}\ \bibnamefont {Dong}}, \bibinfo {author}
  {\bibfnamefont {W.}~\bibnamefont {Schaffer}}, \bibinfo {author}
  {\bibfnamefont {L.~I.}\ \bibnamefont {Unger}}, \ and\ \bibinfo {author}
  {\bibfnamefont {A.}~\bibnamefont {Vaccarino}},\ }\href {\doibase
  10.1103/PhysRevLett.65.2491} {\bibfield  {journal} {\bibinfo  {journal}
  {Phys. Rev. Lett.}\ }\textbf {\bibinfo {volume} {65}},\ \bibinfo {pages}
  {2491} (\bibinfo {year} {1990})}\BibitemShut {NoStop}%
\bibitem [{\citenamefont {Ali~Khan}\ \emph {et~al.}(2000)\citenamefont
  {Ali~Khan} \emph {et~al.}}]{AliKhan:2000wou}%
  \BibitemOpen
  \bibfield  {author} {\bibinfo {author} {\bibfnamefont {A.}~\bibnamefont
  {Ali~Khan}} \emph {et~al.} (\bibinfo {collaboration} {CP-PACS}),\ }\href
  {\doibase 10.1103/PhysRevD.63.034502} {\bibfield  {journal} {\bibinfo
  {journal} {Phys. Rev.}\ }\textbf {\bibinfo {volume} {D63}},\ \bibinfo {pages}
  {034502} (\bibinfo {year} {2000})},\ \Eprint
  {http://arxiv.org/abs/hep-lat/0008011} {arXiv:hep-lat/0008011 [hep-lat]}
  \BibitemShut {NoStop}%
\bibitem [{\citenamefont {Aoki}\ \emph {et~al.}(2006)\citenamefont {Aoki},
  \citenamefont {Endrodi}, \citenamefont {Fodor}, \citenamefont {Katz},\ and\
  \citenamefont {Szabo}}]{Aoki:2006we}%
  \BibitemOpen
  \bibfield  {author} {\bibinfo {author} {\bibfnamefont {Y.}~\bibnamefont
  {Aoki}}, \bibinfo {author} {\bibfnamefont {G.}~\bibnamefont {Endrodi}},
  \bibinfo {author} {\bibfnamefont {Z.}~\bibnamefont {Fodor}}, \bibinfo
  {author} {\bibfnamefont {S.~D.}\ \bibnamefont {Katz}}, \ and\ \bibinfo
  {author} {\bibfnamefont {K.~K.}\ \bibnamefont {Szabo}},\ }\href {\doibase
  10.1038/nature05120} {\bibfield  {journal} {\bibinfo  {journal} {Nature}\
  }\textbf {\bibinfo {volume} {443}},\ \bibinfo {pages} {675} (\bibinfo {year}
  {2006})},\ \Eprint {http://arxiv.org/abs/hep-lat/0611014}
  {arXiv:hep-lat/0611014 [hep-lat]} \BibitemShut {NoStop}%
\bibitem [{\citenamefont {Borsanyi}\ \emph {et~al.}(2014)\citenamefont
  {Borsanyi}, \citenamefont {Fodor}, \citenamefont {Hoelbling}, \citenamefont
  {Katz}, \citenamefont {Krieg},\ and\ \citenamefont
  {Szabo}}]{Borsanyi:2013bia}%
  \BibitemOpen
  \bibfield  {author} {\bibinfo {author} {\bibfnamefont {S.}~\bibnamefont
  {Borsanyi}}, \bibinfo {author} {\bibfnamefont {Z.}~\bibnamefont {Fodor}},
  \bibinfo {author} {\bibfnamefont {C.}~\bibnamefont {Hoelbling}}, \bibinfo
  {author} {\bibfnamefont {S.~D.}\ \bibnamefont {Katz}}, \bibinfo {author}
  {\bibfnamefont {S.}~\bibnamefont {Krieg}}, \ and\ \bibinfo {author}
  {\bibfnamefont {K.~K.}\ \bibnamefont {Szabo}},\ }\href {\doibase
  10.1016/j.physletb.2014.01.007} {\bibfield  {journal} {\bibinfo  {journal}
  {Phys. Lett.}\ }\textbf {\bibinfo {volume} {B730}},\ \bibinfo {pages} {99}
  (\bibinfo {year} {2014})},\ \Eprint {http://arxiv.org/abs/1309.5258}
  {arXiv:1309.5258 [hep-lat]} \BibitemShut {NoStop}%
\bibitem [{\citenamefont {Bazavov}\ \emph {et~al.}(2014)\citenamefont {Bazavov}
  \emph {et~al.}}]{Bazavov:2014pvz}%
  \BibitemOpen
  \bibfield  {author} {\bibinfo {author} {\bibfnamefont {A.}~\bibnamefont
  {Bazavov}} \emph {et~al.} (\bibinfo {collaboration} {HotQCD}),\ }\href
  {\doibase 10.1103/PhysRevD.90.094503} {\bibfield  {journal} {\bibinfo
  {journal} {Phys. Rev.}\ }\textbf {\bibinfo {volume} {D90}},\ \bibinfo {pages}
  {094503} (\bibinfo {year} {2014})},\ \Eprint {http://arxiv.org/abs/1407.6387}
  {arXiv:1407.6387 [hep-lat]} \BibitemShut {NoStop}%
\bibitem [{\citenamefont {de~Forcrand}(2009)}]{deForcrand:2010ys}%
  \BibitemOpen
  \bibfield  {author} {\bibinfo {author} {\bibfnamefont {P.}~\bibnamefont
  {de~Forcrand}},\ }\bibfield  {booktitle} {\emph {\bibinfo {booktitle}
  {{Proceedings, 27th International Symposium on Lattice field theory (Lattice
  2009): Beijing, P.R. China, July 26-31, 2009}}},\ }\href {\doibase
  10.22323/1.091.0010} {\bibfield  {journal} {\bibinfo  {journal} {PoS}\
  }\textbf {\bibinfo {volume} {LAT2009}},\ \bibinfo {pages} {010} (\bibinfo
  {year} {2009})},\ \Eprint {http://arxiv.org/abs/1005.0539} {arXiv:1005.0539
  [hep-lat]} \BibitemShut {NoStop}%
\bibitem [{\citenamefont {Gavai}\ and\ \citenamefont
  {Gupta}(2001)}]{Gavai:2001fr}%
  \BibitemOpen
  \bibfield  {author} {\bibinfo {author} {\bibfnamefont {R.~V.}\ \bibnamefont
  {Gavai}}\ and\ \bibinfo {author} {\bibfnamefont {S.}~\bibnamefont {Gupta}},\
  }\href {\doibase 10.1103/PhysRevD.64.074506} {\bibfield  {journal} {\bibinfo
  {journal} {Phys. Rev.}\ }\textbf {\bibinfo {volume} {D64}},\ \bibinfo {pages}
  {074506} (\bibinfo {year} {2001})},\ \Eprint
  {http://arxiv.org/abs/hep-lat/0103013} {arXiv:hep-lat/0103013 [hep-lat]}
  \BibitemShut {NoStop}%
\bibitem [{\citenamefont {Allton}\ \emph {et~al.}(2002)\citenamefont {Allton},
  \citenamefont {Ejiri}, \citenamefont {Hands}, \citenamefont {Kaczmarek},
  \citenamefont {Karsch}, \citenamefont {Laermann}, \citenamefont {Schmidt},\
  and\ \citenamefont {Scorzato}}]{Allton:2002zi}%
  \BibitemOpen
  \bibfield  {author} {\bibinfo {author} {\bibfnamefont {C.~R.}\ \bibnamefont
  {Allton}}, \bibinfo {author} {\bibfnamefont {S.}~\bibnamefont {Ejiri}},
  \bibinfo {author} {\bibfnamefont {S.~J.}\ \bibnamefont {Hands}}, \bibinfo
  {author} {\bibfnamefont {O.}~\bibnamefont {Kaczmarek}}, \bibinfo {author}
  {\bibfnamefont {F.}~\bibnamefont {Karsch}}, \bibinfo {author} {\bibfnamefont
  {E.}~\bibnamefont {Laermann}}, \bibinfo {author} {\bibfnamefont
  {C.}~\bibnamefont {Schmidt}}, \ and\ \bibinfo {author} {\bibfnamefont
  {L.}~\bibnamefont {Scorzato}},\ }\href {\doibase 10.1103/PhysRevD.66.074507}
  {\bibfield  {journal} {\bibinfo  {journal} {Phys. Rev.}\ }\textbf {\bibinfo
  {volume} {D66}},\ \bibinfo {pages} {074507} (\bibinfo {year} {2002})},\
  \Eprint {http://arxiv.org/abs/hep-lat/0204010} {arXiv:hep-lat/0204010
  [hep-lat]} \BibitemShut {NoStop}%
\bibitem [{\citenamefont {de~Forcrand}\ and\ \citenamefont
  {Philipsen}(2002)}]{deForcrand:2002hgr}%
  \BibitemOpen
  \bibfield  {author} {\bibinfo {author} {\bibfnamefont {P.}~\bibnamefont
  {de~Forcrand}}\ and\ \bibinfo {author} {\bibfnamefont {O.}~\bibnamefont
  {Philipsen}},\ }\href {\doibase 10.1016/S0550-3213(02)00626-0} {\bibfield
  {journal} {\bibinfo  {journal} {Nucl. Phys.}\ }\textbf {\bibinfo {volume}
  {B642}},\ \bibinfo {pages} {290} (\bibinfo {year} {2002})},\ \Eprint
  {http://arxiv.org/abs/hep-lat/0205016} {arXiv:hep-lat/0205016 [hep-lat]}
  \BibitemShut {NoStop}%
\bibitem [{\citenamefont {D'Elia}\ and\ \citenamefont
  {Lombardo}(2003)}]{DElia:2002tig}%
  \BibitemOpen
  \bibfield  {author} {\bibinfo {author} {\bibfnamefont {M.}~\bibnamefont
  {D'Elia}}\ and\ \bibinfo {author} {\bibfnamefont {M.-P.}\ \bibnamefont
  {Lombardo}},\ }\href {\doibase 10.1103/PhysRevD.67.014505} {\bibfield
  {journal} {\bibinfo  {journal} {Phys. Rev.}\ }\textbf {\bibinfo {volume}
  {D67}},\ \bibinfo {pages} {014505} (\bibinfo {year} {2003})},\ \Eprint
  {http://arxiv.org/abs/hep-lat/0209146} {arXiv:hep-lat/0209146 [hep-lat]}
  \BibitemShut {NoStop}%
\bibitem [{\citenamefont {Guenther}\ \emph {et~al.}(2017)\citenamefont
  {Guenther}, \citenamefont {Bellwied}, \citenamefont {Borsanyi}, \citenamefont
  {Fodor}, \citenamefont {Katz}, \citenamefont {Pasztor}, \citenamefont
  {Ratti},\ and\ \citenamefont {Szab^^c3^^83^^c2^^b3}}]{Gunther:2016vcp}%
  \BibitemOpen
  \bibfield  {author} {\bibinfo {author} {\bibfnamefont {J.~N.}\ \bibnamefont
  {Guenther}}, \bibinfo {author} {\bibfnamefont {R.}~\bibnamefont {Bellwied}},
  \bibinfo {author} {\bibfnamefont {S.}~\bibnamefont {Borsanyi}}, \bibinfo
  {author} {\bibfnamefont {Z.}~\bibnamefont {Fodor}}, \bibinfo {author}
  {\bibfnamefont {S.~D.}\ \bibnamefont {Katz}}, \bibinfo {author}
  {\bibfnamefont {A.}~\bibnamefont {Pasztor}}, \bibinfo {author} {\bibfnamefont
  {C.}~\bibnamefont {Ratti}}, \ and\ \bibinfo {author} {\bibfnamefont {K.~K.}\
  \bibnamefont {Szab^^c3^^83^^c2^^b3}},\ }\bibfield  {booktitle} {\emph
  {\bibinfo {booktitle} {{Proceedings, 26th International Conference on
  Ultra-relativistic Nucleus-Nucleus Collisions (Quark Matter 2017): Chicago,
  Illinois, USA, February 5-11, 2017}}},\ }\href {\doibase
  10.1016/j.nuclphysa.2017.05.044} {\bibfield  {journal} {\bibinfo  {journal}
  {Nucl. Phys.}\ }\textbf {\bibinfo {volume} {A967}},\ \bibinfo {pages} {720}
  (\bibinfo {year} {2017})},\ \Eprint {http://arxiv.org/abs/1607.02493}
  {arXiv:1607.02493 [hep-lat]} \BibitemShut {NoStop}%
\bibitem [{\citenamefont {Pham}(1983)}]{Pham:1983}%
  \BibitemOpen
  \bibfield  {author} {\bibinfo {author} {\bibfnamefont {F.}~\bibnamefont
  {Pham}},\ }\href@noop {} {\bibfield  {journal} {\bibinfo  {journal} {Proc.
  Symp. Pure Math.}\ }\textbf {\bibinfo {volume} {40}},\ \bibinfo {pages} {319}
  (\bibinfo {year} {1983})}\BibitemShut {NoStop}%
\bibitem [{\citenamefont {Witten}(2011)}]{Witten:2010cx}%
  \BibitemOpen
  \bibfield  {author} {\bibinfo {author} {\bibfnamefont {E.}~\bibnamefont
  {Witten}},\ }\bibfield  {booktitle} {\emph {\bibinfo {booktitle}
  {{Chern-Simons gauge theory: 20 years after. Proceedings, Workshop, Bonn,
  Germany, August 3-7, 2009}}},\ }\href@noop {} {\bibfield  {journal} {\bibinfo
   {journal} {AMS/IP Stud. Adv. Math.}\ }\textbf {\bibinfo {volume} {50}},\
  \bibinfo {pages} {347} (\bibinfo {year} {2011})},\ \Eprint
  {http://arxiv.org/abs/1001.2933} {arXiv:1001.2933 [hep-th]} \BibitemShut
  {NoStop}%
\bibitem [{\citenamefont {Parisi}(1983)}]{Parisi:1984cs}%
  \BibitemOpen
  \bibfield  {author} {\bibinfo {author} {\bibfnamefont {G.}~\bibnamefont
  {Parisi}},\ }\href {\doibase 10.1016/0370-2693(83)90525-7} {\bibfield
  {journal} {\bibinfo  {journal} {Phys. Lett.}\ }\textbf {\bibinfo {volume}
  {131B}},\ \bibinfo {pages} {393} (\bibinfo {year} {1983})}\BibitemShut
  {NoStop}%
\bibitem [{\citenamefont {Klauder}\ and\ \citenamefont
  {Petersen}(1985)}]{Klauder:1985}%
  \BibitemOpen
  \bibfield  {author} {\bibinfo {author} {\bibfnamefont {J.~R.}\ \bibnamefont
  {Klauder}}\ and\ \bibinfo {author} {\bibfnamefont {W.~P.}\ \bibnamefont
  {Petersen}},\ }\href {\doibase https://doi.org/10.1007/BF01007974} {\bibfield
   {journal} {\bibinfo  {journal} {J. Stat. Phys.}\ }\textbf {\bibinfo {volume}
  {39}},\ \bibinfo {pages} {53} (\bibinfo {year} {1985})}\BibitemShut {NoStop}%
\bibitem [{\citenamefont {Ambjorn}\ and\ \citenamefont
  {Yang}(1985)}]{Ambjorn:1985iw}%
  \BibitemOpen
  \bibfield  {author} {\bibinfo {author} {\bibfnamefont {J.}~\bibnamefont
  {Ambjorn}}\ and\ \bibinfo {author} {\bibfnamefont {S.~K.}\ \bibnamefont
  {Yang}},\ }\href {\doibase 10.1016/0370-2693(85)90708-7} {\bibfield
  {journal} {\bibinfo  {journal} {Phys. Lett.}\ }\textbf {\bibinfo {volume}
  {165B}},\ \bibinfo {pages} {140} (\bibinfo {year} {1985})}\BibitemShut
  {NoStop}%
\bibitem [{\citenamefont {Fukushima}\ and\ \citenamefont
  {Hatsuda}(2011)}]{Fukushima:2010bq}%
  \BibitemOpen
  \bibfield  {author} {\bibinfo {author} {\bibfnamefont {K.}~\bibnamefont
  {Fukushima}}\ and\ \bibinfo {author} {\bibfnamefont {T.}~\bibnamefont
  {Hatsuda}},\ }\href {\doibase 10.1088/0034-4885/74/1/014001} {\bibfield
  {journal} {\bibinfo  {journal} {Rept. Prog. Phys.}\ }\textbf {\bibinfo
  {volume} {74}},\ \bibinfo {pages} {014001} (\bibinfo {year} {2011})},\
  \Eprint {http://arxiv.org/abs/1005.4814} {arXiv:1005.4814 [hep-ph]}
  \BibitemShut {NoStop}%
\bibitem [{\citenamefont {Asakawa}\ and\ \citenamefont
  {Yazaki}(1989)}]{Asakawa:1989bq}%
  \BibitemOpen
  \bibfield  {author} {\bibinfo {author} {\bibfnamefont {M.}~\bibnamefont
  {Asakawa}}\ and\ \bibinfo {author} {\bibfnamefont {K.}~\bibnamefont
  {Yazaki}},\ }\href {\doibase 10.1016/0375-9474(89)90002-X} {\bibfield
  {journal} {\bibinfo  {journal} {Nucl. Phys.}\ }\textbf {\bibinfo {volume}
  {A504}},\ \bibinfo {pages} {668} (\bibinfo {year} {1989})}\BibitemShut
  {NoStop}%
\bibitem [{\citenamefont {Pratt}\ \emph {et~al.}(2015)\citenamefont {Pratt},
  \citenamefont {Sangaline}, \citenamefont {Sorensen},\ and\ \citenamefont
  {Wang}}]{Pratt:2015zsa}%
  \BibitemOpen
  \bibfield  {author} {\bibinfo {author} {\bibfnamefont {S.}~\bibnamefont
  {Pratt}}, \bibinfo {author} {\bibfnamefont {E.}~\bibnamefont {Sangaline}},
  \bibinfo {author} {\bibfnamefont {P.}~\bibnamefont {Sorensen}}, \ and\
  \bibinfo {author} {\bibfnamefont {H.}~\bibnamefont {Wang}},\ }\href {\doibase
  10.1103/PhysRevLett.114.202301} {\bibfield  {journal} {\bibinfo  {journal}
  {Phys. Rev. Lett.}\ }\textbf {\bibinfo {volume} {114}},\ \bibinfo {pages}
  {202301} (\bibinfo {year} {2015})},\ \Eprint
  {http://arxiv.org/abs/1501.04042} {arXiv:1501.04042 [nucl-th]} \BibitemShut
  {NoStop}%
\bibitem [{\citenamefont {Sangaline}\ and\ \citenamefont
  {Pratt}(2016)}]{Sangaline:2015isa}%
  \BibitemOpen
  \bibfield  {author} {\bibinfo {author} {\bibfnamefont {E.}~\bibnamefont
  {Sangaline}}\ and\ \bibinfo {author} {\bibfnamefont {S.}~\bibnamefont
  {Pratt}},\ }\href {\doibase 10.1103/PhysRevC.93.024908} {\bibfield  {journal}
  {\bibinfo  {journal} {Phys. Rev.}\ }\textbf {\bibinfo {volume} {C93}},\
  \bibinfo {pages} {024908} (\bibinfo {year} {2016})},\ \Eprint
  {http://arxiv.org/abs/1508.07017} {arXiv:1508.07017 [nucl-th]} \BibitemShut
  {NoStop}%
\bibitem [{\citenamefont {Bernhard}\ \emph {et~al.}(2016)\citenamefont
  {Bernhard}, \citenamefont {Moreland}, \citenamefont {Bass}, \citenamefont
  {Liu},\ and\ \citenamefont {Heinz}}]{Bernhard:2016tnd}%
  \BibitemOpen
  \bibfield  {author} {\bibinfo {author} {\bibfnamefont {J.~E.}\ \bibnamefont
  {Bernhard}}, \bibinfo {author} {\bibfnamefont {J.~S.}\ \bibnamefont
  {Moreland}}, \bibinfo {author} {\bibfnamefont {S.~A.}\ \bibnamefont {Bass}},
  \bibinfo {author} {\bibfnamefont {J.}~\bibnamefont {Liu}}, \ and\ \bibinfo
  {author} {\bibfnamefont {U.}~\bibnamefont {Heinz}},\ }\href {\doibase
  10.1103/PhysRevC.94.024907} {\bibfield  {journal} {\bibinfo  {journal} {Phys.
  Rev.}\ }\textbf {\bibinfo {volume} {C94}},\ \bibinfo {pages} {024907}
  (\bibinfo {year} {2016})},\ \Eprint {http://arxiv.org/abs/1605.03954}
  {arXiv:1605.03954 [nucl-th]} \BibitemShut {NoStop}%
\bibitem [{\citenamefont {Pang}\ \emph {et~al.}(2018)\citenamefont {Pang},
  \citenamefont {Zhou}, \citenamefont {Su}, \citenamefont {Petersen},
  \citenamefont {Stoecker},\ and\ \citenamefont {Wang}}]{Pang:2016vdc}%
  \BibitemOpen
  \bibfield  {author} {\bibinfo {author} {\bibfnamefont {L.-G.}\ \bibnamefont
  {Pang}}, \bibinfo {author} {\bibfnamefont {K.}~\bibnamefont {Zhou}}, \bibinfo
  {author} {\bibfnamefont {N.}~\bibnamefont {Su}}, \bibinfo {author}
  {\bibfnamefont {H.}~\bibnamefont {Petersen}}, \bibinfo {author}
  {\bibfnamefont {H.}~\bibnamefont {Stoecker}}, \ and\ \bibinfo {author}
  {\bibfnamefont {X.-N.}\ \bibnamefont {Wang}},\ }\href {\doibase
  10.1038/s41467-017-02726-3} {\bibfield  {journal} {\bibinfo  {journal}
  {Nature Commun.}\ }\textbf {\bibinfo {volume} {9}},\ \bibinfo {pages} {210}
  (\bibinfo {year} {2018})},\ \Eprint {http://arxiv.org/abs/1612.04262}
  {arXiv:1612.04262 [hep-ph]} \BibitemShut {NoStop}%
\bibitem [{\citenamefont {Monnai}\ and\ \citenamefont
  {Ollitrault}(2017)}]{Monnai:2017cbv}%
  \BibitemOpen
  \bibfield  {author} {\bibinfo {author} {\bibfnamefont {A.}~\bibnamefont
  {Monnai}}\ and\ \bibinfo {author} {\bibfnamefont {J.-Y.}\ \bibnamefont
  {Ollitrault}},\ }\href {\doibase 10.1103/PhysRevC.96.044902} {\bibfield
  {journal} {\bibinfo  {journal} {Phys. Rev.}\ }\textbf {\bibinfo {volume}
  {C96}},\ \bibinfo {pages} {044902} (\bibinfo {year} {2017})},\ \Eprint
  {http://arxiv.org/abs/1707.08466} {arXiv:1707.08466 [nucl-th]} \BibitemShut
  {NoStop}%
\bibitem [{\citenamefont {Paquet}\ \emph {et~al.}(2017)\citenamefont {Paquet},
  \citenamefont {Shen}, \citenamefont {Denicol}, \citenamefont {Jeon},\ and\
  \citenamefont {Gale}}]{Paquet:2017mny}%
  \BibitemOpen
  \bibfield  {author} {\bibinfo {author} {\bibfnamefont {J.-F.}\ \bibnamefont
  {Paquet}}, \bibinfo {author} {\bibfnamefont {C.}~\bibnamefont {Shen}},
  \bibinfo {author} {\bibfnamefont {G.}~\bibnamefont {Denicol}}, \bibinfo
  {author} {\bibfnamefont {S.}~\bibnamefont {Jeon}}, \ and\ \bibinfo {author}
  {\bibfnamefont {C.}~\bibnamefont {Gale}},\ }\bibfield  {booktitle} {\emph
  {\bibinfo {booktitle} {{Proceedings, 26th International Conference on
  Ultra-relativistic Nucleus-Nucleus Collisions (Quark Matter 2017): Chicago,
  Illinois, USA, February 5-11, 2017}}},\ }\href {\doibase
  10.1016/j.nuclphysa.2017.06.024} {\bibfield  {journal} {\bibinfo  {journal}
  {Nucl. Phys.}\ }\textbf {\bibinfo {volume} {A967}},\ \bibinfo {pages} {429}
  (\bibinfo {year} {2017})}\BibitemShut {NoStop}%
\bibitem [{\citenamefont {Nonaka}\ and\ \citenamefont
  {Asakawa}(2005)}]{Nonaka:2004pg}%
  \BibitemOpen
  \bibfield  {author} {\bibinfo {author} {\bibfnamefont {C.}~\bibnamefont
  {Nonaka}}\ and\ \bibinfo {author} {\bibfnamefont {M.}~\bibnamefont
  {Asakawa}},\ }\href {\doibase 10.1103/PhysRevC.71.044904} {\bibfield
  {journal} {\bibinfo  {journal} {Phys. Rev.}\ }\textbf {\bibinfo {volume}
  {C71}},\ \bibinfo {pages} {044904} (\bibinfo {year} {2005})},\ \Eprint
  {http://arxiv.org/abs/nucl-th/0410078} {arXiv:nucl-th/0410078 [nucl-th]}
  \BibitemShut {NoStop}%
\bibitem [{\citenamefont {Bluhm}\ \emph {et~al.}(2005)\citenamefont {Bluhm},
  \citenamefont {Kampfer},\ and\ \citenamefont {Soff}}]{Bluhm:2004xn}%
  \BibitemOpen
  \bibfield  {author} {\bibinfo {author} {\bibfnamefont {M.}~\bibnamefont
  {Bluhm}}, \bibinfo {author} {\bibfnamefont {B.}~\bibnamefont {Kampfer}}, \
  and\ \bibinfo {author} {\bibfnamefont {G.}~\bibnamefont {Soff}},\ }\href
  {\doibase 10.1016/j.physletb.2005.05.083} {\bibfield  {journal} {\bibinfo
  {journal} {Phys. Lett.}\ }\textbf {\bibinfo {volume} {B620}},\ \bibinfo
  {pages} {131} (\bibinfo {year} {2005})},\ \Eprint
  {http://arxiv.org/abs/hep-ph/0411106} {arXiv:hep-ph/0411106 [hep-ph]}
  \BibitemShut {NoStop}%
\bibitem [{\citenamefont {Bluhm}\ \emph {et~al.}(2007)\citenamefont {Bluhm},
  \citenamefont {Kampfer}, \citenamefont {Schulze}, \citenamefont {Seipt},\
  and\ \citenamefont {Heinz}}]{Bluhm:2007nu}%
  \BibitemOpen
  \bibfield  {author} {\bibinfo {author} {\bibfnamefont {M.}~\bibnamefont
  {Bluhm}}, \bibinfo {author} {\bibfnamefont {B.}~\bibnamefont {Kampfer}},
  \bibinfo {author} {\bibfnamefont {R.}~\bibnamefont {Schulze}}, \bibinfo
  {author} {\bibfnamefont {D.}~\bibnamefont {Seipt}}, \ and\ \bibinfo {author}
  {\bibfnamefont {U.}~\bibnamefont {Heinz}},\ }\href {\doibase
  10.1103/PhysRevC.76.034901} {\bibfield  {journal} {\bibinfo  {journal} {Phys.
  Rev.}\ }\textbf {\bibinfo {volume} {C76}},\ \bibinfo {pages} {034901}
  (\bibinfo {year} {2007})},\ \Eprint {http://arxiv.org/abs/0705.0397}
  {arXiv:0705.0397 [hep-ph]} \BibitemShut {NoStop}%
\bibitem [{\citenamefont {Steinheimer}\ \emph {et~al.}(2011)\citenamefont
  {Steinheimer}, \citenamefont {Schramm},\ and\ \citenamefont
  {Stocker}}]{Steinheimer:2010ib}%
  \BibitemOpen
  \bibfield  {author} {\bibinfo {author} {\bibfnamefont {J.}~\bibnamefont
  {Steinheimer}}, \bibinfo {author} {\bibfnamefont {S.}~\bibnamefont
  {Schramm}}, \ and\ \bibinfo {author} {\bibfnamefont {H.}~\bibnamefont
  {Stocker}},\ }\href {\doibase 10.1088/0954-3899/38/3/035001} {\bibfield
  {journal} {\bibinfo  {journal} {J. Phys.}\ }\textbf {\bibinfo {volume}
  {G38}},\ \bibinfo {pages} {035001} (\bibinfo {year} {2011})},\ \Eprint
  {http://arxiv.org/abs/1009.5239} {arXiv:1009.5239 [hep-ph]} \BibitemShut
  {NoStop}%
\bibitem [{\citenamefont {Huovinen}\ and\ \citenamefont
  {Petreczky}(2011)}]{Huovinen:2011xc}%
  \BibitemOpen
  \bibfield  {author} {\bibinfo {author} {\bibfnamefont {P.}~\bibnamefont
  {Huovinen}}\ and\ \bibinfo {author} {\bibfnamefont {P.}~\bibnamefont
  {Petreczky}},\ }\bibfield  {booktitle} {\emph {\bibinfo {booktitle} {{Quark
  matter. Proceedings, 22nd International Conference on Ultra-Relativistic
  Nucleus-Nucleus Collisions, Quark Matter 2011, Annecy, France, May 23-28,
  2011}}},\ }\href {\doibase 10.1088/0954-3899/38/12/124103} {\bibfield
  {journal} {\bibinfo  {journal} {J. Phys.}\ }\textbf {\bibinfo {volume}
  {G38}},\ \bibinfo {pages} {124103} (\bibinfo {year} {2011})},\ \Eprint
  {http://arxiv.org/abs/1106.6227} {arXiv:1106.6227 [nucl-th]} \BibitemShut
  {NoStop}%
\bibitem [{\citenamefont {Hempel}\ \emph {et~al.}(2013)\citenamefont {Hempel},
  \citenamefont {Dexheimer}, \citenamefont {Schramm},\ and\ \citenamefont
  {Iosilevskiy}}]{Hempel:2013tfa}%
  \BibitemOpen
  \bibfield  {author} {\bibinfo {author} {\bibfnamefont {M.}~\bibnamefont
  {Hempel}}, \bibinfo {author} {\bibfnamefont {V.}~\bibnamefont {Dexheimer}},
  \bibinfo {author} {\bibfnamefont {S.}~\bibnamefont {Schramm}}, \ and\
  \bibinfo {author} {\bibfnamefont {I.}~\bibnamefont {Iosilevskiy}},\ }\href
  {\doibase 10.1103/PhysRevC.88.014906} {\bibfield  {journal} {\bibinfo
  {journal} {Phys. Rev.}\ }\textbf {\bibinfo {volume} {C88}},\ \bibinfo {pages}
  {014906} (\bibinfo {year} {2013})},\ \Eprint {http://arxiv.org/abs/1302.2835}
  {arXiv:1302.2835 [nucl-th]} \BibitemShut {NoStop}%
\bibitem [{\citenamefont {Albright}\ \emph {et~al.}(2014)\citenamefont
  {Albright}, \citenamefont {Kapusta},\ and\ \citenamefont
  {Young}}]{Albright:2014gva}%
  \BibitemOpen
  \bibfield  {author} {\bibinfo {author} {\bibfnamefont {M.}~\bibnamefont
  {Albright}}, \bibinfo {author} {\bibfnamefont {J.}~\bibnamefont {Kapusta}}, \
  and\ \bibinfo {author} {\bibfnamefont {C.}~\bibnamefont {Young}},\ }\href
  {\doibase 10.1103/PhysRevC.90.024915} {\bibfield  {journal} {\bibinfo
  {journal} {Phys. Rev.}\ }\textbf {\bibinfo {volume} {C90}},\ \bibinfo {pages}
  {024915} (\bibinfo {year} {2014})},\ \Eprint {http://arxiv.org/abs/1404.7540}
  {arXiv:1404.7540 [nucl-th]} \BibitemShut {NoStop}%
\bibitem [{\citenamefont {Albright}\ \emph {et~al.}(2015)\citenamefont
  {Albright}, \citenamefont {Kapusta},\ and\ \citenamefont
  {Young}}]{Albright:2015uua}%
  \BibitemOpen
  \bibfield  {author} {\bibinfo {author} {\bibfnamefont {M.}~\bibnamefont
  {Albright}}, \bibinfo {author} {\bibfnamefont {J.}~\bibnamefont {Kapusta}}, \
  and\ \bibinfo {author} {\bibfnamefont {C.}~\bibnamefont {Young}},\ }\href
  {\doibase 10.1103/PhysRevC.92.044904} {\bibfield  {journal} {\bibinfo
  {journal} {Phys. Rev.}\ }\textbf {\bibinfo {volume} {C92}},\ \bibinfo {pages}
  {044904} (\bibinfo {year} {2015})},\ \Eprint
  {http://arxiv.org/abs/1506.03408} {arXiv:1506.03408 [nucl-th]} \BibitemShut
  {NoStop}%
\bibitem [{\citenamefont {Rougemont}\ \emph {et~al.}(2017)\citenamefont
  {Rougemont}, \citenamefont {Critelli}, \citenamefont {Noronha-Hostler},
  \citenamefont {Noronha},\ and\ \citenamefont {Ratti}}]{Rougemont:2017tlu}%
  \BibitemOpen
  \bibfield  {author} {\bibinfo {author} {\bibfnamefont {R.}~\bibnamefont
  {Rougemont}}, \bibinfo {author} {\bibfnamefont {R.}~\bibnamefont {Critelli}},
  \bibinfo {author} {\bibfnamefont {J.}~\bibnamefont {Noronha-Hostler}},
  \bibinfo {author} {\bibfnamefont {J.}~\bibnamefont {Noronha}}, \ and\
  \bibinfo {author} {\bibfnamefont {C.}~\bibnamefont {Ratti}},\ }\href
  {\doibase 10.1103/PhysRevD.96.014032} {\bibfield  {journal} {\bibinfo
  {journal} {Phys. Rev.}\ }\textbf {\bibinfo {volume} {D96}},\ \bibinfo {pages}
  {014032} (\bibinfo {year} {2017})},\ \Eprint
  {http://arxiv.org/abs/1704.05558} {arXiv:1704.05558 [hep-ph]} \BibitemShut
  {NoStop}%
\bibitem [{\citenamefont {Critelli}\ \emph {et~al.}(2017)\citenamefont
  {Critelli}, \citenamefont {Noronha}, \citenamefont {Noronha-Hostler},
  \citenamefont {Portillo}, \citenamefont {Ratti},\ and\ \citenamefont
  {Rougemont}}]{Critelli:2017oub}%
  \BibitemOpen
  \bibfield  {author} {\bibinfo {author} {\bibfnamefont {R.}~\bibnamefont
  {Critelli}}, \bibinfo {author} {\bibfnamefont {J.}~\bibnamefont {Noronha}},
  \bibinfo {author} {\bibfnamefont {J.}~\bibnamefont {Noronha-Hostler}},
  \bibinfo {author} {\bibfnamefont {I.}~\bibnamefont {Portillo}}, \bibinfo
  {author} {\bibfnamefont {C.}~\bibnamefont {Ratti}}, \ and\ \bibinfo {author}
  {\bibfnamefont {R.}~\bibnamefont {Rougemont}},\ }\href {\doibase
  10.1103/PhysRevD.96.096026} {\bibfield  {journal} {\bibinfo  {journal} {Phys.
  Rev.}\ }\textbf {\bibinfo {volume} {D96}},\ \bibinfo {pages} {096026}
  (\bibinfo {year} {2017})},\ \Eprint {http://arxiv.org/abs/1706.00455}
  {arXiv:1706.00455 [nucl-th]} \BibitemShut {NoStop}%
\bibitem [{\citenamefont {Vovchenko}\ \emph {et~al.}(2018)\citenamefont
  {Vovchenko}, \citenamefont {Steinheimer}, \citenamefont {Philipsen},\ and\
  \citenamefont {Stoecker}}]{Vovchenko:2017gkg}%
  \BibitemOpen
  \bibfield  {author} {\bibinfo {author} {\bibfnamefont {V.}~\bibnamefont
  {Vovchenko}}, \bibinfo {author} {\bibfnamefont {J.}~\bibnamefont
  {Steinheimer}}, \bibinfo {author} {\bibfnamefont {O.}~\bibnamefont
  {Philipsen}}, \ and\ \bibinfo {author} {\bibfnamefont {H.}~\bibnamefont
  {Stoecker}},\ }\href {\doibase 10.1103/PhysRevD.97.114030} {\bibfield
  {journal} {\bibinfo  {journal} {Phys. Rev.}\ }\textbf {\bibinfo {volume}
  {D97}},\ \bibinfo {pages} {114030} (\bibinfo {year} {2018})},\ \Eprint
  {http://arxiv.org/abs/1711.01261} {arXiv:1711.01261 [hep-ph]} \BibitemShut
  {NoStop}%
\bibitem [{\citenamefont {Parotto}\ \emph {et~al.}(2018)\citenamefont
  {Parotto}, \citenamefont {Bluhm}, \citenamefont {Mroczek}, \citenamefont
  {Nahrgang}, \citenamefont {Noronha-Hostler}, \citenamefont {Rajagopal},
  \citenamefont {Ratti}, \citenamefont {Schaefer},\ and\ \citenamefont
  {Stephanov}}]{Parotto:2018pwx}%
  \BibitemOpen
  \bibfield  {author} {\bibinfo {author} {\bibfnamefont {P.}~\bibnamefont
  {Parotto}}, \bibinfo {author} {\bibfnamefont {M.}~\bibnamefont {Bluhm}},
  \bibinfo {author} {\bibfnamefont {D.}~\bibnamefont {Mroczek}}, \bibinfo
  {author} {\bibfnamefont {M.}~\bibnamefont {Nahrgang}}, \bibinfo {author}
  {\bibfnamefont {J.}~\bibnamefont {Noronha-Hostler}}, \bibinfo {author}
  {\bibfnamefont {K.}~\bibnamefont {Rajagopal}}, \bibinfo {author}
  {\bibfnamefont {C.}~\bibnamefont {Ratti}}, \bibinfo {author} {\bibfnamefont
  {T.}~\bibnamefont {Schaefer}}, \ and\ \bibinfo {author} {\bibfnamefont
  {M.}~\bibnamefont {Stephanov}},\ }\href@noop {} {\  (\bibinfo {year}
  {2018})},\ \Eprint {http://arxiv.org/abs/1805.05249} {arXiv:1805.05249
  [hep-ph]} \BibitemShut {NoStop}%
\bibitem [{\citenamefont {Vovchenko}\ \emph {et~al.}(2019)\citenamefont
  {Vovchenko}, \citenamefont {Steinheimer}, \citenamefont {Philipsen},
  \citenamefont {Pasztor}, \citenamefont {Fodor}, \citenamefont {Katz},\ and\
  \citenamefont {Stoecker}}]{Vovchenko:2018zgt}%
  \BibitemOpen
  \bibfield  {author} {\bibinfo {author} {\bibfnamefont {V.}~\bibnamefont
  {Vovchenko}}, \bibinfo {author} {\bibfnamefont {J.}~\bibnamefont
  {Steinheimer}}, \bibinfo {author} {\bibfnamefont {O.}~\bibnamefont
  {Philipsen}}, \bibinfo {author} {\bibfnamefont {A.}~\bibnamefont {Pasztor}},
  \bibinfo {author} {\bibfnamefont {Z.}~\bibnamefont {Fodor}}, \bibinfo
  {author} {\bibfnamefont {S.~D.}\ \bibnamefont {Katz}}, \ and\ \bibinfo
  {author} {\bibfnamefont {H.}~\bibnamefont {Stoecker}},\ }\bibfield
  {booktitle} {\emph {\bibinfo {booktitle} {{Proceedings, 27th International
  Conference on Ultrarelativistic Nucleus-Nucleus Collisions (Quark Matter
  2018): Venice, Italy, May 14-19, 2018}}},\ }\href {\doibase
  10.1016/j.nuclphysa.2018.10.068} {\bibfield  {journal} {\bibinfo  {journal}
  {Nucl. Phys.}\ }\textbf {\bibinfo {volume} {A982}},\ \bibinfo {pages} {859}
  (\bibinfo {year} {2019})},\ \Eprint {http://arxiv.org/abs/1807.06472}
  {arXiv:1807.06472 [hep-lat]} \BibitemShut {NoStop}%
\bibitem [{\citenamefont {Fu}\ \emph {et~al.}(2018{\natexlab{a}})\citenamefont
  {Fu}, \citenamefont {Pawlowski},\ and\ \citenamefont
  {Rennecke}}]{Fu:2018qsk}%
  \BibitemOpen
  \bibfield  {author} {\bibinfo {author} {\bibfnamefont {W.-j.}\ \bibnamefont
  {Fu}}, \bibinfo {author} {\bibfnamefont {J.~M.}\ \bibnamefont {Pawlowski}}, \
  and\ \bibinfo {author} {\bibfnamefont {F.}~\bibnamefont {Rennecke}},\
  }\href@noop {} {\  (\bibinfo {year} {2018}{\natexlab{a}})},\ \Eprint
  {http://arxiv.org/abs/1808.00410} {arXiv:1808.00410 [hep-ph]} \BibitemShut
  {NoStop}%
\bibitem [{\citenamefont {Fu}\ \emph {et~al.}(2018{\natexlab{b}})\citenamefont
  {Fu}, \citenamefont {Pawlowski},\ and\ \citenamefont
  {Rennecke}}]{Fu:2018swz}%
  \BibitemOpen
  \bibfield  {author} {\bibinfo {author} {\bibfnamefont {W.-j.}\ \bibnamefont
  {Fu}}, \bibinfo {author} {\bibfnamefont {J.~M.}\ \bibnamefont {Pawlowski}}, \
  and\ \bibinfo {author} {\bibfnamefont {F.}~\bibnamefont {Rennecke}},\
  }\href@noop {} {\  (\bibinfo {year} {2018}{\natexlab{b}})},\ \Eprint
  {http://arxiv.org/abs/1809.01594} {arXiv:1809.01594 [hep-ph]} \BibitemShut
  {NoStop}%
\bibitem [{\citenamefont {Motornenko}\ \emph {et~al.}(2019)\citenamefont
  {Motornenko}, \citenamefont {Vovchenko}, \citenamefont {Steinheimer},
  \citenamefont {Schramm},\ and\ \citenamefont
  {Stoecker}}]{Motornenko:2018hjw}%
  \BibitemOpen
  \bibfield  {author} {\bibinfo {author} {\bibfnamefont {A.}~\bibnamefont
  {Motornenko}}, \bibinfo {author} {\bibfnamefont {V.}~\bibnamefont
  {Vovchenko}}, \bibinfo {author} {\bibfnamefont {J.}~\bibnamefont
  {Steinheimer}}, \bibinfo {author} {\bibfnamefont {S.}~\bibnamefont
  {Schramm}}, \ and\ \bibinfo {author} {\bibfnamefont {H.}~\bibnamefont
  {Stoecker}},\ }\bibfield  {booktitle} {\emph {\bibinfo {booktitle}
  {{Proceedings, 27th International Conference on Ultrarelativistic
  Nucleus-Nucleus Collisions (Quark Matter 2018): Venice, Italy, May 14-19,
  2018}}},\ }\href {\doibase 10.1016/j.nuclphysa.2018.11.028} {\bibfield
  {journal} {\bibinfo  {journal} {Nucl. Phys.}\ }\textbf {\bibinfo {volume}
  {A982}},\ \bibinfo {pages} {891} (\bibinfo {year} {2019})},\ \Eprint
  {http://arxiv.org/abs/1809.02000} {arXiv:1809.02000 [hep-ph]} \BibitemShut
  {NoStop}%
\bibitem [{\citenamefont {Plumberg}\ \emph {et~al.}(2018)\citenamefont
  {Plumberg}, \citenamefont {Welle},\ and\ \citenamefont
  {Kapusta}}]{Plumberg:2018fxo}%
  \BibitemOpen
  \bibfield  {author} {\bibinfo {author} {\bibfnamefont {C.~J.}\ \bibnamefont
  {Plumberg}}, \bibinfo {author} {\bibfnamefont {T.}~\bibnamefont {Welle}}, \
  and\ \bibinfo {author} {\bibfnamefont {J.~I.}\ \bibnamefont {Kapusta}},\ }in\
  \href@noop {} {\emph {\bibinfo {booktitle} {{12th International Workshop on
  Critical Point and Onset of Deconfinement (CPOD 2018) Corfu, Greece,
  September 24-28, 2018}}}}\ (\bibinfo {year} {2018})\ \Eprint
  {http://arxiv.org/abs/1812.01684} {arXiv:1812.01684 [nucl-th]} \BibitemShut
  {NoStop}%
\bibitem [{\citenamefont {Borsanyi}\ \emph {et~al.}(2012)\citenamefont
  {Borsanyi}, \citenamefont {Fodor}, \citenamefont {Katz}, \citenamefont
  {Krieg}, \citenamefont {Ratti},\ and\ \citenamefont
  {Szabo}}]{Borsanyi:2011sw}%
  \BibitemOpen
  \bibfield  {author} {\bibinfo {author} {\bibfnamefont {S.}~\bibnamefont
  {Borsanyi}}, \bibinfo {author} {\bibfnamefont {Z.}~\bibnamefont {Fodor}},
  \bibinfo {author} {\bibfnamefont {S.~D.}\ \bibnamefont {Katz}}, \bibinfo
  {author} {\bibfnamefont {S.}~\bibnamefont {Krieg}}, \bibinfo {author}
  {\bibfnamefont {C.}~\bibnamefont {Ratti}}, \ and\ \bibinfo {author}
  {\bibfnamefont {K.}~\bibnamefont {Szabo}},\ }\href {\doibase
  10.1007/JHEP01(2012)138} {\bibfield  {journal} {\bibinfo  {journal} {JHEP}\
  }\textbf {\bibinfo {volume} {01}},\ \bibinfo {pages} {138} (\bibinfo {year}
  {2012})},\ \Eprint {http://arxiv.org/abs/1112.4416} {arXiv:1112.4416
  [hep-lat]} \BibitemShut {NoStop}%
\bibitem [{\citenamefont {Bellwied}\ \emph {et~al.}(2015)\citenamefont
  {Bellwied}, \citenamefont {Borsanyi}, \citenamefont {Fodor}, \citenamefont
  {Katz}, \citenamefont {Pasztor}, \citenamefont {Ratti},\ and\ \citenamefont
  {Szabo}}]{Bellwied:2015lba}%
  \BibitemOpen
  \bibfield  {author} {\bibinfo {author} {\bibfnamefont {R.}~\bibnamefont
  {Bellwied}}, \bibinfo {author} {\bibfnamefont {S.}~\bibnamefont {Borsanyi}},
  \bibinfo {author} {\bibfnamefont {Z.}~\bibnamefont {Fodor}}, \bibinfo
  {author} {\bibfnamefont {S.~D.}\ \bibnamefont {Katz}}, \bibinfo {author}
  {\bibfnamefont {A.}~\bibnamefont {Pasztor}}, \bibinfo {author} {\bibfnamefont
  {C.}~\bibnamefont {Ratti}}, \ and\ \bibinfo {author} {\bibfnamefont {K.~K.}\
  \bibnamefont {Szabo}},\ }\href {\doibase 10.1103/PhysRevD.92.114505}
  {\bibfield  {journal} {\bibinfo  {journal} {Phys. Rev.}\ }\textbf {\bibinfo
  {volume} {D92}},\ \bibinfo {pages} {114505} (\bibinfo {year} {2015})},\
  \Eprint {http://arxiv.org/abs/1507.04627} {arXiv:1507.04627 [hep-lat]}
  \BibitemShut {NoStop}%
\bibitem [{\citenamefont {Borsanyi}\ \emph {et~al.}(2018)\citenamefont
  {Borsanyi}, \citenamefont {Fodor}, \citenamefont {Guenther}, \citenamefont
  {Katz}, \citenamefont {Szabo}, \citenamefont {Pasztor}, \citenamefont
  {Portillo},\ and\ \citenamefont {Ratti}}]{Borsanyi:2018grb}%
  \BibitemOpen
  \bibfield  {author} {\bibinfo {author} {\bibfnamefont {S.}~\bibnamefont
  {Borsanyi}}, \bibinfo {author} {\bibfnamefont {Z.}~\bibnamefont {Fodor}},
  \bibinfo {author} {\bibfnamefont {J.~N.}\ \bibnamefont {Guenther}}, \bibinfo
  {author} {\bibfnamefont {S.~K.}\ \bibnamefont {Katz}}, \bibinfo {author}
  {\bibfnamefont {K.~K.}\ \bibnamefont {Szabo}}, \bibinfo {author}
  {\bibfnamefont {A.}~\bibnamefont {Pasztor}}, \bibinfo {author} {\bibfnamefont
  {I.}~\bibnamefont {Portillo}}, \ and\ \bibinfo {author} {\bibfnamefont
  {C.}~\bibnamefont {Ratti}},\ }\href {\doibase 10.1007/JHEP10(2018)205}
  {\bibfield  {journal} {\bibinfo  {journal} {JHEP}\ }\textbf {\bibinfo
  {volume} {10}},\ \bibinfo {pages} {205} (\bibinfo {year} {2018})},\ \Eprint
  {http://arxiv.org/abs/1805.04445} {arXiv:1805.04445 [hep-lat]} \BibitemShut
  {NoStop}%
\bibitem [{\citenamefont {Bazavov}\ \emph
  {et~al.}(2012{\natexlab{a}})\citenamefont {Bazavov} \emph
  {et~al.}}]{Bazavov:2012jq}%
  \BibitemOpen
  \bibfield  {author} {\bibinfo {author} {\bibfnamefont {A.}~\bibnamefont
  {Bazavov}} \emph {et~al.} (\bibinfo {collaboration} {HotQCD}),\ }\href
  {\doibase 10.1103/PhysRevD.86.034509} {\bibfield  {journal} {\bibinfo
  {journal} {Phys. Rev.}\ }\textbf {\bibinfo {volume} {D86}},\ \bibinfo {pages}
  {034509} (\bibinfo {year} {2012}{\natexlab{a}})},\ \Eprint
  {http://arxiv.org/abs/1203.0784} {arXiv:1203.0784 [hep-lat]} \BibitemShut
  {NoStop}%
\bibitem [{\citenamefont {Ding}\ \emph {et~al.}(2015)\citenamefont {Ding},
  \citenamefont {Mukherjee}, \citenamefont {Ohno}, \citenamefont {Petreczky},\
  and\ \citenamefont {Schadler}}]{Ding:2015fca}%
  \BibitemOpen
  \bibfield  {author} {\bibinfo {author} {\bibfnamefont {H.~T.}\ \bibnamefont
  {Ding}}, \bibinfo {author} {\bibfnamefont {S.}~\bibnamefont {Mukherjee}},
  \bibinfo {author} {\bibfnamefont {H.}~\bibnamefont {Ohno}}, \bibinfo {author}
  {\bibfnamefont {P.}~\bibnamefont {Petreczky}}, \ and\ \bibinfo {author}
  {\bibfnamefont {H.~P.}\ \bibnamefont {Schadler}},\ }\href {\doibase
  10.1103/PhysRevD.92.074043} {\bibfield  {journal} {\bibinfo  {journal} {Phys.
  Rev.}\ }\textbf {\bibinfo {volume} {D92}},\ \bibinfo {pages} {074043}
  (\bibinfo {year} {2015})},\ \Eprint {http://arxiv.org/abs/1507.06637}
  {arXiv:1507.06637 [hep-lat]} \BibitemShut {NoStop}%
\bibitem [{\citenamefont {Bazavov}\ \emph {et~al.}(2017)\citenamefont {Bazavov}
  \emph {et~al.}}]{Bazavov:2017dus}%
  \BibitemOpen
  \bibfield  {author} {\bibinfo {author} {\bibfnamefont {A.}~\bibnamefont
  {Bazavov}} \emph {et~al.},\ }\href {\doibase 10.1103/PhysRevD.95.054504}
  {\bibfield  {journal} {\bibinfo  {journal} {Phys. Rev.}\ }\textbf {\bibinfo
  {volume} {D95}},\ \bibinfo {pages} {054504} (\bibinfo {year} {2017})},\
  \Eprint {http://arxiv.org/abs/1701.04325} {arXiv:1701.04325 [hep-lat]}
  \BibitemShut {NoStop}%
\bibitem [{\citenamefont {Denicol}\ \emph {et~al.}(2016)\citenamefont
  {Denicol}, \citenamefont {Monnai},\ and\ \citenamefont
  {Schenke}}]{Denicol:2015nhu}%
  \BibitemOpen
  \bibfield  {author} {\bibinfo {author} {\bibfnamefont {G.}~\bibnamefont
  {Denicol}}, \bibinfo {author} {\bibfnamefont {A.}~\bibnamefont {Monnai}}, \
  and\ \bibinfo {author} {\bibfnamefont {B.}~\bibnamefont {Schenke}},\ }\href
  {\doibase 10.1103/PhysRevLett.116.212301} {\bibfield  {journal} {\bibinfo
  {journal} {Phys. Rev. Lett.}\ }\textbf {\bibinfo {volume} {116}},\ \bibinfo
  {pages} {212301} (\bibinfo {year} {2016})},\ \Eprint
  {http://arxiv.org/abs/1512.01538} {arXiv:1512.01538 [nucl-th]} \BibitemShut
  {NoStop}%
\bibitem [{\citenamefont {Monnai}\ and\ \citenamefont
  {Schenke}(2016)}]{Monnai:2015sca}%
  \BibitemOpen
  \bibfield  {author} {\bibinfo {author} {\bibfnamefont {A.}~\bibnamefont
  {Monnai}}\ and\ \bibinfo {author} {\bibfnamefont {B.}~\bibnamefont
  {Schenke}},\ }\href {\doibase 10.1016/j.physletb.2015.11.063} {\bibfield
  {journal} {\bibinfo  {journal} {Phys. Lett.}\ }\textbf {\bibinfo {volume}
  {B752}},\ \bibinfo {pages} {317} (\bibinfo {year} {2016})},\ \Eprint
  {http://arxiv.org/abs/1509.04103} {arXiv:1509.04103 [nucl-th]} \BibitemShut
  {NoStop}%
\bibitem [{\citenamefont {Shen}\ \emph {et~al.}(2017)\citenamefont {Shen},
  \citenamefont {Denicol}, \citenamefont {Gale}, \citenamefont {Jeon},
  \citenamefont {Monnai},\ and\ \citenamefont {Schenke}}]{Shen:2017ruz}%
  \BibitemOpen
  \bibfield  {author} {\bibinfo {author} {\bibfnamefont {C.}~\bibnamefont
  {Shen}}, \bibinfo {author} {\bibfnamefont {G.}~\bibnamefont {Denicol}},
  \bibinfo {author} {\bibfnamefont {C.}~\bibnamefont {Gale}}, \bibinfo {author}
  {\bibfnamefont {S.}~\bibnamefont {Jeon}}, \bibinfo {author} {\bibfnamefont
  {A.}~\bibnamefont {Monnai}}, \ and\ \bibinfo {author} {\bibfnamefont
  {B.}~\bibnamefont {Schenke}},\ }\bibfield  {booktitle} {\emph {\bibinfo
  {booktitle} {{Proceedings, 26th International Conference on
  Ultra-relativistic Nucleus-Nucleus Collisions (Quark Matter 2017): Chicago,
  Illinois, USA, February 5-11, 2017}}},\ }\href {\doibase
  10.1016/j.nuclphysa.2017.06.008} {\bibfield  {journal} {\bibinfo  {journal}
  {Nucl. Phys.}\ }\textbf {\bibinfo {volume} {A967}},\ \bibinfo {pages} {796}
  (\bibinfo {year} {2017})},\ \Eprint {http://arxiv.org/abs/1704.04109}
  {arXiv:1704.04109 [nucl-th]} \BibitemShut {NoStop}%
\bibitem [{\citenamefont {Shen}\ and\ \citenamefont
  {Schenke}(2018)}]{Shen:2017bsr}%
  \BibitemOpen
  \bibfield  {author} {\bibinfo {author} {\bibfnamefont {C.}~\bibnamefont
  {Shen}}\ and\ \bibinfo {author} {\bibfnamefont {B.}~\bibnamefont {Schenke}},\
  }\href {\doibase 10.1103/PhysRevC.97.024907} {\bibfield  {journal} {\bibinfo
  {journal} {Phys. Rev.}\ }\textbf {\bibinfo {volume} {C97}},\ \bibinfo {pages}
  {024907} (\bibinfo {year} {2018})},\ \Eprint
  {http://arxiv.org/abs/1710.00881} {arXiv:1710.00881 [nucl-th]} \BibitemShut
  {NoStop}%
\bibitem [{\citenamefont {Denicol}\ \emph {et~al.}(2018)\citenamefont
  {Denicol}, \citenamefont {Gale}, \citenamefont {Jeon}, \citenamefont
  {Monnai}, \citenamefont {Schenke},\ and\ \citenamefont
  {Shen}}]{Denicol:2018wdp}%
  \BibitemOpen
  \bibfield  {author} {\bibinfo {author} {\bibfnamefont {G.~S.}\ \bibnamefont
  {Denicol}}, \bibinfo {author} {\bibfnamefont {C.}~\bibnamefont {Gale}},
  \bibinfo {author} {\bibfnamefont {S.}~\bibnamefont {Jeon}}, \bibinfo {author}
  {\bibfnamefont {A.}~\bibnamefont {Monnai}}, \bibinfo {author} {\bibfnamefont
  {B.}~\bibnamefont {Schenke}}, \ and\ \bibinfo {author} {\bibfnamefont
  {C.}~\bibnamefont {Shen}},\ }\href {\doibase 10.1103/PhysRevC.98.034916}
  {\bibfield  {journal} {\bibinfo  {journal} {Phys. Rev.}\ }\textbf {\bibinfo
  {volume} {C98}},\ \bibinfo {pages} {034916} (\bibinfo {year} {2018})},\
  \Eprint {http://arxiv.org/abs/1804.10557} {arXiv:1804.10557 [nucl-th]}
  \BibitemShut {NoStop}%
\bibitem [{\citenamefont {Shen}\ and\ \citenamefont
  {Schenke}(2019)}]{Shen:2018pty}%
  \BibitemOpen
  \bibfield  {author} {\bibinfo {author} {\bibfnamefont {C.}~\bibnamefont
  {Shen}}\ and\ \bibinfo {author} {\bibfnamefont {B.}~\bibnamefont {Schenke}},\
  }\bibfield  {booktitle} {\emph {\bibinfo {booktitle} {{Proceedings, 27th
  International Conference on Ultrarelativistic Nucleus-Nucleus Collisions
  (Quark Matter 2018): Venice, Italy, May 14-19, 2018}}},\ }\href {\doibase
  10.1016/j.nuclphysa.2018.08.007} {\bibfield  {journal} {\bibinfo  {journal}
  {Nucl. Phys.}\ }\textbf {\bibinfo {volume} {A982}},\ \bibinfo {pages} {411}
  (\bibinfo {year} {2019})},\ \Eprint {http://arxiv.org/abs/1807.05141}
  {arXiv:1807.05141 [nucl-th]} \BibitemShut {NoStop}%
\bibitem [{\citenamefont {Gale}\ \emph {et~al.}(2019)\citenamefont {Gale},
  \citenamefont {Jeon}, \citenamefont {McDonald}, \citenamefont {Paquet},\ and\
  \citenamefont {Shen}}]{Gale:2018vuh}%
  \BibitemOpen
  \bibfield  {author} {\bibinfo {author} {\bibfnamefont {C.}~\bibnamefont
  {Gale}}, \bibinfo {author} {\bibfnamefont {S.}~\bibnamefont {Jeon}}, \bibinfo
  {author} {\bibfnamefont {S.}~\bibnamefont {McDonald}}, \bibinfo {author}
  {\bibfnamefont {J.-F.}\ \bibnamefont {Paquet}}, \ and\ \bibinfo {author}
  {\bibfnamefont {C.}~\bibnamefont {Shen}},\ }\bibfield  {booktitle} {\emph
  {\bibinfo {booktitle} {{Proceedings, 27th International Conference on
  Ultrarelativistic Nucleus-Nucleus Collisions (Quark Matter 2018): Venice,
  Italy, May 14-19, 2018}}},\ }\href {\doibase 10.1016/j.nuclphysa.2018.08.005}
  {\bibfield  {journal} {\bibinfo  {journal} {Nucl. Phys.}\ }\textbf {\bibinfo
  {volume} {A982}},\ \bibinfo {pages} {767} (\bibinfo {year} {2019})},\ \Eprint
  {http://arxiv.org/abs/1807.09326} {arXiv:1807.09326 [nucl-th]} \BibitemShut
  {NoStop}%
\bibitem [{\citenamefont {Li}\ and\ \citenamefont {Shen}(2018)}]{Li:2018fow}%
  \BibitemOpen
  \bibfield  {author} {\bibinfo {author} {\bibfnamefont {M.}~\bibnamefont
  {Li}}\ and\ \bibinfo {author} {\bibfnamefont {C.}~\bibnamefont {Shen}},\
  }\href {\doibase 10.1103/PhysRevC.98.064908} {\bibfield  {journal} {\bibinfo
  {journal} {Phys. Rev.}\ }\textbf {\bibinfo {volume} {C98}},\ \bibinfo {pages}
  {064908} (\bibinfo {year} {2018})},\ \Eprint
  {http://arxiv.org/abs/1809.04034} {arXiv:1809.04034 [nucl-th]} \BibitemShut
  {NoStop}%
\bibitem [{\citenamefont {Monnai}(2012)}]{Monnai:2012jc}%
  \BibitemOpen
  \bibfield  {author} {\bibinfo {author} {\bibfnamefont {A.}~\bibnamefont
  {Monnai}},\ }\href {\doibase 10.1103/PhysRevC.86.014908} {\bibfield
  {journal} {\bibinfo  {journal} {Phys. Rev.}\ }\textbf {\bibinfo {volume}
  {C86}},\ \bibinfo {pages} {014908} (\bibinfo {year} {2012})},\ \Eprint
  {http://arxiv.org/abs/1204.4713} {arXiv:1204.4713 [nucl-th]} \BibitemShut
  {NoStop}%
\bibitem [{\citenamefont {Andronic}\ \emph {et~al.}(2006)\citenamefont
  {Andronic}, \citenamefont {Braun-Munzinger},\ and\ \citenamefont
  {Stachel}}]{Andronic:2005yp}%
  \BibitemOpen
  \bibfield  {author} {\bibinfo {author} {\bibfnamefont {A.}~\bibnamefont
  {Andronic}}, \bibinfo {author} {\bibfnamefont {P.}~\bibnamefont
  {Braun-Munzinger}}, \ and\ \bibinfo {author} {\bibfnamefont {J.}~\bibnamefont
  {Stachel}},\ }\href {\doibase 10.1016/j.nuclphysa.2006.03.012} {\bibfield
  {journal} {\bibinfo  {journal} {Nucl. Phys.}\ }\textbf {\bibinfo {volume}
  {A772}},\ \bibinfo {pages} {167} (\bibinfo {year} {2006})},\ \Eprint
  {http://arxiv.org/abs/nucl-th/0511071} {arXiv:nucl-th/0511071 [nucl-th]}
  \BibitemShut {NoStop}%
\bibitem [{\citenamefont {Karpenko}\ \emph {et~al.}(2015)\citenamefont
  {Karpenko}, \citenamefont {Huovinen}, \citenamefont {Petersen},\ and\
  \citenamefont {Bleicher}}]{Karpenko:2015xea}%
  \BibitemOpen
  \bibfield  {author} {\bibinfo {author} {\bibfnamefont {I.~A.}\ \bibnamefont
  {Karpenko}}, \bibinfo {author} {\bibfnamefont {P.}~\bibnamefont {Huovinen}},
  \bibinfo {author} {\bibfnamefont {H.}~\bibnamefont {Petersen}}, \ and\
  \bibinfo {author} {\bibfnamefont {M.}~\bibnamefont {Bleicher}},\ }\href
  {\doibase 10.1103/PhysRevC.91.064901} {\bibfield  {journal} {\bibinfo
  {journal} {Phys. Rev.}\ }\textbf {\bibinfo {volume} {C91}},\ \bibinfo {pages}
  {064901} (\bibinfo {year} {2015})},\ \Eprint
  {http://arxiv.org/abs/1502.01978} {arXiv:1502.01978 [nucl-th]} \BibitemShut
  {NoStop}%
\bibitem [{\citenamefont {Hatta}\ \emph {et~al.}(2015)\citenamefont {Hatta},
  \citenamefont {Monnai},\ and\ \citenamefont {Xiao}}]{Hatta:2015era}%
  \BibitemOpen
  \bibfield  {author} {\bibinfo {author} {\bibfnamefont {Y.}~\bibnamefont
  {Hatta}}, \bibinfo {author} {\bibfnamefont {A.}~\bibnamefont {Monnai}}, \
  and\ \bibinfo {author} {\bibfnamefont {B.-W.}\ \bibnamefont {Xiao}},\ }\href
  {\doibase 10.1103/PhysRevD.92.114010} {\bibfield  {journal} {\bibinfo
  {journal} {Phys. Rev.}\ }\textbf {\bibinfo {volume} {D92}},\ \bibinfo {pages}
  {114010} (\bibinfo {year} {2015})},\ \Eprint
  {http://arxiv.org/abs/1505.04226} {arXiv:1505.04226 [hep-ph]} \BibitemShut
  {NoStop}%
\bibitem [{\citenamefont {Hatta}\ \emph {et~al.}(2016)\citenamefont {Hatta},
  \citenamefont {Monnai},\ and\ \citenamefont {Xiao}}]{Hatta:2015hca}%
  \BibitemOpen
  \bibfield  {author} {\bibinfo {author} {\bibfnamefont {Y.}~\bibnamefont
  {Hatta}}, \bibinfo {author} {\bibfnamefont {A.}~\bibnamefont {Monnai}}, \
  and\ \bibinfo {author} {\bibfnamefont {B.-W.}\ \bibnamefont {Xiao}},\ }\href
  {\doibase 10.1016/j.nuclphysa.2015.12.009} {\bibfield  {journal} {\bibinfo
  {journal} {Nucl. Phys.}\ }\textbf {\bibinfo {volume} {A947}},\ \bibinfo
  {pages} {155} (\bibinfo {year} {2016})},\ \Eprint
  {http://arxiv.org/abs/1507.04690} {arXiv:1507.04690 [hep-ph]} \BibitemShut
  {NoStop}%
\bibitem [{\citenamefont {Toublan}\ and\ \citenamefont
  {Kogut}(2005)}]{Toublan:2004ks}%
  \BibitemOpen
  \bibfield  {author} {\bibinfo {author} {\bibfnamefont {D.}~\bibnamefont
  {Toublan}}\ and\ \bibinfo {author} {\bibfnamefont {J.~B.}\ \bibnamefont
  {Kogut}},\ }\href {\doibase 10.1016/j.physletb.2004.11.018} {\bibfield
  {journal} {\bibinfo  {journal} {Phys. Lett.}\ }\textbf {\bibinfo {volume}
  {B605}},\ \bibinfo {pages} {129} (\bibinfo {year} {2005})},\ \Eprint
  {http://arxiv.org/abs/hep-ph/0409310} {arXiv:hep-ph/0409310 [hep-ph]}
  \BibitemShut {NoStop}%
\bibitem [{\citenamefont {Xu}\ \emph {et~al.}(2011)\citenamefont {Xu},
  \citenamefont {Mao}, \citenamefont {Mukherjee},\ and\ \citenamefont
  {Huang}}]{Xu:2011pz}%
  \BibitemOpen
  \bibfield  {author} {\bibinfo {author} {\bibfnamefont {F.}~\bibnamefont
  {Xu}}, \bibinfo {author} {\bibfnamefont {H.}~\bibnamefont {Mao}}, \bibinfo
  {author} {\bibfnamefont {T.~K.}\ \bibnamefont {Mukherjee}}, \ and\ \bibinfo
  {author} {\bibfnamefont {M.}~\bibnamefont {Huang}},\ }\href {\doibase
  10.1103/PhysRevD.84.074009} {\bibfield  {journal} {\bibinfo  {journal} {Phys.
  Rev.}\ }\textbf {\bibinfo {volume} {D84}},\ \bibinfo {pages} {074009}
  (\bibinfo {year} {2011})},\ \Eprint {http://arxiv.org/abs/1104.0873}
  {arXiv:1104.0873 [hep-ph]} \BibitemShut {NoStop}%
\bibitem [{\citenamefont {Kamikado}\ \emph {et~al.}(2013)\citenamefont
  {Kamikado}, \citenamefont {Strodthoff}, \citenamefont {von Smekal},\ and\
  \citenamefont {Wambach}}]{Kamikado:2012bt}%
  \BibitemOpen
  \bibfield  {author} {\bibinfo {author} {\bibfnamefont {K.}~\bibnamefont
  {Kamikado}}, \bibinfo {author} {\bibfnamefont {N.}~\bibnamefont
  {Strodthoff}}, \bibinfo {author} {\bibfnamefont {L.}~\bibnamefont {von
  Smekal}}, \ and\ \bibinfo {author} {\bibfnamefont {J.}~\bibnamefont
  {Wambach}},\ }\href {\doibase 10.1016/j.physletb.2012.11.055} {\bibfield
  {journal} {\bibinfo  {journal} {Phys. Lett.}\ }\textbf {\bibinfo {volume}
  {B718}},\ \bibinfo {pages} {1044} (\bibinfo {year} {2013})},\ \Eprint
  {http://arxiv.org/abs/1207.0400} {arXiv:1207.0400 [hep-ph]} \BibitemShut
  {NoStop}%
\bibitem [{\citenamefont {Ueda}\ \emph {et~al.}(2013)\citenamefont {Ueda},
  \citenamefont {Nakano}, \citenamefont {Ohnishi}, \citenamefont {Ruggieri},\
  and\ \citenamefont {Sumiyoshi}}]{Ueda:2013sia}%
  \BibitemOpen
  \bibfield  {author} {\bibinfo {author} {\bibfnamefont {H.}~\bibnamefont
  {Ueda}}, \bibinfo {author} {\bibfnamefont {T.~Z.}\ \bibnamefont {Nakano}},
  \bibinfo {author} {\bibfnamefont {A.}~\bibnamefont {Ohnishi}}, \bibinfo
  {author} {\bibfnamefont {M.}~\bibnamefont {Ruggieri}}, \ and\ \bibinfo
  {author} {\bibfnamefont {K.}~\bibnamefont {Sumiyoshi}},\ }\href {\doibase
  10.1103/PhysRevD.88.074006} {\bibfield  {journal} {\bibinfo  {journal} {Phys.
  Rev.}\ }\textbf {\bibinfo {volume} {D88}},\ \bibinfo {pages} {074006}
  (\bibinfo {year} {2013})},\ \Eprint {http://arxiv.org/abs/1304.4331}
  {arXiv:1304.4331 [nucl-th]} \BibitemShut {NoStop}%
\bibitem [{\citenamefont {Barducci}\ \emph {et~al.}(2004)\citenamefont
  {Barducci}, \citenamefont {Casalbuoni}, \citenamefont {Pettini},\ and\
  \citenamefont {Ravagli}}]{Barducci:2004tt}%
  \BibitemOpen
  \bibfield  {author} {\bibinfo {author} {\bibfnamefont {A.}~\bibnamefont
  {Barducci}}, \bibinfo {author} {\bibfnamefont {R.}~\bibnamefont
  {Casalbuoni}}, \bibinfo {author} {\bibfnamefont {G.}~\bibnamefont {Pettini}},
  \ and\ \bibinfo {author} {\bibfnamefont {L.}~\bibnamefont {Ravagli}},\ }\href
  {\doibase 10.1103/PhysRevD.69.096004} {\bibfield  {journal} {\bibinfo
  {journal} {Phys. Rev.}\ }\textbf {\bibinfo {volume} {D69}},\ \bibinfo {pages}
  {096004} (\bibinfo {year} {2004})},\ \Eprint
  {http://arxiv.org/abs/hep-ph/0402104} {arXiv:hep-ph/0402104 [hep-ph]}
  \BibitemShut {NoStop}%
\bibitem [{\citenamefont {Nishida}(2004)}]{Nishida:2003fb}%
  \BibitemOpen
  \bibfield  {author} {\bibinfo {author} {\bibfnamefont {Y.}~\bibnamefont
  {Nishida}},\ }\href {\doibase 10.1103/PhysRevD.69.094501} {\bibfield
  {journal} {\bibinfo  {journal} {Phys. Rev.}\ }\textbf {\bibinfo {volume}
  {D69}},\ \bibinfo {pages} {094501} (\bibinfo {year} {2004})},\ \Eprint
  {http://arxiv.org/abs/hep-ph/0312371} {arXiv:hep-ph/0312371 [hep-ph]}
  \BibitemShut {NoStop}%
\bibitem [{\citenamefont {Son}\ and\ \citenamefont
  {Stephanov}(2001)}]{Son:2000xc}%
  \BibitemOpen
  \bibfield  {author} {\bibinfo {author} {\bibfnamefont {D.~T.}\ \bibnamefont
  {Son}}\ and\ \bibinfo {author} {\bibfnamefont {M.~A.}\ \bibnamefont
  {Stephanov}},\ }\href {\doibase 10.1103/PhysRevLett.86.592} {\bibfield
  {journal} {\bibinfo  {journal} {Phys. Rev. Lett.}\ }\textbf {\bibinfo
  {volume} {86}},\ \bibinfo {pages} {592} (\bibinfo {year} {2001})},\ \Eprint
  {http://arxiv.org/abs/hep-ph/0005225} {arXiv:hep-ph/0005225 [hep-ph]}
  \BibitemShut {NoStop}%
\bibitem [{\citenamefont {Cooper}\ and\ \citenamefont
  {Frye}(1974)}]{Cooper:1974mv}%
  \BibitemOpen
  \bibfield  {author} {\bibinfo {author} {\bibfnamefont {F.}~\bibnamefont
  {Cooper}}\ and\ \bibinfo {author} {\bibfnamefont {G.}~\bibnamefont {Frye}},\
  }\href {\doibase 10.1103/PhysRevD.10.186} {\bibfield  {journal} {\bibinfo
  {journal} {Phys. Rev.}\ }\textbf {\bibinfo {volume} {D10}},\ \bibinfo {pages}
  {186} (\bibinfo {year} {1974})}\BibitemShut {NoStop}%
\bibitem [{\citenamefont {Huovinen}\ and\ \citenamefont
  {Petreczky}(2010)}]{Huovinen:2009yb}%
  \BibitemOpen
  \bibfield  {author} {\bibinfo {author} {\bibfnamefont {P.}~\bibnamefont
  {Huovinen}}\ and\ \bibinfo {author} {\bibfnamefont {P.}~\bibnamefont
  {Petreczky}},\ }\href {\doibase 10.1016/j.nuclphysa.2010.02.015} {\bibfield
  {journal} {\bibinfo  {journal} {Nucl. Phys.}\ }\textbf {\bibinfo {volume}
  {A837}},\ \bibinfo {pages} {26} (\bibinfo {year} {2010})},\ \Eprint
  {http://arxiv.org/abs/0912.2541} {arXiv:0912.2541 [hep-ph]} \BibitemShut
  {NoStop}%
\bibitem [{\citenamefont {Sharma}()}]{Sharma}%
  \BibitemOpen
  \bibfield  {author} {\bibinfo {author} {\bibfnamefont {S.}~\bibnamefont
  {Sharma}},\ }\href@noop {} {\bibinfo  {journal} {private communications}\
  }\BibitemShut {NoStop}%
\bibitem [{\citenamefont {Tanabashi}\ \emph {et~al.}(2018)\citenamefont
  {Tanabashi} \emph {et~al.}}]{Tanabashi:2018oca}%
  \BibitemOpen
\bibfield  {journal} {  }\bibfield  {author} {\bibinfo {author} {\bibfnamefont
  {M.}~\bibnamefont {Tanabashi}} \emph {et~al.} (\bibinfo {collaboration}
  {Particle Data Group}),\ }\href {\doibase 10.1103/PhysRevD.98.030001}
  {\bibfield  {journal} {\bibinfo  {journal} {Phys. Rev.}\ }\textbf {\bibinfo
  {volume} {D98}},\ \bibinfo {pages} {030001} (\bibinfo {year}
  {2018})}\BibitemShut {NoStop}%
\bibitem [{\citenamefont {Cleymans}\ \emph {et~al.}(2006)\citenamefont
  {Cleymans}, \citenamefont {Oeschler}, \citenamefont {Redlich},\ and\
  \citenamefont {Wheaton}}]{Cleymans:2005xv}%
  \BibitemOpen
  \bibfield  {author} {\bibinfo {author} {\bibfnamefont {J.}~\bibnamefont
  {Cleymans}}, \bibinfo {author} {\bibfnamefont {H.}~\bibnamefont {Oeschler}},
  \bibinfo {author} {\bibfnamefont {K.}~\bibnamefont {Redlich}}, \ and\
  \bibinfo {author} {\bibfnamefont {S.}~\bibnamefont {Wheaton}},\ }\href
  {\doibase 10.1103/PhysRevC.73.034905} {\bibfield  {journal} {\bibinfo
  {journal} {Phys. Rev.}\ }\textbf {\bibinfo {volume} {C73}},\ \bibinfo {pages}
  {034905} (\bibinfo {year} {2006})},\ \Eprint
  {http://arxiv.org/abs/hep-ph/0511094} {arXiv:hep-ph/0511094 [hep-ph]}
  \BibitemShut {NoStop}%
\bibitem [{\citenamefont {Bazavov}\ \emph
  {et~al.}(2012{\natexlab{b}})\citenamefont {Bazavov} \emph
  {et~al.}}]{Bazavov:2012vg}%
  \BibitemOpen
  \bibfield  {author} {\bibinfo {author} {\bibfnamefont {A.}~\bibnamefont
  {Bazavov}} \emph {et~al.},\ }\href {\doibase 10.1103/PhysRevLett.109.192302}
  {\bibfield  {journal} {\bibinfo  {journal} {Phys. Rev. Lett.}\ }\textbf
  {\bibinfo {volume} {109}},\ \bibinfo {pages} {192302} (\bibinfo {year}
  {2012}{\natexlab{b}})},\ \Eprint {http://arxiv.org/abs/1208.1220}
  {arXiv:1208.1220 [hep-lat]} \BibitemShut {NoStop}%
\bibitem [{\citenamefont {Vovchenko}\ \emph {et~al.}(2017)\citenamefont
  {Vovchenko}, \citenamefont {Pasztor}, \citenamefont {Fodor}, \citenamefont
  {Katz},\ and\ \citenamefont {Stoecker}}]{Vovchenko:2017xad}%
  \BibitemOpen
  \bibfield  {author} {\bibinfo {author} {\bibfnamefont {V.}~\bibnamefont
  {Vovchenko}}, \bibinfo {author} {\bibfnamefont {A.}~\bibnamefont {Pasztor}},
  \bibinfo {author} {\bibfnamefont {Z.}~\bibnamefont {Fodor}}, \bibinfo
  {author} {\bibfnamefont {S.~D.}\ \bibnamefont {Katz}}, \ and\ \bibinfo
  {author} {\bibfnamefont {H.}~\bibnamefont {Stoecker}},\ }\href {\doibase
  10.1016/j.physletb.2017.10.042} {\bibfield  {journal} {\bibinfo  {journal}
  {Phys. Lett.}\ }\textbf {\bibinfo {volume} {B775}},\ \bibinfo {pages} {71}
  (\bibinfo {year} {2017})},\ \Eprint {http://arxiv.org/abs/1708.02852}
  {arXiv:1708.02852 [hep-ph]} \BibitemShut {NoStop}%
\bibitem [{\citenamefont {Schenke}\ \emph {et~al.}(2019)\citenamefont
  {Schenke}, \citenamefont {Shen},\ and\ \citenamefont
  {Tribedy}}]{Schenke:2019ruo}%
  \BibitemOpen
  \bibfield  {author} {\bibinfo {author} {\bibfnamefont {B.}~\bibnamefont
  {Schenke}}, \bibinfo {author} {\bibfnamefont {C.}~\bibnamefont {Shen}}, \
  and\ \bibinfo {author} {\bibfnamefont {P.}~\bibnamefont {Tribedy}},\
  }\href@noop {} {\  (\bibinfo {year} {2019})},\ \Eprint
  {http://arxiv.org/abs/1901.04378} {arXiv:1901.04378 [nucl-th]} \BibitemShut
  {NoStop}%
\bibitem [{\citenamefont {Afanasiev}\ \emph {et~al.}(2002)\citenamefont
  {Afanasiev} \emph {et~al.}}]{Afanasiev:2002mx}%
  \BibitemOpen
  \bibfield  {author} {\bibinfo {author} {\bibfnamefont {S.~V.}\ \bibnamefont
  {Afanasiev}} \emph {et~al.} (\bibinfo {collaboration} {NA49}),\ }\href
  {\doibase 10.1103/PhysRevC.66.054902} {\bibfield  {journal} {\bibinfo
  {journal} {Phys. Rev.}\ }\textbf {\bibinfo {volume} {C66}},\ \bibinfo {pages}
  {054902} (\bibinfo {year} {2002})},\ \Eprint
  {http://arxiv.org/abs/nucl-ex/0205002} {arXiv:nucl-ex/0205002 [nucl-ex]}
  \BibitemShut {NoStop}%
\bibitem [{\citenamefont {Alt}\ \emph {et~al.}(2005)\citenamefont {Alt} \emph
  {et~al.}}]{Alt:2004kq}%
  \BibitemOpen
  \bibfield  {author} {\bibinfo {author} {\bibfnamefont {C.}~\bibnamefont
  {Alt}} \emph {et~al.} (\bibinfo {collaboration} {NA49}),\ }\href {\doibase
  10.1103/PhysRevLett.94.192301} {\bibfield  {journal} {\bibinfo  {journal}
  {Phys. Rev. Lett.}\ }\textbf {\bibinfo {volume} {94}},\ \bibinfo {pages}
  {192301} (\bibinfo {year} {2005})},\ \Eprint
  {http://arxiv.org/abs/nucl-ex/0409004} {arXiv:nucl-ex/0409004 [nucl-ex]}
  \BibitemShut {NoStop}%
\bibitem [{\citenamefont {Alt}\ \emph {et~al.}(2006)\citenamefont {Alt} \emph
  {et~al.}}]{Alt:2006dk}%
  \BibitemOpen
  \bibfield  {author} {\bibinfo {author} {\bibfnamefont {C.}~\bibnamefont
  {Alt}} \emph {et~al.} (\bibinfo {collaboration} {NA49}),\ }\href {\doibase
  10.1103/PhysRevC.73.044910} {\bibfield  {journal} {\bibinfo  {journal} {Phys.
  Rev.}\ }\textbf {\bibinfo {volume} {C73}},\ \bibinfo {pages} {044910}
  (\bibinfo {year} {2006})}\BibitemShut {NoStop}%
\bibitem [{\citenamefont {Alt}\ \emph {et~al.}(2008{\natexlab{a}})\citenamefont
  {Alt} \emph {et~al.}}]{Alt:2007aa}%
  \BibitemOpen
  \bibfield  {author} {\bibinfo {author} {\bibfnamefont {C.}~\bibnamefont
  {Alt}} \emph {et~al.} (\bibinfo {collaboration} {NA49}),\ }\href {\doibase
  10.1103/PhysRevC.77.024903} {\bibfield  {journal} {\bibinfo  {journal} {Phys.
  Rev.}\ }\textbf {\bibinfo {volume} {C77}},\ \bibinfo {pages} {024903}
  (\bibinfo {year} {2008}{\natexlab{a}})},\ \Eprint
  {http://arxiv.org/abs/0710.0118} {arXiv:0710.0118 [nucl-ex]} \BibitemShut
  {NoStop}%
\bibitem [{\citenamefont {Alt}\ \emph {et~al.}(2008{\natexlab{b}})\citenamefont
  {Alt} \emph {et~al.}}]{Alt:2008qm}%
  \BibitemOpen
  \bibfield  {author} {\bibinfo {author} {\bibfnamefont {C.}~\bibnamefont
  {Alt}} \emph {et~al.} (\bibinfo {collaboration} {NA49}),\ }\href {\doibase
  10.1103/PhysRevC.78.034918} {\bibfield  {journal} {\bibinfo  {journal} {Phys.
  Rev.}\ }\textbf {\bibinfo {volume} {C78}},\ \bibinfo {pages} {034918}
  (\bibinfo {year} {2008}{\natexlab{b}})},\ \Eprint
  {http://arxiv.org/abs/0804.3770} {arXiv:0804.3770 [nucl-ex]} \BibitemShut
  {NoStop}%
\bibitem [{\citenamefont {Alt}\ \emph {et~al.}(2008{\natexlab{c}})\citenamefont
  {Alt} \emph {et~al.}}]{Alt:2008iv}%
  \BibitemOpen
  \bibfield  {author} {\bibinfo {author} {\bibfnamefont {C.}~\bibnamefont
  {Alt}} \emph {et~al.} (\bibinfo {collaboration} {NA49}),\ }\href {\doibase
  10.1103/PhysRevC.78.044907} {\bibfield  {journal} {\bibinfo  {journal} {Phys.
  Rev.}\ }\textbf {\bibinfo {volume} {C78}},\ \bibinfo {pages} {044907}
  (\bibinfo {year} {2008}{\natexlab{c}})},\ \Eprint
  {http://arxiv.org/abs/0806.1937} {arXiv:0806.1937 [nucl-ex]} \BibitemShut
  {NoStop}%
\bibitem [{NA4()}]{NA49data}%
  \BibitemOpen
  \href@noop {} {\bibinfo  {journal} {https://edms.cern.ch/document/1075059}\
  }\BibitemShut {NoStop}%
\bibitem [{\citenamefont {Schenke}\ \emph {et~al.}(2010)\citenamefont
  {Schenke}, \citenamefont {Jeon},\ and\ \citenamefont
  {Gale}}]{Schenke:2010nt}%
  \BibitemOpen
\bibfield  {journal} {  }\bibfield  {author} {\bibinfo {author} {\bibfnamefont
  {B.}~\bibnamefont {Schenke}}, \bibinfo {author} {\bibfnamefont
  {S.}~\bibnamefont {Jeon}}, \ and\ \bibinfo {author} {\bibfnamefont
  {C.}~\bibnamefont {Gale}},\ }\href {\doibase 10.1103/PhysRevC.82.014903}
  {\bibfield  {journal} {\bibinfo  {journal} {Phys. Rev.}\ }\textbf {\bibinfo
  {volume} {C82}},\ \bibinfo {pages} {014903} (\bibinfo {year} {2010})},\
  \Eprint {http://arxiv.org/abs/1004.1408} {arXiv:1004.1408 [hep-ph]}
  \BibitemShut {NoStop}%
\bibitem [{\citenamefont {Schenke}\ \emph {et~al.}(2011)\citenamefont
  {Schenke}, \citenamefont {Jeon},\ and\ \citenamefont
  {Gale}}]{Schenke:2010rr}%
  \BibitemOpen
  \bibfield  {author} {\bibinfo {author} {\bibfnamefont {B.}~\bibnamefont
  {Schenke}}, \bibinfo {author} {\bibfnamefont {S.}~\bibnamefont {Jeon}}, \
  and\ \bibinfo {author} {\bibfnamefont {C.}~\bibnamefont {Gale}},\ }\href
  {\doibase 10.1103/PhysRevLett.106.042301} {\bibfield  {journal} {\bibinfo
  {journal} {Phys. Rev. Lett.}\ }\textbf {\bibinfo {volume} {106}},\ \bibinfo
  {pages} {042301} (\bibinfo {year} {2011})},\ \Eprint
  {http://arxiv.org/abs/1009.3244} {arXiv:1009.3244 [hep-ph]} \BibitemShut
  {NoStop}%
\bibitem [{\citenamefont {Schenke}\ \emph {et~al.}(2012)\citenamefont
  {Schenke}, \citenamefont {Jeon},\ and\ \citenamefont
  {Gale}}]{Schenke:2011bn}%
  \BibitemOpen
  \bibfield  {author} {\bibinfo {author} {\bibfnamefont {B.}~\bibnamefont
  {Schenke}}, \bibinfo {author} {\bibfnamefont {S.}~\bibnamefont {Jeon}}, \
  and\ \bibinfo {author} {\bibfnamefont {C.}~\bibnamefont {Gale}},\ }\href
  {\doibase 10.1103/PhysRevC.85.024901} {\bibfield  {journal} {\bibinfo
  {journal} {Phys. Rev.}\ }\textbf {\bibinfo {volume} {C85}},\ \bibinfo {pages}
  {024901} (\bibinfo {year} {2012})},\ \Eprint {http://arxiv.org/abs/1109.6289}
  {arXiv:1109.6289 [hep-ph]} \BibitemShut {NoStop}%
\bibitem [{\citenamefont {Bass}\ \emph {et~al.}(1998)\citenamefont {Bass} \emph
  {et~al.}}]{Bass:1998ca}%
  \BibitemOpen
  \bibfield  {author} {\bibinfo {author} {\bibfnamefont {S.~A.}\ \bibnamefont
  {Bass}} \emph {et~al.},\ }\href@noop {} {\bibfield  {journal} {\bibinfo
  {journal} {Prog. Part. Nucl. Phys.}\ }\textbf {\bibinfo {volume} {41}},\
  \bibinfo {pages} {255} (\bibinfo {year} {1998})}\BibitemShut {NoStop}%
\bibitem [{\citenamefont {Bleicher}\ \emph {et~al.}(1999)\citenamefont
  {Bleicher} \emph {et~al.}}]{Bleicher:1999xi}%
  \BibitemOpen
  \bibfield  {author} {\bibinfo {author} {\bibfnamefont {M.}~\bibnamefont
  {Bleicher}} \emph {et~al.},\ }\href {\doibase 10.1088/0954-3899/25/9/308}
  {\bibfield  {journal} {\bibinfo  {journal} {J. Phys.}\ }\textbf {\bibinfo
  {volume} {G25}},\ \bibinfo {pages} {1859} (\bibinfo {year} {1999})},\ \Eprint
  {http://arxiv.org/abs/hep-ph/9909407} {arXiv:hep-ph/9909407 [hep-ph]}
  \BibitemShut {NoStop}%
\bibitem [{neo()}]{neos}%
  \BibitemOpen
  \href@noop {} {\bibinfo  {journal} {https://sites.google.com/view/qcdneos/}\
  }\BibitemShut {NoStop}%
\end{thebibliography}%

\end{document}